\documentclass[a4paper]{article}
\usepackage{amsmath}
\usepackage{amssymb}
\usepackage{amsfonts}
\usepackage{amsthm}
\usepackage{amsopn}
\usepackage{graphicx}
\usepackage{tikz}
\usepackage{url}
\usepackage{float}
\usepackage{hyphenat}
\usepackage[normalem]{ulem}

\def\be{\begin{eqnarray}}
\def\ee{\end{eqnarray}}
\def\nn{\nonumber}

\def\p{\partial}

\def\Tr{{\rm tr}\,}

\def\l[{\phantom.[}

\newtheorem{rmk}{Remark}
\newcommand\Theremark[1]{\begin{rmk} #1 \end{rmk}}

\hoffset= - 1.0in         

\begin{document}

\title{{\bf {Matrix model and dimensions at hypercube vertices
}\vspace{.2cm}}
\author{{\bf A.Morozov$^{a,b,c}$},\ {\bf And.Morozov$^{a,b,c,d}$} \ and \ {\bf A.Popolitov$^{a,b,e}$}}
\date{ }
}

\maketitle

\vspace{-4.5cm}

\begin{center}
\hfill IITP/TH-09/15
\end{center}

\vspace{3.3cm}

\begin{center}

$^a$ {\small {\it ITEP, Moscow 117218, Russia}}\\
$^b$ {\small {\it Institute for Information Transmission Problems, Moscow 127994, Russia}}\\
$^c$ {\small {\it National Research Nuclear University MEPhI, Moscow 115409, Russia }}\\
$^d$ {\small {\it Laboratory of Quantum Topology, Chelyabinsk State University, Chelyabinsk 454001, Russia}} \\
$^e$ {\small {\it Korteweg-de Vries Institute for Mathematics, University of
Amsterdam, P.O. Box 94248, 1090 GE Amsterdam, The Netherlands}}
\end{center}


\vspace{1cm}

\centerline{ABSTRACT}

\bigskip

{\footnotesize
In hypercube approach to correlation functions in Chern-Simons theory
(knot polynomials) the central role is played by the numbers of cycles,
in which the link diagram is decomposed under different resolutions.
Certain functions of these numbers  are further
interpreted as dimensions of graded spaces, associated with hypercube vertices.
Finding these functions is, however,  a somewhat non-trivial problem.
In \cite{MMP1}  it was suggested to solve it with the help of
the matrix model technique, in the spirit of AMM/EO topological recursion.
In this paper we further elaborate on this idea and provide a vast
collection of non-trivial examples, related both to ordinary and virtual
links and knots.
Remarkably, most powerful versions of the formalism freely convert
ordinary knots/links to virtual and back -- moreover, go beyond the
knot-related set of the (2,2)-valent graphs.
}

\bigskip

\bigskip

\newpage

\tableofcontents

\newpage

\section{Introduction}

Knot polynomials \cite{knotpols}
(Wilson-loop averages in Chern-Simons theory \cite{CS}
and their various deformations \cite{virt}-\cite{ArtSha})
are currently among the hot topics in theoretical physics.
They belong to the important class of Hurwitz $\tau$-functions \cite{Hurtau},
and will supposedly inherit the main properties of the matrix-model $\tau$-functions
\cite{mamo,AMMEO}, which currently unify all the known special functions and provide
a basis for analytical description of Nature.
There is a number of ways to study and even define the knot polynomials,
coming from different overlapping sciences, which they are supposed to put together:
representation theory, topology, modular transformations of conformal blocks,
integrable systems, Morse theory etc.
The most intriguing of these is categorification program \cite{cate}, originated in this
context by M.Khovanov \cite{Kh}, which is physically interpreted as certain lifting to
higher dimensions (when quantities in question are associated with stable points
of some new evolution \cite{KW}).
Detailed presentation of these ideas is far beyond the scope of the present text.
Here we concentrate on technical issues, needed to building such kind of a formalism
and making it computationally effective.
For this purpose we rely upon the two relatively old advances -- hypercube approach
\cite{KRma}-\cite{DM12} and its considerable simplification in \cite{DM3} -- complemented
by the recently suggested \cite{MMP1} matrix-model-style reinterpretation.
This allows us to practically solve the problem of calculating quantum dimensions
of "spaces" at hypercube vertices,
which are certain clever constituents of HOMFLY polynomials.
This opens a way to a simple interpretation of cut-and-join morphisms
and consequently -- to a simple homology calculus, giving rise to Khovanov-Rozansky
superpolynomials, -- though this part of the story is also a subject for future
investigations.
However, already in this text we reveal new far-going relations between different branches
of knot theory itself and establish new connections to other sections of mathematical
physics.
Most spectacular, the what looks like a relevant matrix-model recursion
is, first, immediately related to Reidemeister invariance, when restricted to
ordinary knots, while in general seems to mix not only ordinary knots with virtual ones,
but lifts the entire theory to graphs of rather general type, far beyond the usual
valence four.
Amusingly, this appears not just as an attractive fantasy or a vision
(as it was in original \cite{DMnla} and \cite{DM3}),
but as a necessary thing to do in construction of a practically working computer program,
aimed at calculating knot polynomials by the hypercube method.
Further work in this directions will definitely lead to new insights, both conceptual
and practical.

\section{The DM approach}

In this section we remind the main ingredients of the new version \cite{DM3}
of the hypercube approach \cite{KRma}-\cite{DM12}, which was further developed in
\cite{AnoMKhR}, \cite{virtDM3} and \cite{MMP1}.

\subsection{Seifert decomposition of link diagram and fat graphs \label{graphs}}

Despite their name, knot/link polynomials are actually
associated not so much with links and knots as with graphs --
and this emphasizes that the story is closely related to lattice theories.
The main difference is that graphs are of rather general type, far from regular
periodic lattices.
The possibility of  interpretation in terms of knots and links is provided by Reidemeister invariances,
of which the main is Yang-Baxter relation -- and therefore relevant are
the so called {\it integrable} lattice theories.
Directly related to links are oriented $(2,2)$-valent graphs,
which are interpreted as link diagrams ${\cal L}$ -- projections of oriented lines
from three to two dimensions.
For links in $R_3$ ({\it ordinary} links)  the link diagrams are planar,
non-planar diagrams are associated with {\it virtual} links -- those in
non-simply-connected $3d$ space-times.
As clear from their interpretations as projections,
link diagrams can have vertices of two types (colors) -- black and white.
In the Reshetikhin-Turaev (RT) formalism \cite{RT}-\cite{RTmod}
the two colors are associated with ${\cal R}$-matrix and its inverse,
and in original version of \cite{RT} what also matters is a choice of direction in the plane --
then "turning matrices" $q^{\pm \rho}$ are inserted at the turning points.
However, in fact these insertions are rather associated with Seifert cycles and are taken into account
by the properly defined traces (or dimensions -- in the context of the present paper)
so that distinguished direction does not show up in modern versions of RT formalism.
As to Seifert cycles, they appear when each vertex of ${\cal L}$ is resolved in the
following way:
\be
\begin{picture}(300,75)(-50,-40)
\put(0,0){\vector(1,1){30}}
\put(30,0){\line(-1,1){12}}
\put(12,18){\vector(-1,1){12}}
\put(40,13){\mbox{$=$}}
\put(70,15){\circle*{6}}
\put(55,0){\vector(1,1){12}}
\put(73,18){\vector(1,1){12}}
\put(85,0){\vector(-1,1){12}}
\put(67,18){\vector(-1,1){12}}
\put(0,-40){\line(1,1){12}}
\put(30,-40){\vector(-1,1){30}}
\put(18,-22){\vector(1,1){12}}
\put(40,-27){\mbox{$=$}}
\put(70,-25){\circle{6}}
\put(55,-40){\vector(1,1){12}}
\put(73,-22){\vector(1,1){12}}
\put(85,-40){\vector(-1,1){12}}
\put(67,-22){\vector(-1,1){12}}
\put(150,-7){\vector(1,0){30}}
\put(130,5){\mbox{Seifert resolution}}
\put(10,0){
\qbezier(230,-20)(250,-5)(230,10)
\qbezier(260,-20)(240,-5)(260,10)
\put(240,-5){\vector(0,1){2}}
\put(250,-5){\vector(0,1){2}}
}
\end{picture}
\label{Seireso}
\ee
\noindent
Then the valence-$(2,2)$ graph turns into a collection of $\nu^{\cal L}$ non-intersecting
oriented closed lines, which we call Seifert cycles.
In the following picture the link diagrams for  two component links,
(virtual in the upper part of the picture and ordinary -- in the lower one)
are decomposed respectively into two and three Seifert cycles:

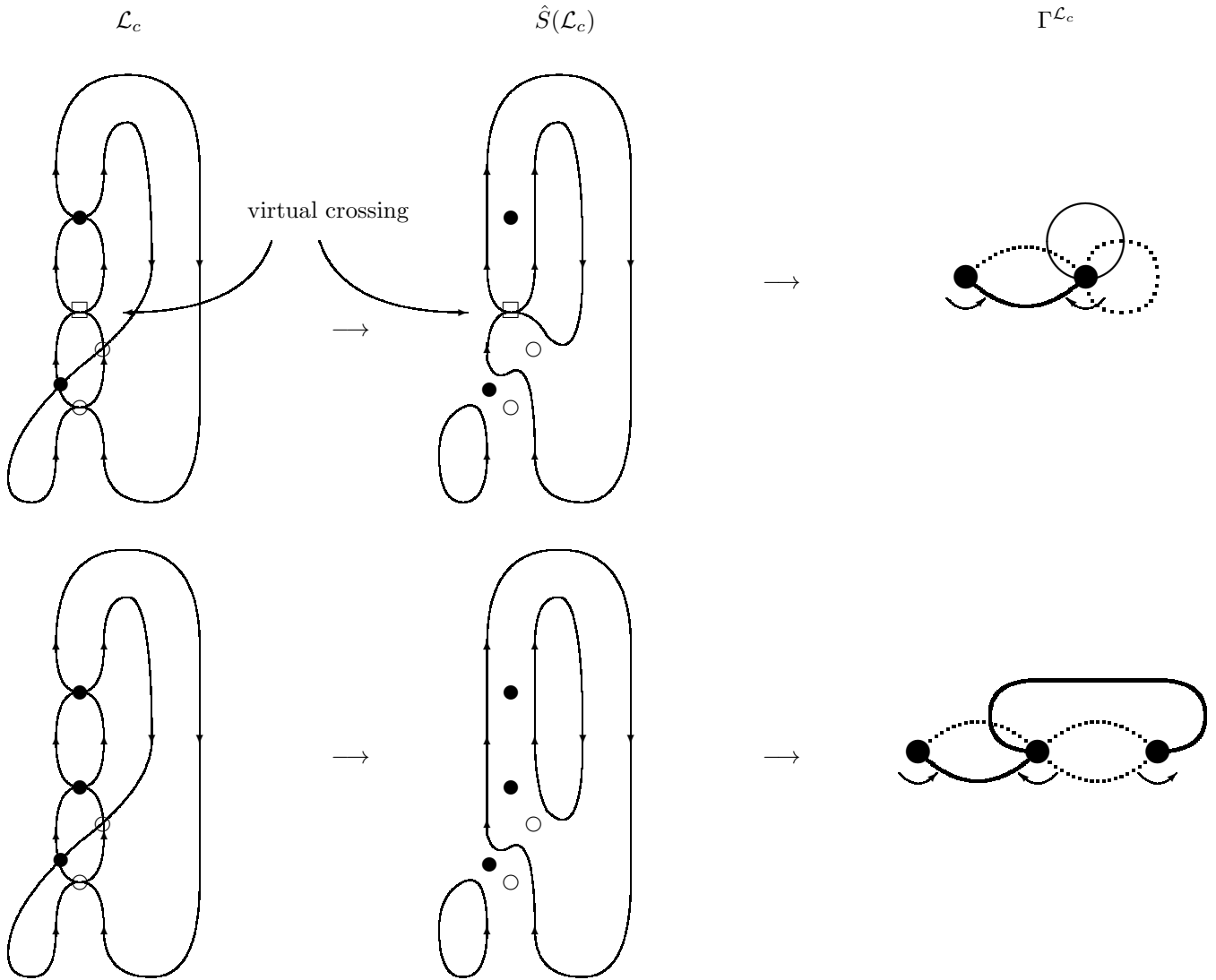
\newcommand\GLC[0]{\Gamma^{{\cal L}_c}}

\begin{picture}(300,430)(-20,-250)
%
%
\put(15,160){\mbox{${\cal L}_c$}}
\put(190,160){\mbox{$\hat S({\cal L}_c)$}}
\put(400,160){\mbox{$\GLC$}}
\qbezier(-10,-20)(-10,0)(0,0)
\qbezier(10,-20)(10,0)(0,0)
\qbezier(-10,20)(-10,0)(0,0)
\qbezier(10,20)(10,0)(0,0)
\qbezier(-10,20)(-10,40)(0,40)
\qbezier(10,20)(10,40)(0,40)
\qbezier(-10,60)(-10,40)(0,40)
\qbezier(10,60)(10,40)(0,40)
\qbezier(-10,60)(-10,80)(0,80)
\qbezier(10,60)(10,80)(0,80)
\qbezier(-10,100)(-10,80)(0,80)
\qbezier(10,100)(10,80)(0,80)
\qbezier(10,100)(10,120)(20,120)
\qbezier(20,120)(30,120)(30,60)
\qbezier(-10,-20)(-10,-40)(-20,-40)
\qbezier(-20,-40)(-30,-40)(-30,-30)
\qbezier(30,60)(30,40)(0,17)
\qbezier(-30,-30)(-30,-10)(0,17)
\qbezier(10,-20)(10,-40)(30,-40)
\qbezier(30,-40)(50,-40)(50,0)
\qbezier(-10,100)(-10,140)(20,140)
\qbezier(20,140)(50,140)(50,100)
\put(50,0){\line(0,1){100}}
\put(20,40){
\qbezier(0,0)(50,0)(60,30)\put(0,0){\vector(-1,0){2}}
\qbezier(80,30)(90,0)(140,0)\put(140,0){\vector(1,0){2}}
\put(50,40){\mbox{virtual crossing}}
\put(-24,-2.5){\mbox{$\Box$}}\put(156,-2.5){\mbox{$\Box$}}
}
\put(0,0){\circle{6}}
\put(0,80){\circle*{6}}
\put(-8,9.5){\circle*{6}}
\put(9.5,24.5){\circle{6}}
\put(-10,20){\vector(0,1){2}}
\put(-10,60){\vector(0,1){2}}
\put(-10,100){\vector(0,1){2}}
\put(10,20){\vector(0,1){2}}
\put(10,60){\vector(0,1){2}}
\put(-10,100){\vector(0,1){2}}
\put(10,100){\vector(0,1){2}}
\put(-10,-20){\vector(0,1){2}}
\put(10,-20){\vector(0,1){2}}
\put(30,60){\vector(0,-1){2}}
\put(50,60){\vector(0,-1){2}}
\put(105,30){\mbox{$\longrightarrow$}}
\put(180,0){
\qbezier(-10,-20)(-10,5)(-20,0)
\qbezier(10,-20)(10,20)(0,15)
\qbezier(-10,20)(-7,10)(0,15)
\qbezier(-10,20)(-10,40)(0,40)
\qbezier(15,30)(10,40)(0,40)
\qbezier(-10,60)(-10,40)(0,40)
\qbezier(10,60)(10,40)(0,40)
\qbezier(-10,60)(-10,80)(-10,100)
\qbezier(10,60)(10,80)(10,100)
%
\qbezier(10,100)(10,120)(20,120)
\qbezier(20,120)(30,120)(30,60)
\qbezier(-10,-20)(-10,-40)(-20,-40)
\qbezier(-20,-40)(-30,-40)(-30,-20)
\qbezier(30,60)(30,15)(15,30)
\qbezier(-30,-20)(-30,-5)(-20,0)
\qbezier(10,-20)(10,-40)(30,-40)
\qbezier(30,-40)(50,-40)(50,0)
\qbezier(-10,100)(-10,140)(20,140)
\qbezier(20,140)(50,140)(50,100)
\put(50,0){\line(0,1){100}}
\put(0,0){\circle{6}}
\put(0,80){\circle*{6}}
\put(-9,7.5){\circle*{6}}
\put(9.5,24.5){\circle{6}}
\put(-10,25){\vector(0,1){2}}
\put(-10,60){\vector(0,1){2}}
\put(-10,100){\vector(0,1){2}}
\put(10,60){\vector(0,1){2}}
\put(-10,100){\vector(0,1){2}}
\put(10,100){\vector(0,1){2}}
\put(-10,-20){\vector(0,1){2}}
\put(10,-20){\vector(0,1){2}}
\put(30,60){\vector(0,-1){2}}
\put(50,60){\vector(0,-1){2}}
}
\put(0,-200){
\qbezier(-10,-20)(-10,0)(0,0)
\qbezier(10,-20)(10,0)(0,0)
\qbezier(-10,20)(-10,0)(0,0)
\qbezier(10,20)(10,0)(0,0)
\qbezier(-10,20)(-10,40)(0,40)
\qbezier(10,20)(10,40)(0,40)
\qbezier(-10,60)(-10,40)(0,40)
\qbezier(10,60)(10,40)(0,40)
\qbezier(-10,60)(-10,80)(0,80)
\qbezier(10,60)(10,80)(0,80)
\qbezier(-10,100)(-10,80)(0,80)
\qbezier(10,100)(10,80)(0,80)
\qbezier(10,100)(10,120)(20,120)
\qbezier(20,120)(30,120)(30,60)
\qbezier(-10,-20)(-10,-40)(-20,-40)
\qbezier(-20,-40)(-30,-40)(-30,-30)
\qbezier(30,60)(30,40)(0,17)
\qbezier(-30,-30)(-30,-10)(0,17)
\qbezier(10,-20)(10,-40)(30,-40)
\qbezier(30,-40)(50,-40)(50,0)
\qbezier(-10,100)(-10,140)(20,140)
\qbezier(20,140)(50,140)(50,100)
\put(50,0){\line(0,1){100}}
\put(0,0){\circle{6}}
\put(0,80){\circle*{6}}
\put(-8,9.5){\circle*{6}}
\put(9.5,24.5){\circle{6}}
\put(0,40){\circle*{6}}
\put(-10,20){\vector(0,1){2}}
\put(-10,60){\vector(0,1){2}}
\put(-10,100){\vector(0,1){2}}
\put(10,20){\vector(0,1){2}}
\put(10,60){\vector(0,1){2}}
\put(-10,100){\vector(0,1){2}}
\put(10,100){\vector(0,1){2}}
\put(-10,-20){\vector(0,1){2}}
\put(10,-20){\vector(0,1){2}}
\put(30,60){\vector(0,-1){2}}
\put(50,60){\vector(0,-1){2}}
}
\put(0,-200){
\put(105,50){\mbox{$\longrightarrow$}}
\put(180,0){
\qbezier(-10,-20)(-10,5)(-20,0)
\qbezier(10,-20)(10,20)(0,15)
\qbezier(-10,20)(-7,10)(0,15)
\qbezier(-10,20)(-10,40)(-10,100)
\qbezier(10,60)(10,40)(15,30)
\qbezier(10,60)(10,80)(10,100)
%
\qbezier(10,100)(10,120)(20,120)
\qbezier(20,120)(30,120)(30,60)
\qbezier(-10,-20)(-10,-40)(-20,-40)
\qbezier(-20,-40)(-30,-40)(-30,-20)
\qbezier(30,60)(30,15)(15,30)
\qbezier(-30,-20)(-30,-5)(-20,0)
\qbezier(10,-20)(10,-40)(30,-40)
\qbezier(30,-40)(50,-40)(50,0)
\qbezier(-10,100)(-10,140)(20,140)
\qbezier(20,140)(50,140)(50,100)
\put(50,0){\line(0,1){100}}
\put(0,0){\circle{6}}
\put(0,80){\circle*{6}}
\put(-9,7.5){\circle*{6}}
\put(9.5,24.5){\circle{6}}
\put(0,40){\circle*{6}}
\put(-10,25){\vector(0,1){2}}
\put(-10,60){\vector(0,1){2}}
\put(-10,100){\vector(0,1){2}}
\put(10,60){\vector(0,1){2}}
\put(-10,100){\vector(0,1){2}}
\put(10,100){\vector(0,1){2}}
\put(-10,-20){\vector(0,1){2}}
\put(10,-20){\vector(0,1){2}}
\put(30,60){\vector(0,-1){2}}
\put(50,60){\vector(0,-1){2}}
}
}
\put(285,50){\mbox{$\longrightarrow$}}
\put(285,-150){\mbox{$\longrightarrow$}}
\put(370,55){
\qbezier(-8,-9)(0,-18)(8,-9)\put(6,-11){\vector(1,1){2}}
\qbezier(42,-9)(50,-18)(58,-9)\put(44.5,-11){\vector(-1,1){2}}
}
\put(350,-145){
\qbezier(-8,-9)(0,-18)(8,-9)\put(6,-11){\vector(1,1){2}}
\qbezier(42,-9)(50,-18)(58,-9)\put(44.5,-11){\vector(-1,1){2}}
\qbezier(92,-9)(100,-18)(108,-9)\put(106,-11){\vector(1,1){2}}
}
\thicklines \linethickness{0.4mm}
\put(370,55){
\put(0,0){\circle*{10}}
\put(50,0){\circle*{10}}
\qbezier[20](0,0)(25,25)(50,0)
\qbezier(0,0)(25,-25)(50,0)
\put(50,15){\circle{30}}
%
%
\qbezier[7](50,0)(50,15)(65,15)\qbezier[7](65,15)(80,15)(80,0)
\qbezier[7](50,0)(50,-15)(65,-15)\qbezier[7](65,-15)(80,-15)(80,0)
%
}
\put(350,-145){
  \put(0,0){\circle*{10}}
  \put(50,0){\circle*{10}}
  \put(100,0){\circle*{10}}
  \qbezier[20](0,0)(25,25)(50,0)
  \qbezier(0,0)(25,-25)(50,0)
  \qbezier[20](100,0)(75,25)(50,0)
  \qbezier[20](100,0)(75,-25)(50,0)
  \qbezier(50,0)(30,0)(30,15)
  \qbezier(30,15)(30,30)(50,30)
  \qbezier(100,0)(120,0)(120,15)
  \qbezier(120,15)(120,30)(100,30)
  \qbezier(50,30)(75,30)(100,30)
  %
}
\end{picture}

\noindent
Note, that according to our definition, Seifert decomposition does not depend on the colors of vertices,
while virtual (sterile) crossings are not resolved, i.e. for non-planar (virtual) diagrams
Seifert cycles are also non-planar.
In this picture and in what follows we mark virtual crossings by boxes.

Following \cite{MMP1}, in the third column we converted the Seifert-resolved diagram $\hat S({\cal L}_c)$
into a graph $\GLC$, which has $\, \nu^{\cal L}\,$ Seifert cycles as vertices and
original $n^{\cal L}$ crossings as edges.
To preserve information about ${\cal L}$, we preserve the order, in which edges are attached to the
vertices, i.e. $\Gamma$ is actually a {\it fat} graph ({\it dessin d'enfant}).
In the second column one could actually substitute dots at positions of resolved vertices
by strips -- edges of the fat graph $\Gamma$.
In other words, black and white crossings of the first column, i.e.
black and white dots  at positions of resolved vertices in the second column,
become respectively dotted and straight  thick lines (strips) in the third one.
To preserve the continuity of arrows,
these strips are actually twisted (Mobius-like), what is a peculiar feature of the
corresponding matrix model, see \cite{MMP1} and sec.\ref{mamoc} below.
In above pictures one can substitute twisting by orientation of vertices,
but this does not work in more complicated examples.
Fat graphs $\Gamma$ will appear in  eq.(\ref{HMMP}) below and start playing a big role afterwards.

For alternative (perhaps, parallel) appearance of fat graphs in knot theory see
\cite{fatgraphs}.

\subsection{Colorings and hypercube}

We denote link diagram with colored vertices through ${\cal L}_c$ -- a diagram ${\cal L}$ with $n^{\cal L}$
vertices can have $2^{n}$ different colorings $c$
(from now on the quantities, which depend on the link diagram,
will carry ${\cal L}$ as a superscript, but sometimes
we omit it to make formulas readable).
One can also consider two different resolutions of vertices
-- again there will be $2^{n}$ variants to resolve/decompose ${\cal L}$.
With ${\cal L}$ we associate an $n^{\cal L}$-dimensional hypercube ${\cal H}^{\cal L}$
whose $2^{n}$ vertices correspond to different resolutions and each of the
$2^{n-1}n$ edges -- to a flip between two different resolutions of particular
vertex of ${\cal L}$.
Hypercube has a distinguished {\it Seifert} vertex $s$, when all vertices are resolved as (\ref{Seireso})
and ${\cal L}$ is decomposed into $\nu^{\cal L}$ Seifert cycles.
The opposite hypercube vertex $\bar s$ is called anti-Seifert.

Note that our hypercube ${\cal H}^{\cal L}$ is independent of the coloring $c$,
its vertices describe different resolutions of ${\cal L}$.
Coloring $c$ is associated with particular vertex of the hypercube (named {\it initial} vertex
for a given knot/link) --
by formally identifying two vertex colors with the two resolutions.

\subsection{Resolutions and dimensions}

The idea of hypercube formalism is to put at the hypercube vertices some objects,
defined by the corresponding resolutions, and associate morphisms between these objects
with the resolution flips along the edges of the hypercube.
If these morphisms are commuting, the hypercube turns into Abelian quiver,
which can be further reduced to a complex and one can construct its Euler and Poincare
polynomials.
If the entire construction respects Reidemeister moves, these quantities can be
interpreted as  HOMFLY and Khovanov-Rozansky   polynomials for the knot/link,
which was described by the link diagram ${\cal L}_c$.

In the original version of \cite{Kh}, for ordinary (non-virtual) links and for the group
$SL(2)$ (i.e. for $N=2$), the {\it objects} were just $q$-graded vector spaces
and {\it morphisms} were defined as cut-and-join operators, associated with the flips

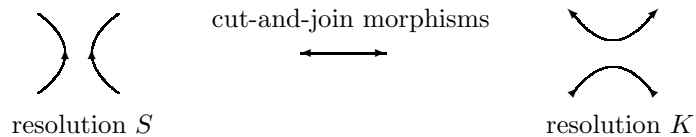
\begin{picture}(300,60)(-130,-45)
\put(-240,0){
\qbezier(230,-20)(250,-5)(230,10)
\qbezier(260,-20)(240,-5)(260,10)
\put(240,-5){\vector(0,1){2}}
\put(250,-5){\vector(0,1){2}}
\put(330,-5){\vector(1,0){30}}\put(330,-5){\vector(-1,0){2}}
\put(50,0){
\qbezier(380,-20)(395,0)(410,-20)
\qbezier(380,10)(395,-10)(410,10)
\put(380,10){\vector(-1,1){2}}
\put(410,10){\vector(1,1){2}}
\put(380,-20){\vector(1,1){2}}
\put(410,-20){\vector(-1,1){2}}
}
\put(220,-35){\mbox{resolution $S$}}
\put(420,-35){\mbox{resolution $K$}}
\put(295,5){\mbox{cut-and-join morphisms}}
}
\end{picture}

\noindent
Each hypercube vertex $c\in {\cal H}^{\cal L}$ corresponds to some coloring of the diagram ${\cal L}$.
If black and white vertices are resolved as $S$ and $K$, then the diagram decomposes into $\nu_c^K$
cycles, and the corresponding vector space has quantum ($q$-graded) dimension
$\ [N]^{\nu_c^K} \stackrel{N=2}{=}\ [2]^{\nu_c^K}$,
where $[N] = \frac{q^N-q^{-N}}{q-q^{-1}}= \frac{\{q^N\}}{\{q\}}$ is the quantum number $N$.
Morphisms are linear maps of  grade $-1$ and can be easily constructed in explicit form,
leading (for $N=2$) to Reidemeister-invariant Jones and Khovanov polynomials.

However, it is clear from above picture that the resolution $K$ does not respect arrows,
thus it can be used only for $N=2$, when orientation does not matter.
For generic $N$ the suggestion of \cite{DM3} was to substitute $K$ by a linear combination:

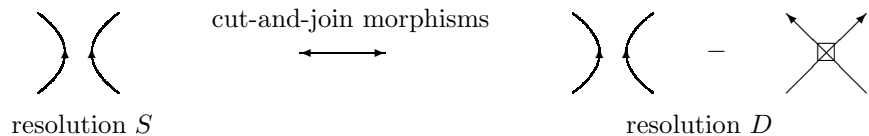
\begin{picture}(300,65)(-100,-45)
\put(-240,0){
\qbezier(230,-20)(250,-5)(230,10)
\qbezier(260,-20)(240,-5)(260,10)
\put(240,-5){\vector(0,1){2}}
\put(250,-5){\vector(0,1){2}}
\put(330,-5){\vector(1,0){30}}\put(330,-5){\vector(-1,0){2}}
\put(200,0){
\qbezier(230,-20)(250,-5)(230,10)
\qbezier(260,-20)(240,-5)(260,10)
\put(240,-5){\vector(0,1){2}}
\put(250,-5){\vector(0,1){2}}
}
\put(480,-7){\mbox{$-$}}
\put(100,0){
\put(410,-20){\vector(1,1){30}}
\put(440,-20){\vector(-1,1){30}}
\put(421,-8){\mbox{$\Box$}}
}
\put(220,-35){\mbox{resolution $S$}}
\put(450,-35){\mbox{resolution $D$}}
\put(295,5){\mbox{cut-and-join morphisms}}
}
\end{picture}

\noindent
and there is already a lot of evidence that this indeed provides Reidemeister-invariant
HOMFLY and Khovanov-Rozansky polynomials.
Unfortunately, exhaustive description is not so straightforward,
because there is a number of technical complications as compared to the case of $N=2$ --
which, however, seem conceptually important and unavoidable.

First, only at the Seifert vertex $v=s$ dimension remains a monomial $d_s^{\cal L} = [N]^{\nu^{\cal L}}$.
Whenever the second resolution ($D$) is involved, dimension $d_v^{\cal L}$
is a polynomial in $N$, even at $q=1$.

Second, for virtual knots and links this polynomial need not be positive,
and negative dimensions can require interpretation in terms of K-theory.

Third, associated spaces are now similar to factor-spaces,
with quantization (gradings) and morphisms somewhat more involved than they were for
structure-less graded vector spaces.

All this makes description of dimensions and morphisms a challenging problem,
and the present paper is one more step towards its resolution.

\subsection{HOMFLY polynomial}

According to \cite{DM3} HOMFLY in the fundamental representation is given by the sum
\be
\label{eq:hypercube-sum-with-distinguished-vertex}
H^{{\cal L}_c}_{_\Box}=
{\cal C}^{{\cal L}_c}
\sum_{v\in {\cal H}^{\cal L}}^{2^n}
(-q)^{h(v,c)}\cdot d_v^{{\cal L}}(q,N)
\label{HDM3}
\ee
over all vertices $v$ of the hypercube ${\cal H}^{\cal L}$.
While hypercube and dimensions depend only on {\it un}colored link diagram ${\cal L}$,
HOMFLY depends on its coloring
-- but in (\ref{HDM3}) this dependence appears only through the weight factor $(-q)^{h(v,c)}$
where exponent is the number of hypercube edges in the shortest path from "initial" vertex $c$ to $v$
-- the only non-trivial element in the r.h.s. which depends on $c$.
In topological framing there is also a color-dependent common factor in front of the sum
over $v$:
\be
{\cal C}^{{\cal L}_c} = q^{(N-1)n_\bullet}\cdot (-q^{-N})^{n_\circ},
\ee
where $n_\bullet$ and $n_\circ$ are total number of black and white vertices
of link diagram ${\cal L}_c$, respectively.

\bigskip

At $q=1$ the explicit form of the resolution $D$ implies that dimension
\be
\left.d_v^{\cal L}\right|_{q=1} = \sum_{s\leq w \leq v} (-)^{h(w,s)} N^{\nu_w^\times}
\label{cladim}
\ee
is a sum over all vertices of a sub-cube in ${\cal H}^{\cal L}$,
located between $v$ and the Seifert vertex $s$, and $\nu_w^\times$ is a number of cycles,
arising when white vertices for the coloring $w$ are resolved as
\begin{picture}(70,16)(-12,-3)
\put(0,2){
\put(-10,-10){\vector(1,1){20}}\put(10,-10){\vector(-1,1){20}}\put(0,0){\circle{6}}
\put(15,-2){\mbox{$\longrightarrow$}}
\put(45,0){
\put(-10,-10){\vector(1,1){20}}\put(10,-10){\vector(-1,1){20}}\put(-4,-3){\mbox{$\Box$}}
}
}
\end{picture}.
Note that $\nu_w^\times \neq \nu_w^K$, and there is nothing like the number $\nu_w^D$
of cycles, associated with the composite resolution $D$ of the white vertices at $w$.
After substitution of (\ref{cladim}), expression (\ref{HDM3}) becomes a double sum
\be
\left.H^{{\cal L}_c}_{_\Box}\right|_{q=1} =
\sum_{v\in {\cal H}^{\cal L}} (-)^{h(v,c)}\sum_{s\leq w \leq v} (-)^{h(w,s)} N^{\nu_w^\times}
=\sum_{w\in {\cal H}^{\cal L}}  (-)^{h(w,s)} N^{\nu_w^\times} \sum_{w\leq v\leq \bar s}(-)^{h(v,c)}
= (-)^{h(\bar s,c)+h(\bar s,s)} N^{\nu_{\bar s}^\times}
\ee
Since the sum over $v$ is non-vanishing only for $w=\bar s$,
the double sum reduces to a single item  at anti-Seifert vertex,
where $\nu_{\bar s}^\times=l^{\cal L}$ is actually the number of link components in ${\cal L}$
However, at $q\neq 1$ the two sums are quantized ($q$-deformed) in a very different way,
cancelations are far less pronounced and double sum provides HOMFLY polynomial
as a non-trivial quantization of a trivial $N^{l}$.
Of course, the key point here is  the quantization of dimensions (\ref{cladim}).

\subsection{Quantization of $d_v^{\cal L}$}

There are several levels of understanding for this procedure.

First, \cite{DM3}, in the simplest cases the sums (\ref{cladim}) provide
polynomials in $N$ with all roots integer, i.e. factorize in the products of
monomials $N-k$. Naturally, such products are trivially $q$-deformed into
the same products of quantum numbers $[N-k]$.

Second, \cite{DM3}, one can use the fact, that dimensions are independent of coloring.
One and the same diagram ${\cal L}$ for different colorings can describe
both complicated and relatively simple knots/links, sometimes as simple as the unknot.
If one assumes topological invariance of (\ref{HDM3}) this idea provides a recursion
procedure for making more complicated dimensions from the simpler ones.

Third, \cite{AnoMKhR}, one can deduce dimensions from RT formalism \cite{RT}-\cite{RTmod},
where they are actually traces of combinations of projectors $P_{[11]}$ on antisymmetric
representation in the product of two fundamental representations.
Though there is no {\it a priori} reason for this formalism to be applicable
to virtual knots, it perfectly works there.
Moreover, the results imply, that representation theory turns
inconsistent with virtual crossings only in symmetric channel,
while no problems occur in the antisymmetric one -- see \cite{virtDM3} for impressive
examples.

Fourth, \cite{MMP1}, classical dimensions (\ref{cladim}) depend only on the
sub-cube between $v$ and $s$, not on the entire hypercube -- and therefore not
on entire ${\cal L}$, and  quantization preserves this property.
A formalism, which takes this fact into account and allows to properly separate
the variables, makes use of the fat graphs $\GLC$, introduced in above sec.\ref{graphs}.
As additional bonus, this formalism introduces the matrix-model language and intuition,
allowing to systematize various recursion ideas and observations.

\subsection{HOMFLY through fat graph \label{sec:fat-graph->homfly}}

In \cite{MMP1} we suggested to convert (\ref{HDM3}) into a different sum --
over {\it all} $2^{n^{\cal L}}$ subgraphs $\bar\gamma$ of the fat graph $\GLC$
with $n^{\cal L}$ edges:
\be
\label{eq:hypercube-sum-over-fatgraphs}
H^{{\cal L}_c}_{_\Box}
= {\cal C}^{{\cal L}_c}\!\!\!\!
\sum_{\bar\gamma \in \{\text{flipped}\ \GLC\}}
\!\!\!\! (-q)^{\# \text{of flips}}
\cdot D_{\bar\gamma}(q,N)
\label{HMMP}
\ee
We remind from sec.\ref{graphs} that the vertices of $\GLC$ are the $\nu^{\cal L}$
Seifert cycles of ${\cal L}$, while its $n^{\cal L}$ edges are the non-virtual crossings
in ${\cal L}$.
Coloring $c$ makes some of the edges dotted.
The sum in (\ref{HMMP}) is  over all color-flips: substitution of straight edges
by dotted ones and vice-versa -- of dotted by straight.
However, dimensions do not feel dotted lines:
\be \label{eq:amp-convention}
D_{\bar\gamma} = D_{\bar\gamma_{amp}}
\ee
where $\bar\gamma_{amp}$ is  the graph $\bar\gamma$ with all dotted lines erased (amputated).

\bigskip

An elementary combinatorial exercise shows, that this color-flip sum can
be rewritten as a somewhat different edge-amputation sum:
\be
\boxed{
\  H^{{\cal L}_c}_{_\Box}
=   q^{(N-1)\left( n_{..}(\GLC)-n_-(\GLC)\right)}
\sum_{\gamma\subseteq \GLC} (-q)^{n_{..}(\gamma) - n_-(\gamma)}
\cdot D_\gamma(q,N)
\phantom{5^{5^{5^{5^{5^5}}}}}\!\!\!\!\!\!\!\!\!\!\!\!
}
\label{HMMPmod}
\ee
Here $n_-(\gamma)$ and $n_{..}(\gamma)$ denote the numbers of straight and dotted lines in
the graph $\gamma$.
In variance with ${\cal C}^{{\cal L}_c}$,  exponent in the
prefactor in this formula is just antisymmetric in the number of straight and dotted edges,
$n_-(\GLC)= n_\bullet(\GLC)$ and $n_{..}(\GLC) = n_\circ(\GLC)$.
This time the sum is over
all subgraphs $\gamma$ of $\GLC$ with the same set of vertices, but some edges
removed, while the type of remaining edges is kept -- it enters the
formula through $n_-(\gamma)$ and $n_{..}(\gamma)$.
Finally, instead of not feeling dotted lines at all,
dimensions $D_\gamma$ now feel them exactly as straight ones
\be \label{eq:colorless-convention}
D_{\gamma} = D_{\gamma_{colorless}},
\ee
where $\gamma_{colorless}$ is $\gamma$ with color of its edges forgotten.

This seemingly innocuous re-summation and change of convention has important
consequence: contribution of a particular subgraph $\gamma$ to the HOMFLY
polynomial, i.e. {\it both} the $(-q)$-prefactor and quantum dimension, are completely
determined by $\gamma$ itself and not by original $\GLC$.
If the same $\gamma$ shows up
in two sums for different link diagrams ${\cal L}_1$ and ${\cal L}_2$,
its contribution is exactly \textit{the same} in both cases.
This is not the case for the original   formula \eqref{HMMP},
where quantum dimension are the same, but the $(-q)$-factor can be different
-- it depends on which edges were color-flipped.

\bigskip

\noindent
\textbf{Example: Seifert vertex for (2,4) torus knot}

\noindent color-flip summation:

\begin{picture}(300,35)(20,-15)
%
\put(0,0){\circle*{6}}
\put(20,0){\circle*{6}}
\qbezier[15](0,0)(10,21)(20,0)
\qbezier[15](0,0)(10,7)(20,0)
\qbezier[15](0,0)(10,-7)(20,0)
\qbezier[15](0,0)(10,-21)(20,0)
%
\put(26,-2){\mbox{$-4q$}}
\put(50,0){
\put(0,0){\circle*{6}}
\put(20,0){\circle*{6}}
\qbezier[15](0,0)(10,21)(20,0)
\qbezier[15](0,0)(10,7)(20,0)
\qbezier[15](0,0)(10,-7)(20,0)
\qbezier(0,0)(10,-21)(20,0)
}
\put(76,-2){\mbox{$+6q^2$}}
\put(100,0){
\put(0,0){\circle*{6}}
\put(20,0){\circle*{6}}
\qbezier[15](0,0)(10,21)(20,0)
\qbezier[15](0,0)(10,7)(20,0)
\qbezier(0,0)(10,-7)(20,0)
\qbezier(0,0)(10,-21)(20,0)
}
%
\put(127,-3){\mbox{$-4q^3$}}
\put(150,0){
\put(0,0){\circle*{6}}
\put(20,0){\circle*{6}}
\qbezier[15](0,0)(10,21)(20,0)
\qbezier(0,0)(10,7)(20,0)
\qbezier(0,0)(10,-7)(20,0)
\qbezier(0,0)(10,-21)(20,0)
}
\put(178,-2){\mbox{$+q^4$}}
\put(200,0){
\put(0,0){\circle*{6}}
\put(20,0){\circle*{6}}
\qbezier(0,0)(10,21)(20,0)
\qbezier(0,0)(10,7)(20,0)
\qbezier(0,0)(10,-7)(20,0)
\qbezier(0,0)(10,-21)(20,0)
}
\put(240,-2){\mbox{$\stackrel{(\ref{eq:amp-convention})}{=}$}}
\put(270,0){
\put(0,0){\circle*{6}}
\put(20,0){\circle*{6}}
%
\put(26,-2){\mbox{$-4q$}}
\put(50,0){
\put(0,0){\circle*{6}}
\put(20,0){\circle*{6}}
\qbezier(0,0)(10,-21)(20,0)
}
\put(76,-2){\mbox{$+6q^2$}}
\put(100,0){
\put(0,0){\circle*{6}}
\put(20,0){\circle*{6}}
\qbezier(0,0)(10,-7)(20,0)
\qbezier(0,0)(10,-21)(20,0)
}
\put(126,-2){\mbox{$-4q^3$}}
\put(150,0){
\put(0,0){\circle*{6}}
\put(20,0){\circle*{6}}
\qbezier(0,0)(10,7)(20,0)
\qbezier(0,0)(10,-7)(20,0)
\qbezier(0,0)(10,-21)(20,0)
}
\put(178,-2){\mbox{$+q^4$}}
\put(200,0){
\put(0,0){\circle*{6}}
\put(20,0){\circle*{6}}
\qbezier(0,0)(10,21)(20,0)
\qbezier(0,0)(10,7)(20,0)
\qbezier(0,0)(10,-7)(20,0)
\qbezier(0,0)(10,-21)(20,0)
}
}
\end{picture}

\noindent edge-amputation summation (for all-black edges it is identical to color-flip summation):

\begin{picture}(300,35)(-120,-15)

\put(-140,0){
\put(-12,-2){\mbox{$q^4$}}
\put(0,0){\circle*{6}}
\put(20,0){\circle*{6}}
\qbezier[15](0,0)(10,21)(20,0)
\qbezier[15](0,0)(10,7)(20,0)
\qbezier[15](0,0)(10,-7)(20,0)
\qbezier[15](0,0)(10,-21)(20,0)
\put(26,-2){\mbox{$-4q^3$}}
\put(50,0){
\put(0,0){\circle*{6}}
\put(20,0){\circle*{6}}
\qbezier[15](0,0)(10,21)(20,0)
\qbezier[15](0,0)(10,7)(20,0)
\qbezier[15](0,0)(10,-7)(20,0)
}
\put(76,-2){\mbox{$+6q^2$}}
\put(100,0){
\put(0,0){\circle*{6}}
\put(20,0){\circle*{6}}
\qbezier[15](0,0)(10,21)(20,0)
\qbezier[15](0,0)(10,7)(20,0)
}
\put(127,-2){\mbox{$-4q$}}
\put(150,0){
\put(0,0){\circle*{6}}
\put(20,0){\circle*{6}}
\qbezier[15](0,0)(10,21)(20,0)
}
\put(178,-2){\mbox{$+$}}
\put(200,0){
\put(0,0){\circle*{6}}
\put(20,0){\circle*{6}}
}
}

\put(100,-2){\mbox{$\stackrel{(\ref{eq:colorless-convention})}{=}$}}

\put(140,0){
\put(-12,-2){\mbox{$q^4$}}
\put(0,0){\circle*{6}}
\put(20,0){\circle*{6}}
\qbezier(0,0)(10,21)(20,0)
\qbezier(0,0)(10,7)(20,0)
\qbezier(0,0)(10,-7)(20,0)
\qbezier(0,0)(10,-21)(20,0)
\put(26,-2){\mbox{$-4q^3$}}
\put(50,0){
\put(0,0){\circle*{6}}
\put(20,0){\circle*{6}}
\qbezier(0,0)(10,21)(20,0)
\qbezier(0,0)(10,7)(20,0)
\qbezier(0,0)(10,-7)(20,0)
}
\put(76,-2){\mbox{$+6q^2$}}
\put(100,0){
\put(0,0){\circle*{6}}
\put(20,0){\circle*{6}}
\qbezier(0,0)(10,21)(20,0)
\qbezier(0,0)(10,7)(20,0)
}
\put(127,-2){\mbox{$-4q$}}
\put(150,0){
\put(0,0){\circle*{6}}
\put(20,0){\circle*{6}}
\qbezier(0,0)(10,21)(20,0)
}
\put(178,-2){\mbox{$+$}}
\put(200,0){
\put(0,0){\circle*{6}}
\put(20,0){\circle*{6}}
}}
\end{picture}

\bigskip
\noindent \textbf{Another example: non-Seifert vertex of (2,4) torus knot}

\noindent color-flip summation (vertical slices have the same $(-q)$-prefactors):

%
\begin{picture}(300,90)(20,-39)
%
\put(0,0){\circle*{6}}
\put(20,0){\circle*{6}}
\qbezier(0,0)(10,21)(20,0)
\qbezier[15](0,0)(10,7)(20,0)
\qbezier[15](0,0)(10,-7)(20,0)
\qbezier[15](0,0)(10,-21)(20,0)
%
\put(30,-2){\mbox{$-q$}}
\put(50,15){
\put(0,0){\circle*{6}}
\put(20,0){\circle*{6}}
\qbezier[15](0,0)(10,21)(20,0)
\qbezier[15](0,0)(10,7)(20,0)
\qbezier[15](0,0)(10,-7)(20,0)
\qbezier[15](0,0)(10,-21)(20,0)
}
\put(40,-17){\mbox{$3$}}
\put(50,-15){
\put(0,0){\circle*{6}}
\put(20,0){\circle*{6}}
\qbezier(0,0)(10,21)(20,0)
\qbezier[15](0,0)(10,7)(20,0)
\qbezier[15](0,0)(10,-7)(20,0)
\qbezier(0,0)(10,-21)(20,0)
}
\put(80,-2){\mbox{$+q^2$}}
\put(90,13){\mbox{$3$}}
\put(90,-17){\mbox{$3$}}
\put(100,15){
\put(0,0){\circle*{6}}
\put(20,0){\circle*{6}}
\qbezier[15](0,0)(10,21)(20,0)
\qbezier[15](0,0)(10,7)(20,0)
\qbezier[15](0,0)(10,-7)(20,0)
\qbezier(0,0)(10,-21)(20,0)
}
\put(100,-15){
\put(0,0){\circle*{6}}
\put(20,0){\circle*{6}}
\qbezier[15](0,0)(10,21)(20,0)
\qbezier(0,0)(10,7)(20,0)
\qbezier(0,0)(10,-7)(20,0)
\qbezier(0,0)(10,-21)(20,0)
}
\put(130,-2){\mbox{$-q^3$}}
\put(140,13){\mbox{$3$}}
\put(150,15){
\put(0,0){\circle*{6}}
\put(20,0){\circle*{6}}
\qbezier[15](0,0)(10,21)(20,0)
\qbezier[15](0,0)(10,7)(20,0)
\qbezier(0,0)(10,-7)(20,0)
\qbezier(0,0)(10,-21)(20,0)
}
\put(150,-15){
\put(0,0){\circle*{6}}
\put(20,0){\circle*{6}}
\qbezier(0,0)(10,21)(20,0)
\qbezier(0,0)(10,7)(20,0)
\qbezier(0,0)(10,-7)(20,0)
\qbezier(0,0)(10,-21)(20,0)
}
\put(180,-2){\mbox{$+q^4$}}
\put(200,0){
\put(0,0){\circle*{6}}
\put(20,0){\circle*{6}}
\qbezier[15](0,0)(10,21)(20,0)
\qbezier(0,0)(10,7)(20,0)
\qbezier(0,0)(10,-7)(20,0)
\qbezier(0,0)(10,-21)(20,0)
}
\put(240,-2){\mbox{$\stackrel{(\ref{eq:amp-convention})}{=}$}}
\put(270,0){
\put(0,0){\circle*{6}}
\put(20,0){\circle*{6}}
\qbezier(0,0)(10,21)(20,0)
%
\put(27,-2){\mbox{$-q$}}
\put(50,15){
\put(0,0){\circle*{6}}
\put(20,0){\circle*{6}}
}
\put(40,-17){\mbox{$3$}}
\put(50,-15){
\put(0,0){\circle*{6}}
\put(20,0){\circle*{6}}
\qbezier(0,0)(10,21)(20,0)
\qbezier(0,0)(10,-21)(20,0)
}
\put(80,-2){\mbox{$+q^2$}}
\put(90,13){\mbox{$3$}}
\put(90,-17){\mbox{$3$}}
\put(100,15){
\put(0,0){\circle*{6}}
\put(20,0){\circle*{6}}
\qbezier(0,0)(10,-21)(20,0)
}
\put(100,-15){
\put(0,0){\circle*{6}}
\put(20,0){\circle*{6}}
\qbezier(0,0)(10,7)(20,0)
\qbezier(0,0)(10,-7)(20,0)
\qbezier(0,0)(10,-21)(20,0)
}
\put(130,-2){\mbox{$-q^3$}}
\put(140,13){\mbox{$3$}}
\put(150,15){
\put(0,0){\circle*{6}}
\put(20,0){\circle*{6}}
\qbezier(0,0)(10,-7)(20,0)
\qbezier(0,0)(10,-21)(20,0)
}
\put(150,-15){
\put(0,0){\circle*{6}}
\put(20,0){\circle*{6}}
\qbezier(0,0)(10,21)(20,0)
\qbezier(0,0)(10,7)(20,0)
\qbezier(0,0)(10,-7)(20,0)
\qbezier(0,0)(10,-21)(20,0)
}
\put(180,-2){\mbox{$+q^4$}}
\put(200,0){
\put(0,0){\circle*{6}}
\put(20,0){\circle*{6}}
\qbezier(0,0)(10,7)(20,0)
\qbezier(0,0)(10,-7)(20,0)
\qbezier(0,0)(10,-21)(20,0)
}
}
\end{picture}

\noindent edge-amputation summation (vertical slices have same number of edges removed):

\begin{picture}(300,180)(-100,-130)
\put(-30,0){
\put(-12,-2){\mbox{$(-q)^{2}$}}
\put(15,0) {
  \put(0,0){\circle*{6}}
  \put(20,0){\circle*{6}}
  \qbezier(0,0)(10,21)(20,0)
  \qbezier[15](0,0)(10,7)(20,0)
  \qbezier[15](0,0)(10,-7)(20,0)
  \qbezier[15](0,0)(10,-21)(20,0)
}
\put(0,30) {
  \put(40,-17){\mbox{$3 (-q)^{1}$}}
  \put(75,-15){
    \put(0,0){\circle*{6}}
    \put(20,0){\circle*{6}}
    \qbezier(0,0)(10,21)(20,0)
    \qbezier[15](0,0)(10,7)(20,0)
    \qbezier[15](0,0)(10,-7)(20,0)
  }
}
\put(0,0) {
  \put(40,-17){\mbox{$(-q)^{3}$}}
  \put(70,-15){
    \put(0,0){\circle*{6}}
    \put(20,0){\circle*{6}}
    \qbezier[15](0,0)(10,7)(20,0)
    \qbezier[15](0,0)(10,-7)(20,0)
    \qbezier[15](0,0)(10,-21)(20,0)
  }
}
\put(70,0) {
  \put(0,30) {
    \put(40,-17){\mbox{$3 (-q)^0$}}
    \put(75,-15){
      \put(0,0){\circle*{6}}
      \put(20,0){\circle*{6}}
      \qbezier(0,0)(10,21)(20,0)
      \qbezier[15](0,0)(10,7)(20,0)
    }
  }
  \put(0,0) {
    \put(40,-17){\mbox{$3 (-q)^{2}$}}
    \put(75,-15){
      \put(0,0){\circle*{6}}
      \put(20,0){\circle*{6}}
      \qbezier[15](0,0)(10,7)(20,0)
      \qbezier[15](0,0)(10,-7)(20,0)
    }
  }
}
\put(140,0) {
  \put(0,30) {
    \put(40,-17){\mbox{$(-q)^{-1}$}}
    \put(75,-15){
      \put(0,0){\circle*{6}}
      \put(20,0){\circle*{6}}
      \qbezier(0,0)(10,21)(20,0)
    }
  }
  \put(0,0) {
    \put(40,-17){\mbox{$3 (-q)^{1}$}}
    \put(75,-15){
      \put(0,0){\circle*{6}}
      \put(20,0){\circle*{6}}
      \qbezier[15](0,0)(10,7)(20,0)
    }
  }
}
\put(210,0) {
  \put(0,15) {
    \put(40,-17){\mbox{$(-q)^0$}}
    \put(75,-15){
      \put(0,0){\circle*{6}}
      \put(20,0){\circle*{6}}
    }
  }
}

\put(330,-2){\mbox{$\stackrel{(\ref{eq:colorless-convention})}{=}$}}
}

\put(5,-90){
\put(-45,-2){\mbox{$\stackrel{(\ref{eq:colorless-convention})}{=}$}}
\put(-12,-2){\mbox{$(-q)^{2}$}}
\put(15,0) {
  \put(0,0){\circle*{6}}
  \put(20,0){\circle*{6}}
  \qbezier(0,0)(10,21)(20,0)
  \qbezier(0,0)(10,7)(20,0)
  \qbezier(0,0)(10,-7)(20,0)
  \qbezier(0,0)(10,-21)(20,0)
}
\put(0,30) {
  \put(40,-17){\mbox{$3 (-q)^{1}$}}
  \put(75,-15){
    \put(0,0){\circle*{6}}
    \put(20,0){\circle*{6}}
    \qbezier(0,0)(10,21)(20,0)
    \qbezier(0,0)(10,7)(20,0)
    \qbezier(0,0)(10,-7)(20,0)
  }
}
\put(0,0) {
  \put(40,-17){\mbox{$(-q)^{3}$}}
  \put(70,-15){
    \put(0,0){\circle*{6}}
    \put(20,0){\circle*{6}}
    \qbezier(0,0)(10,7)(20,0)
    \qbezier(0,0)(10,-7)(20,0)
    \qbezier(0,0)(10,-21)(20,0)
  }
}
\put(70,0) {
  \put(0,30) {
    \put(40,-17){\mbox{$3 (-q)^0$}}
    \put(75,-15){
      \put(0,0){\circle*{6}}
      \put(20,0){\circle*{6}}
      \qbezier(0,0)(10,21)(20,0)
      \qbezier(0,0)(10,7)(20,0)
    }
  }
  \put(0,0) {
    \put(40,-17){\mbox{$3 (-q)^{2}$}}
    \put(75,-15){
      \put(0,0){\circle*{6}}
      \put(20,0){\circle*{6}}
      \qbezier(0,0)(10,7)(20,0)
      \qbezier(0,0)(10,-7)(20,0)
    }
  }
}
\put(140,0) {
  \put(0,30) {
    \put(40,-17){\mbox{$(-q)^{-1}$}}
    \put(75,-15){
      \put(0,0){\circle*{6}}
      \put(20,0){\circle*{6}}
      \qbezier(0,0)(10,21)(20,0)
    }
  }
  \put(0,0) {
    \put(40,-17){\mbox{$3 (-q)^{1}$}}
    \put(75,-15){
      \put(0,0){\circle*{6}}
      \put(20,0){\circle*{6}}
      \qbezier(0,0)(10,7)(20,0)
    }
  }
}
\put(210,0) {
  \put(0,15) {
    \put(40,-17){\mbox{$(-q)^0$}}
    \put(75,-15){
      \put(0,0){\circle*{6}}
      \put(20,0){\circle*{6}}
    }
  }
}
}
\end{picture}

\noindent The extra factor of $-1/q$ in the edge-amputation summation is compensated by
the difference between ${\cal C}^{{\cal L}_c}$ and $q^{(N-1)(n_{..} - n_{-})}$.

\bigskip

In this version of fat-graph/matrix-model formalism it is manifest, that
both quantum dimensions $D_\gamma$ and $q$-weights $(-q)^{n_{..} - n_-}$ depend
only on  $\gamma$, which is by itself a fat graph, and neither on ${\cal L}_c$,
nor even on $\GLC$, i.e. that hypercube construction for HOMFLY polynomial
is \textit{local}. This locality property, which is crucial for the proof
 of Reidemeister invariance in the present text, may or may not survive
when going to superpolynomials (Khovanov-Rozansky).
At least naively KR construction is \textit{non-local}, as whole
layers of vertices (as opposed to individual vertices) of hypercube
explicitly appear in its formulation. Hence, reexpressing KR construction
in local terms becomes an important direction of the future research.

The two summation formulas (\ref{HDM3}) and (\ref{HMMPmod}) are closely related,
and quantum dimensions $D_\gamma$ and $d_v$ are essentially the same quantities,
still (\ref{HMMPmod}) allows to emphasize the special properties of quantum dimensions
like (\ref{fact-reduced}),
which are not so easy to express in (\ref{HDM3}) -- because in the two formulas
$D_\gamma$ and $d_v$ are {\it functions} of different variables.
Properties of this type give hypercube formalism an extra layer of {\it locality}
(dimensions $D_\gamma$ change nicely under local transformations of the fatgraph),
which is, together with above-mentioned vertex-wise locality,
 crucial for Reidemeister invariance.
Moreover, there are hints (see section \ref{sec:really-recursive-relations})
that Reidemeister invariance is only the tip of the iceberg
implied by locality and in fact a far richer structure
(similar or even exactly like AMM/EO topological recursion \cite{AMMEO}) hides beneath the water.
Of course, this is expected from matrix-model point of view.

\subsection{Factorization properties}

Dimensions satisfy factorization property:
for disconnected graph
\be
D_\gamma = D_{\gamma_1}\cdot D_{\gamma_2} \ \ \ \ \
\text{for} \ \ \ \gamma = \gamma_1\cup \gamma_2
\ \ \text{ with} \ \ \gamma_1\cap\gamma_2 = \emptyset,
\label{fact}
\ee
which is exactly the property that seems natural to postulate, if
one remembers trace contraction way of calculation.

Moreover, there is a much stronger factorization property:
if the graph consists of two parts, connected only through a single vertex,
where edges are {\it not intertwined}
then {\it reduced} dimension $D(\gamma)/[N]$ also factorizes:

\be
\begin{picture}(300,30)(60,-15)
\put(0,-2){$[N] \cdot D \Bigg ($}
\put(60,0) {
  \put(0,0){\circle{30}}\put(-4,-2){\mbox{$\gamma_1$}}
  \put(80,0){\circle{30}}\put(76,-2){\mbox{$\gamma_2$}}
  \thicklines \linethickness{0.2mm}
  \put(40,0){\circle*{8}}
  \thinlines \linethickness{0.125mm}
  \qbezier(36,0)(35,0)(16,0)
  \qbezier(37,3)(20,20)(13,10)
  \qbezier(37,-3)(20,-20)(13,-10)
  \qbezier(42,0)(45,0)(64,0)
  \qbezier(43,3)(60,20)(67,10)
  \qbezier(43,-3)(60,-20)(67,-10)
}
\put(160,-2){$\Bigg )\ =\ D\Bigg ($}
\put(225,0) {
  \put(0,0){\circle{30}}\put(-4,-2){\mbox{$\gamma_1$}}
  \thicklines \linethickness{0.2mm}
  \put(40,0){\circle*{8}}
  \thinlines \linethickness{0.125mm}
  \qbezier(36,0)(35,0)(16,0)
  \qbezier(37,3)(20,20)(13,10)
  \qbezier(37,-3)(20,-20)(13,-10)
}
\put(275,-2) {$\Bigg )\ \cdot\ D\Bigg ($}
\put(280,0) {
  \put(80,0){\circle{30}}\put(76,-2){\mbox{$\gamma_2$}}
  \thicklines \linethickness{0.2mm}
  \put(40,0){\circle*{8}}
  \thinlines \linethickness{0.125mm}
  \qbezier(43,0)(45,0)(64,0)
  \qbezier(43,3)(60,20)(67,10)
  \qbezier(43,-3)(60,-20)(67,-10)
}
\put(380,-2){$\Bigg )$}
\end{picture}
\label{fact-reduced}
\ee
If combined with explicit expression for
\be
D\left(\begin{picture}(56,20)(-3,0) \put(0,0){\circle*{6}}\put(50,0){\circle*{6}}
\qbezier(0,0)(25,30)(50,0)\qbezier(0,0)(25,-30)(50,0)\qbezier(0,0)(25,15)(50,0)\qbezier(0,0)(25,-15)(50,0)
\put(18,-2){\makebox{$\ldots$}}
\put(0,-5){
\qbezier(-5,-5)(0,-10)(5,-5)\put(5,-5){\vector(1,1){2}}
\qbezier(45,-5)(50,-10)(55,-5)\put(45,-5){\vector(-1,1){2}}
}
\end{picture} \right) \ = \ [2]^{n-1}[N][N-1]
\ee
where $n\geq 1$ is the number of edges in this simple fat diagram
(for the sake of simplicity of drawing, vertices are drawn contra-oriented and edges are, hence, not twisted),
this can be further enhanced to
\be
\begin{picture}(300,30)(40,-15)
\put(-45,-2){$[N] \cdot D \Bigg ($}
\put(60,0) {
  \put(-50,0){\circle{30}}\put(-54,-2){\mbox{$\gamma_1$}}
  \put(80,0){\circle{30}}\put(76,-2){\mbox{$\gamma_2$}}
  \thicklines \linethickness{0.2mm}
  \put(-10,0){\circle*{8}}
  \put(40,0){\circle*{8}}
  \thinlines \linethickness{0.125mm}
  \qbezier(-14,0)(-15,0)(-34,0)
  \qbezier(-13,3)(-30,20)(-37,10)
  \qbezier(-13,-3)(-30,-20)(-37,-10)
  \put(-10,0){
\qbezier(0,0)(25,30)(50,0)\qbezier(0,0)(25,-30)(50,0)\qbezier(0,0)(25,15)(50,0)\qbezier(0,0)(25,-15)(50,0)
\put(18,-2){\makebox{$\ldots$}}
\put(0,-5){
\qbezier(-5,-5)(0,-10)(5,-5)\put(5,-5){\vector(1,1){2}}
\qbezier(45,-5)(50,-10)(55,-5)\put(45,-5){\vector(-1,1){2}}
}
}
  \qbezier(42,0)(45,0)(64,0)
  \qbezier(43,3)(60,20)(67,10)
  \qbezier(43,-3)(60,-20)(67,-10)
}
\put(160,-2){$\Bigg )\ =\ [2]^{n-1}[N-1] \cdot D\Bigg ($}
\put(60,0){
\put(228,0) {
  \put(0,0){\circle{30}}\put(-4,-2){\mbox{$\gamma_1$}}
  \thicklines \linethickness{0.2mm}
  \put(40,0){\circle*{8}}
  \thinlines \linethickness{0.125mm}
  \qbezier(36,0)(35,0)(16,0)
  \qbezier(37,3)(20,20)(13,10)
  \qbezier(37,-3)(20,-20)(13,-10)
}
\put(275,-2) {$\Bigg )\ \cdot\ D\Bigg ($}
\put(280,0) {
  \put(80,0){\circle{30}}\put(76,-2){\mbox{$\gamma_2$}}
  \thicklines \linethickness{0.2mm}
  \put(40,0){\circle*{8}}
  \thinlines \linethickness{0.125mm}
  \qbezier(43,0)(45,0)(64,0)
  \qbezier(43,3)(60,20)(67,10)
  \qbezier(43,-3)(60,-20)(67,-10)
}
\put(380,-2){$\Bigg )$}
}
\end{picture}
\nn
\ee

Formula \eqref{fact-reduced} can be considered as a manifestation of
factorization property of reduced HOMFLY polynomials for composite knots (both usual and virtual)
into the product of reduced HOMFLY for their components:
it is enough to note that the fat graphs for composites has exactly the form of the l.h.s.
of (\ref{fact-reduced}) -- therefore  in what follows we call such graphs composite.

\Theremark {
  Note
  that formulas \eqref{fact} and \eqref{fact-reduced} together
  fix dimension of an isolated vertex to be $[N]$,  since
  \be \label{eq:n-rule}
  \begin{picture}(5,5)(0,-3)\circle*{6}\end{picture}
  \mathop{=}^{\eqref{fact-reduced}}
  \frac{\displaystyle 1}{\displaystyle [N]}\ \
  \begin{picture}(15,5)(0,-3)\put(0,0){\circle*{6}} \put(10,0){\circle*{6}}\end{picture}
  \mathop{=}^{\eqref{fact}}
  \frac{\displaystyle 1}{\displaystyle [N]}\ \
  \begin{picture}(5,5)(0,-3)\put(0,0){\circle*{6}}\end{picture}^2
  \ \longrightarrow\
  \begin{picture}(5,5)(0,-3)\circle*{6}\end{picture} = [N]
  \ee
}

\subsection{Summary}

After exposition of DM approach in this section the remaining problem -- in the part, concerning
HOMFLY polynomials for ordinary and virtual links and knots -- splits into three
practical topics:

\begin{itemize}
\item building fat graphs for particular knot diagrams;
\item evaluating quantum dimensions $D_\gamma$ for various
(ideally --arbitrary) fat graphs $\gamma$;
\item justifying Reidemeister invariance of the answers for HOMFLY polynomials.
\end{itemize}

\noindent
Algorithm for the first topic should be clear from examples in section \ref{graphs}.
Extra details about injectivity and surjectivity of this map,
together with additional examples,
are provided in appendix \ref{sec:link->graph}.
We elaborate on the second subject in sec.\ref{mamoc},
putting additional emphasis on relation to matrix-model correlators.
Then in sec.\ref{sec:n-rules} we use matrix-model intuition to look for recursion relations
and their generalization to $q \neq 1$.
Easily found are the relations, reflecting
Reidemeister invariance, see sec.\ref{sec:reid}).
Despite they are not quite a {\it recursion}, adding just one more
relation (a "flip") is sufficient to make an efficient computer program, sec.6,
which calculates dimensions $D_\gamma$ and HOMFLY polynomials by hypercube method
for entire Rolfsen table of \cite{katlas} and \cite{virtkatlas}.
The last section 7 describes a search for more sophisticated recursions,
which would improve the computation capacity even further.
Among other things, this involves extension to diagrams of valences different
from $(2,2)$, what -- as anticipated in \cite{DM3} --  opens new horizons for the entire theory.
A serious step forward as compared to \cite{MMP1} is not just a more detailed,
exhaustive and extended presentation, but also a proof of Reidemeister invariance of (\ref{HMMPmod}).
As to the interplay between link diagrams (perhaps, generalized to other valences), and the fat graphs,
it remains partly mysterious: at the moment some things are better formulated in one language,
some -- in another. This is not such a rare things when dual descriptions are available,
still better understanding and adequate reformulation of emerging recursions is certainly required.

\newpage

\section{Matrix-model correlators and their $q$-deformation
\label{mamoc}}

We begin this section from the genus expansions of the correlators
\be
< \prod_{i=1}^m \Tr M^{k_i} >
\label{corr}
\ee
with and without normal ordering,
which are obtained by the Wick rule with the propagator
\be
\pi^{ik}_{jl} =\ <M^i_jM^k_l>^{(0)} = \delta^i_l \delta^k_j
\label{simpro}
\ee
and attribute different terms of these expansions (particular Feynman diagrams/graphs)
to different links (generically, virtual).
This establishes the mapping  \ \ \ \ \ link ${\cal L}$ \ $\longrightarrow$ \
fat graph $\Gamma^{\cal L}$.

After that classical ($q=1$) dimensions $d_v^{\cal L}$ at hypercube vertices $v\in {\cal H}^{\cal L}$
are obtained by substituting (\ref{simpro}) on edges of $\Gamma^{\cal L}$
by the propagator
\be
\Pi^{ik}_{jl} =   \delta^i_j\delta^k_l - \delta^i_l\delta^k_j
\label{propknot}
\ee
what provides expressions for Feynman diagrams
in the form of linear combinations of sub-graphs of $\Gamma^{\cal L}$.

These combinations should further be quantized ($q$-deformed) --
and this is the most difficult part of the story.
Quantization is provided by the rules from \cite{AnoMKhR}, originally deduced
within  Reshetikhin-Turaev (RT) approach \cite{RT}-\cite{RTmod}.
and extended to virtual case in \cite{virtDM3}.
Following the suggestion of \cite{MMP1} we reformulate these rules in terms
of the matrix-model recursion relations -- with the idea that some of recursions,
associated to Reidemeister moves, survive quantization.
In this way we obtain a universal calculus, which can be applied not only to
ordinary, but also to virtual knots, where RT approach is not directly applicable.
Even more important we get an easily computerizable formalism,
which is capable to provide many (if not all)
desired dimensions (in \cite{MMP1} it was applied
for up to 14 intersections in ${\cal L}$, i.e. for hypercubes ${\cal H}^{\cal L}$
with up to $2^{14}$ vertices).

The purpose of this section is to provide the beginning of the  dimension list
(see Appendix \ref{sec:table-of-dimensions} for a little more), while its origins and implications
will be discussed in the following sections.
We provide only the most complicated dimension, at the anti-Seifert vertex $\bar S$
of the hypercube -- dimensions at other vertices correspond to subgraphs of $\Gamma^{\cal L}$
with some amputated edges, which appear in the previous lines of the table.
When graph is made from several disconnected components, dimensions factorize
 in nice accordance with \eqref{fact}.

$$
\begin{array}{c|c|c|c|c|c|ccc}
\text{correlator} & \text{genus} & \Gamma^{\cal L} & {\cal L}& d_{\bar S}^{\cal L}(q=1)& d_{\bar S}^{\cal L}\\
&&&&&&&\\
\hline
<\Tr I>& 0 & \begin{picture}(10,10) \put(3,3){\circle*{8}} \end{picture} & \text{unknot} & N& [N] \\
\hline
<\Tr M^2> & 0
& \begin{picture}(15,20)(0,-2)
    \put(0,0){\circle*{8}}
    \put(8,0){\circle{16}}
  \end{picture}
& \text{virtual Hopf} & N-N^2 &  -[N][N-1] \\
&&&&&&&\\
\hline
&&&&&&&\\
<\Tr M^4> &0 &
\begin{picture}(30,18)(-15,-10)
  \put(0,0){\circle*{8}}
  \put(9,0){
    \put(0,0){\circle{16}}
  }
  \put(-9,0){
    \put(0,0){\circle{16}}
  }
\end{picture}
&
\substack{
  \displaystyle \text{composite of} \\
  \displaystyle \text{two virtual Hopfs}
} & N (N-1)^2 & [N][N-1]^2\\
& 1 &
\begin{picture}(30,20)(-15,-8)
  \put(0,0){\circle*{8}}
  \put(5,0){
    \put(0,0){\circle{12}}
  }
  \put(0,5){
    \put(0,0){\circle{12}}
  }
\end{picture}
& \text{virtual trefoil (2.1)} & - 2 N (N-1) & -[2][N][N-1]
&&\\
&&&&&&&\\
\hline
&&&&&&&\\
<\Tr M^6> & 0 &
\begin{picture}(30,20)(-15,-12)
  \put(0,0){\circle*{8}}
  \put(9,-5){
    \put(0,0){\circle{16}}
  }
  \put(-9,-5){
    \put(0,0){\circle{16}}
  }
  \put(0,9){
    \put(0,0){\circle{16}}
  }
\end{picture}
&
\substack{
  \displaystyle \text{composite of} \\
  \displaystyle \text{three virtual Hopfs}
} & - N (N-1)^3 & -[N][N-1]^3 && \\
& 0 &
\begin{picture}(40,20)(-15,-2)
  \put(0,0){\circle*{8}}
  \put(-9,0){
    \put(0,0){\circle{16}}
  }
  \put(9,0){
    \put(0,0){\circle{16}}
  }
  \put(11,0){
    \put(0,0){\circle{24}}
  }
\end{picture}
&
\substack{
  \displaystyle \text{different} \\
  \displaystyle \text{composite of} \\
  \displaystyle \text{three virtual Hopfs}
} & - N (N-1)^3 & -[N][N-1]^3
&& \\
&&&&&&&\\
& 1 &
\begin{picture}(30,20)(-15,-8)
  \put(0,0){\circle*{8}}
  \put(5,0){
    \put(0,0){\circle{12}}
  }
  \put(0,5){
    \put(0,0){\circle{12}}
  }
  \put(-6,-6){
    \put(0,0){\circle{12}}
  }
\end{picture}
&
\substack{
  \displaystyle \text{composite of} \\
  \displaystyle \text{virtual Hopf} \\
  \displaystyle \text{and virtual 2.1}
} & 2 N (N-1)^2 & [2] [N][N-1]^2
&& \\
& 1 &
\begin{picture}(30,20)(-15,-8)
  \put(0,0){\circle*{8}}
  \put(8,0){
    \put(0,0){\circle{16}}
  }
  \put(5,-5){
    \put(0,0){\circle{16}}
  }
  \put(-3,-6){
    \put(0,0){\circle{16}}
  }
\end{picture}
&
& -4 N (N-1) & - [2]^2 [N][N-1]  \\
& 1 &
\begin{picture}(30,25)(-15,-5)
  \put(0,0){\circle*{8}}
  \put(6,0){
    \put(0,0){\circle{12}}
  }
  \put(-6,0){
    \put(0,0){\circle{12}}
  }
  \qbezier(0,0)(-10,0)(-10,-5) \qbezier(0,0)(10,0)(10,-5)
  \qbezier(-10,-5)(-10,-10)(0,-10) \qbezier(10,-5)(10,-10)(0,-10)
\end{picture}
&
& N (N-1) (N-3) & [N][N-1]\Big([N-2]-1\Big)
& & * \\
&&&&&&&\\
\hline
\end{array}
$$

$$
\begin{array}{c|c|c|c|c|c|ccc}
\hline
<\Tr M \ \Tr M> & 0 &
\begin{picture}(30,15)(-15,-2)
  \put(-10,0){\circle*{8}} \put(10,0){\circle*{8}}
  \put(-10,0){\line(1,0){20}}
\end{picture}
& \text{twisted unknot} & N(N-1) & [N][N-1]
&&\\
&&&&&&&\\
\hline
<\Tr M \ \Tr M^3> & 0 &
\begin{picture}(35,20)(-15,-6)
  \put(-10,0){\circle*{8}}
  \put(10,0){
    \put(0,0){\circle*{8}}
    \put(8,0){\circle{16}}
  }
  \put(-10,0){\line(1,0){20}}
\end{picture}
& \text{virtual Hopf with a twist} & -N(N-1)^2 & -[N][N-1]^2
&&\\
&&&&&&&\\
\hline
<\Tr M^2 \ \Tr M^2> & 0,\ 0 &
\begin{picture}(35,20)(-15,-6)
  \put(-10,0){
    \put(0,0){\circle*{8}}
    \put(6,0){\circle{12}}
  }
  \put(10,0){
    \put(0,0){\circle*{8}}
    \put(6,0){\circle{12}}
  }
\end{picture}
& \text{two virtual Hopfs} & N^2(N-1)^2 & [N]^2[N-1]^2 \\
& 0 &
\begin{picture}(35,20)(-15,-6)
  \put(-10,0){
    \put(0,0){\circle*{8}}
  }
  \put(10,0){
    \put(0,0){\circle*{8}}
  }
  \qbezier(-10,0)(0,10)(10,0) \qbezier(-10,0)(0,-10)(10,0)
\end{picture}
& \text{Hopf} & 2 N (N-1) & [2][N][N-1]
&&\\
&&&&&&&\\
\hline
<\Tr M \ \Tr M^5>
& 0 &
\begin{picture}(30,20)(-15,-2)
  \put(0,0){\circle*{8}}
  \put(-10,0){\circle*{8}}
  \put(-10,0){\line(1,0){10}}
  \put(9,0){
    \put(0,0){\circle{16}}
  }
  \put(-9,0){
    \put(0,0){\circle{16}}
  }
\end{picture}
& \substack{
  \displaystyle \text{composite of} \\
  \displaystyle \text{two virtual Hopfs} \\
  \displaystyle \text{and twisted unknot}
} & N (N-1)^3 & [N][N-1]^3 \\
& 0 &
\begin{picture}(30,20)(-15,-2)
  \put(0,0){\circle*{8}}
  \put(0,10){\circle*{8}}
  \put(0,10){\line(0,-1){10}}
  \put(9,0){
    \put(0,0){\circle{16}}
  }
  \put(-9,0){
    \put(0,0){\circle{16}}
  }
\end{picture}
& \substack{
  \displaystyle \text{different} \\
  \displaystyle \text{composite of} \\
  \displaystyle \text{two virtual Hopfs} \\
  \displaystyle \text{and twisted unknot}
} & N (N-1)^3 & [N][N-1]^3 \\
& 1 &
\begin{picture}(30,30)(-15,-12)
  \put(0,0){\circle*{8}}
  \put(-10,-10){\circle*{8}}
  \put(-10,-10){\line(1,1){10}}
  \put(9,0){
    \put(0,0){\circle{16}}
  }
  \put(0,9){
    \put(0,0){\circle{16}}
  }
\end{picture}
& \substack{
  \displaystyle \text{composite of} \\
  \displaystyle \text{virtual 2.1} \\
  \displaystyle \text{and twisted unknot}
} & -2 N (N-1)^2 & -[2][N][N-1]^2
&&\\
&&&&&&&\\
\hline
<\Tr M^2 \ \Tr M^4> & 0,\ 0
&
\begin{picture}(30,20)(-15,-2)
  \put(-15,0){\circle*{8}} \put(-10,0){\circle{10}}
  \put(10,0){\put(0,0){\circle*{8}} \put(5,0){\circle{10}} \put(-5,0){\circle{10}}}
\end{picture}
& \substack{
  \displaystyle \text{virtual Hopf and} \\
  \displaystyle \text{composite of} \\
  \displaystyle \text{two virtual Hopfs}
} & - N^2 (N-1)^3 & -[N]^2[N-1]^3
&& \\
& 0,\ 1
&
\begin{picture}(30,20)(-15,-2)
  \put(-15,0){\circle*{8}} \put(-10,0){\circle{10}}
  \put(10,0){\put(0,0){\circle*{8}} \put(5,0){\circle{10}} \put(0,5){\circle{10}}}
\end{picture}
& \substack{
  \displaystyle \text{virtual Hopf and} \\
  \displaystyle \text{virtual 2.1}
} & 2 N^2 (N-1)^2 & [2][N]^2[N-1]^2
&&\\
& 0 &
\begin{picture}(35,20)(-15,-6)
  \put(-10,0){
    \put(0,0){\circle*{8}}
  }
  \put(10,0){
    \put(0,0){\circle*{8}}
    \put(6,0){\circle{12}}
  }
  \qbezier(-10,0)(0,10)(10,0) \qbezier(-10,0)(0,-10)(10,0)
\end{picture}
& \substack{
  \displaystyle \text{composite of} \\
  \displaystyle \text{Hopf and virtual Hopf}
} & -2 N (N-1)^2 & -[2][N][N-1]^2
&&\\
& 1 &
\begin{picture}(35,20)(-15,-6)
  \put(-10,0){
    \put(0,0){\circle*{8}}
  }
  \put(10,0){
    \put(0,0){\circle*{8}}
    \put(-5,5){\circle{10}}
  }
  \qbezier(-10,0)(0,10)(10,0) \qbezier(-10,0)(0,-10)(10,0)
\end{picture}
& \substack{
  \displaystyle \text{virtual 3.1, 3.2, 3.3, 3.4}
} & -N (N-1) (N-3) & -[N][N-1]\Big([N-2]-1\Big)
&& * \\
&&&&&&&\\
\hline
<\Tr M^3 \ \Tr M^3>
& 0 &
\begin{picture}(35,20)(-15,-6)
  \put(-10,0){\put(0,0){\circle*{8}} \put(-5,0){\circle{10}}}
  \put(10,0){\put(0,0){\circle*{8}} \put(5,0){\circle{10}}}
  \put(-10,0){\line(1,0){20}}
\end{picture}
& \substack{
  \displaystyle \text{twisted composite of} \\
  \displaystyle \text{two virtual Hopfs}
} & N (N-1)^3 & [N][N-1]^3
&&\\
& 1 &
\begin{picture}(35,20)(-15,-6)
  \put(-10,0){\put(0,0){\circle*{8}}}
  \put(10,0){\put(0,0){\circle*{8}}}
  \put(-10,0){\line(1,0){20}}
  \qbezier(-10,0)(0,10)(10,0) \qbezier(-10,0)(0,-10)(10,0)
  \qbezier(-17,0)(-17,-7)(-10,-7) \put(-7,-7){\vector(1,0){0}}
  \qbezier(17,0)(17,-7)(10,-7) \put(7,-7){\vector(-1,0){0}}
\end{picture}
& \text{trefoil or virtual 3.5, 3.7} & 4 N (N-1) & [2]^2[N][N-1]
&&\\
& 0 &
\begin{picture}(35,20)(-15,-6)
  \put(-10,0){\put(0,0){\circle*{8}}}
  \put(10,0){\put(0,0){\circle*{8}}}
  \put(-10,0){\line(1,0){20}}
  \qbezier(-10,0)(0,10)(10,0) \qbezier(-10,0)(0,-10)(10,0)
  \qbezier(-17,0)(-17,-7)(-10,-7) \put(-7,-7){\vector(1,0){0}}
  \qbezier(17,0)(17,-7)(10,-7) \put(17,3){\vector(0,1){0}}
\end{picture}
&
& -N (N-1) (N-3) & -[N][N-1]\Big([N-2]-1\Big)
&& * \\
&&&&&&&\\
\hline
\end{array}
$$

\bigskip

Even this short table may seem a bit intimidating.
Still, there are several important things in it.

First, the answers (dimensions)
are considerably less diverse than the fat graphs themselves: modulo overall sign
all mutually distinct answers for connected graphs appear in 1-point
correlators.
This is an experimental evidence for flip-rule, described in the next section.

Second, answers for disconnected graphs are products of answers for components
-- this is the evidence for factorization property \eqref{fact}.

Third, fat graphs marked by asterisk (*) violate naive quantization rule:
while classical dimension contains simple linear factor $(N-3)$,
its quantization is the non-trivial $([N-2]-1)$ -- an indication that the
whole story can't somehow trivialize to just some quantization prescription
at the level of polynomials -- some sort of fat graph formalism is really
necessary.

Fourth, it's common for different virtual knots to have same fat graph.
(see also Appendix \ref{sec:link->graph} for an example with celebrated Kishino knot)

Finally, ${\cal L}$-column is empty for some fat graphs. It is
not because there is no link diagram for these fat graphs, but just because
we don't know of \textit{sufficiently simple} link diagrams, which give such
fat graphs. See Appendix \ref{sec:link->graph} for explanation of construction
of (rather complicated) link diagram for arbitrary fat graph.

\bigskip

More formally, generic matrix-model correlator decomposes into a sum of Feynman graphs
$<\Gamma>$, associated with fat graphs $\Gamma$,
\be
\text{Wick theorem:} \ \ \ \ \ \ \
\left< \prod_i \Tr M^{k_i} \right>\ = \
\sum_\Gamma C_\Gamma\{k_i\}\cdot <\Gamma>
\ee
Due to specific form of the propagator (\ref{propknot}) each
\be
<\Gamma>\ = \ \sum_{\gamma\subset\Gamma} (-)^{E_\Gamma - E_\gamma} D_\gamma(q=1)
\ee
i.e. is an alternated sum over all fat subgraphs $\gamma$ of $\Gamma$, where the contribution
of $\gamma$ is the classical dimension $D_\gamma(q=1)$.

Since fat graph $\Gamma$ can be built for any link diagram ${\cal L}$
(though different link diagrams may have same graph -- the map ${\cal L} \twoheadrightarrow \Gamma$
is actually surjective and not injective, see Appendix \ref{sec:link->graph})
this provides all the constituents of the formula (\ref{HMMPmod}) for HOMFLY polynomials at $q=1$.

\bigskip

This result raises the following three issues, which we discuss
in the remaining part of this paper:

$\bullet$ Matrix-model Ward identities for correlators
$\left< \prod_i \Tr M^{k_i} \right>$ provide recursion relations for
classical dimensions $D_\gamma(q=1)$

$\bullet$ Quantization procedure $D_\gamma(q=1) \ \longrightarrow\ D_\gamma(q)$

$\bullet$ The possibility to lift Ward identities and the matrix model itself
to $q\neq 1$

\bigskip

As stated  in \cite{MMP1}, what can be lifted are the three Ward identities,
associated with Reidemesiter invariance.
This is a very important story by itself, because relations are {\it local}
(i.e. are valid whenever the appropriate small fragment of the fat graph is there,
regardless of how complex the entire graph is)  --
while dimensions themselves are defined {\it globally} (depend on entire graph).
However, it is a separate problem to convert this restricted set
of Ward identities into a {\it recursion}.
Current computer program addresses this rather bluntly and ineffectively by
doing a full search at some stage, see section \ref{sec:computer-program}.
But in section \ref{sec:really-recursive-relations} we speculate about
a smarter approach.
Still, even the current blunt approach allows to reproduce fundamental HOMFLY
polynomials for all knots from Rolfsen table.
To be able to calculate even the simplest virtual knot examples,
one needs additional {\it flip transform}, which mysteriously
survives at $q\neq 1$.
Incidentially, it also allows to calculate
HOMFLY for non-virtual knots more effectively.

\section{Local fat graph transformations at $q \neq 1$ \label{sec:n-rules}}

Though in theory the quantum dimension $D_\gamma$ for any fat graph $\gamma$ can
be calculated directly from the RT-inspired definition \cite{AnoMKhR},
(and for some classes of knots/links this method is actually very effective),
in practice it is not the way we calculate them.
Instead, looking at available answers in \cite{AnoMKhR},
 we devised in \cite{MMP1}
a set of relations that dimensions $D_\gamma$ seem to obey.
They are transformations of the fat graph $\gamma$, that preserve its associated quantum dimension
$D_\gamma$.
The crucial point is that these transformations are \textit{local} --
they are about the small sub-graphs (with one, two and three vertices)
and remain true when arbitrarily complicated surrounding is attached to both
sides of the identity.
In other words, though dimensions themselves depend on the entire graph (i.e. are non-local),
relations between them occur when just a small sub-graph is reshuffled,
and reshufflings are always the same, independently of complexity of the original graph.
In pictures below, thick arrows denote pieces of fat graph vertices, while thin lines denote
fat graph edges. It is assumed that the rest of the graph does not change
when transformation is applied, and it is not shown in the picture.
Also, let us stress, that, in accordance with the edge-amputation summation convention
of section \ref{sec:fat-graph->homfly}, any straight edge may be substituted by
a dotted edge (this does not affect the quantum dimension, only the $q$-dependent prefactor).

These relations are purely experimental, but they enjoy several good properties,
which a sort of justify them.
First, they of course are valid at $q = 1$, where they are elementary consequences
of explicit form of propagator \eqref{propknot} and the Wick rule.
Second, calculating dimensions $D_\gamma$ solely from these relations
and factorization properties \eqref{fact} and \eqref{fact-reduced}
and plugging them into formula \eqref{HMMPmod} produces correct answers for
fundamental HOMFLY polynomials for all knots in Rolfsen table
(see section \ref{sec:computer-program} for the explanation of the algorithm).
Third, in the next section we show, that these relations
and factorization properties \eqref{fact} and \eqref{fact-reduced}
are sufficient for Reidemeister invariance.


\bigskip

Anyway, here are these relations (we also list the corresponding $q = 1$
identities for matrix-model correlators):

\bigskip

\noindent $\bullet$ \textbf{$[N-1]$-rule} -- we can contract all the
vertices of valence 1:

\be \label{eq:n-1-rule}
&  \begin{picture}(300,50)(0,-10)
\put(100,0){
    \linethickness{0.4mm} \thicklines
    \qbezier(0,0)(20,20)(0,40)
    \put(0,40){\vector(-1,1){0}}
    \put(35,10){\vector(1,0){0}}
    \put(30,20){\circle{20}}
        {\linethickness{0.15mm} \put(10,20){\line(1,0){10}}}
        \put(50,17) {$\simeq\ \ [N-1]$}
        \put(110,0){\vector(0,1){40}}
}
  \end{picture}
\\ \nn
& < \Tr M \  F(M) >\ =\ (N-1)\delta^k_l  \left<\frac{\p F}{\p M^k_l}\right>
\ee

\noindent $\bullet$ \textbf{$[2]$-rule} -- we can eliminate double edges
(mutual orientation of fat graph vertices is important here):

\be \label{eq:2-rule}
& \begin{picture}(300,60)(-80,-20)
    \linethickness{0.4mm} \thicklines
    \qbezier(0,0)(20,20)(0,40)
    \put(0,40){\vector(-1,1){0}}
    \put(40,0){\qbezier(0,0)(-20,20)(0,40)}
    \put(40,40){\vector(1,1){0}}
        {\linethickness{0.15mm}
          \put(8,30){\line(1,0){24}}
          \put(8,10){\line(1,0){24}}
        }
        \put(50,17) {$\simeq\ \ [2]$}
        \put(80,0){
          \linethickness{0.4mm} \thicklines
          \qbezier(0,0)(20,20)(0,40)
          \put(0,40){\vector(-1,1){0}}
          \put(40,0){\qbezier(0,0)(-20,20)(0,40)}
          \put(40,40){\vector(1,1){0}}
              {\linethickness{0.15mm}
                \put(10,20){\line(1,0){20}}
              }
        }
  \end{picture}
\\ \nn
& < M^i_j M^j_k M^l_m M^m_n F^{k n}_{i l} (M) >\ =\ [2] < M^i_k M^l_n F^{k n}_{i l} (M) >
\begin{picture}(0,0)
  \put(-225,12){\put(0,0){\line(1,0){30}} \put(0,0){\line(0,-1){3}} \put(30,0){\line(0,-1){3}}}
  \put(-210,18){\put(0,0){\line(1,0){30}} \put(0,0){\line(0,-1){8}} \put(30,0){\line(0,-1){8}}}
  \put(-75,12){\put(0,0){\line(1,0){15}}\put(0,0){\line(0,-1){3}} \put(15,0){\line(0,-1){3}}}
\end{picture}
\ee

\noindent $\bullet$ \textbf{$[N-2]$-rule} -- we can eliminate all
2-valent vertices:

\be \label{eq:n-2-rule}
& \begin{picture}(300,60)(-80,-20)
    \linethickness{0.4mm} \thicklines
    \put(-40,0){\qbezier(0,0)(20,20)(0,40) \put(0,40){\vector(-1,1){0}}}
    \put(40,0){\qbezier(0,0)(-20,20)(0,40)}
    \put(40,0){\vector(1,-1){0}}
    \put(-40,0){\linethickness{0.15mm}
      \put(10,20){\line(1,0){20}}
    }
    \put(0,0){\linethickness{0.15mm}
      \put(10,20){\line(1,0){20}}
    }
    \put(0,20){\circle{20} \put(5,10){\vector(1,0){0}} \put(-5,-10){\vector(-1,0){0}}}

    \put(50,17) {$\simeq\ \ [N-2]$}
    \put(100,0) {
      \linethickness{0.4mm} \thicklines
      \qbezier(0,0)(20,20)(0,40)
      \put(0,40){\vector(-1,1){0}}
      \put(40,0){\qbezier(0,0)(-20,20)(0,40)}
      \put(40,0){\vector(1,-1){0}}
    }
    \put(150,17){$+$}
    \put(165,0) {
      \linethickness{0.4mm} \thicklines
      \qbezier(0,0)(20,20)(40,0)
      \put(0,40){\vector(-1,1){0}}
      \put(40,40){\qbezier(0,0)(-20,-20)(-40,0)}
      \put(40,0){\vector(1,-1){0}}
    }
  \end{picture}
\\ \nn
& < :\Tr M^2: \  F(M) >\ =\ \Big((N-2)\delta^k_l\delta^m_n +\delta^k_n\delta^m_l\Big)
\left<\frac{\p^2 F}{\p M^k_l\p M^m_n}\right>
\ee

\noindent $\bullet$ \textbf{$[N-3]$-rule} -- we can ``mutate'' 3-valent vertices:

\be \label{eq:n-3-rule}
& \begin{picture}(300,50)(-10,-10)
    \put(-40,0) {
      \linethickness{0.4mm} \thicklines
      \qbezier(0,0)(0,20)(-20,20) \put(-20,20){\vector(-1,0){0}}
      \put(20,0){\qbezier(0,0)(0,20)(20,20) \put(0,0){\vector(0,-1){0}}}
      \put(20,40){\qbezier(0,0)(-10,-10)(-20,0) \put(0,0){\vector(2,1){0}}}
      \put(10,20){\circle{16}}
      \put(10,0){
        \linethickness{0.15mm} \thinlines
        \qbezier(-12,11)(-12,11)(-6,15)
      }
      \put(10,0){
        \linethickness{0.15mm} \thinlines
        \qbezier(12,11)(12,11)(6,15)
      }
      \put(10,28){
        \linethickness{0.15mm} \thinlines
        \qbezier(0,0)(0,5)(0,7)
      }
    }
    \put(10,17){$-\ [N-3]$}
    \put(80,0) {
      \linethickness{0.4mm} \thicklines
      \qbezier(0,0)(0,20)(-20,20) \put(-20,20){\vector(-1,0){0}}
      \put(20,0){\qbezier(0,0)(0,20)(20,20) \put(0,0){\vector(0,-1){0}}}
      \put(20,40){\qbezier(0,0)(-10,-10)(-20,0) \put(0,0){\vector(2,1){0}}}
    }
    \put(130,17){$\simeq$}
    \put(210,0) {
      \put(-40,0) {
        \linethickness{0.4mm} \thicklines
        \qbezier(0,40)(0,20)(-20,20) \put(-20,20){\vector(-1,0){0}}
        \put(20,0){\qbezier(0,40)(0,20)(20,20) \put(0,40){\vector(0,1){0}}}
        \put(20,0){\qbezier(0,0)(-10,10)(-20,0) \put(0,0){\vector(2,-1){0}}}
        \put(10,20){\circle{16}}
        \put(10,40){
          \linethickness{0.15mm} \thinlines
          \qbezier(-12,-11)(-12,-11)(-6,-15)
        }
        \put(10,40){
          \linethickness{0.15mm} \thinlines
          \qbezier(12,-11)(12,-11)(6,-15)
        }
        \put(10,5){
          \linethickness{0.15mm} \thinlines
          \qbezier(0,0)(0,5)(0,7)
        }
      }
      \put(10,17){$-\ [N-3]$}
      \put(80,0) {
        \linethickness{0.4mm} \thicklines
        \qbezier(0,40)(0,20)(-20,20) \put(-20,20){\vector(-1,0){0}}
        \put(20,0){\qbezier(0,40)(0,20)(20,20) \put(0,40){\vector(0,1){0}}}
        \put(20,0){\qbezier(0,0)(-10,10)(-20,0) \put(0,0){\vector(2,-1){0}}}
      }
    }
  \end{picture}
\\ \nn
& < :\Tr M^3: \  F(M) >\ =\
\Big((N-3)\delta^l_m\delta^n_p\delta^r_s
+\left( \delta^n_m \delta^l_p \delta^r_s + \delta^l_m \delta^r_p \delta^n_s + \delta^n_p \delta^l_s \delta^r_m \right) - \delta^r_m \delta^l_p \delta^n_s \Big)
\left<\frac{\p^3 F}{\p M^l_m\p M^n_p\p M^r_s}\right>
\ee

\noindent $\bullet$ \textbf{$1$-rule} -- we can swap edges between 3 strands
as shown:

\be \label{eq:1-rule}
& \begin{picture}(300,60)(-30,-10)
    \linethickness{0.4mm} \thicklines
    \put(0,0) {
      \linethickness{0.4mm} \thicklines
      \qbezier(0,0)(0,20)(0,40)
      \put(0,40){\vector(0,1){0}}
      \put(-10,10){\linethickness{0.15mm}
        \put(10,20){\line(1,0){20}}
      }
      \put(-10,-10){\linethickness{0.15mm}
        \put(10,20){\line(1,0){20}}
      }
    }
    \put(20,0) {
      \linethickness{0.4mm} \thicklines
      \qbezier(0,0)(0,20)(0,40)
      \put(0,40){\vector(0,1){0}}
      \put(-10,0){\linethickness{0.15mm}
        \put(10,20){\line(1,0){20}}
      }
    }
    \put(40,0) {
      \linethickness{0.4mm} \thicklines
      \qbezier(0,0)(0,20)(0,40)
      \put(0,40){\vector(0,1){0}}
    }
    \put(50,17){$-$}
    \put(65,0) {
      \linethickness{0.4mm} \thicklines
      \put(0,0) {
        \linethickness{0.4mm} \thicklines
        \qbezier(0,0)(0,20)(0,40)
        \put(0,40){\vector(0,1){0}}
        \put(-10,0){\linethickness{0.15mm}
          \put(10,20){\line(1,0){20}}
        }
      }
      \put(20,0) {
        \linethickness{0.4mm} \thicklines
        \qbezier(0,0)(0,20)(0,40)
        \put(0,40){\vector(0,1){0}}
      }
      \put(40,0) {
        \linethickness{0.4mm} \thicklines
        \qbezier(0,0)(0,20)(0,40)
        \put(0,40){\vector(0,1){0}}
      }
    }
    \put(120,17){$\simeq$}
    \put(140,0) {
      \linethickness{0.4mm} \thicklines
      \put(0,0) {
        \linethickness{0.4mm} \thicklines
        \qbezier(0,0)(0,20)(0,40)
        \put(0,40){\vector(0,1){0}}
        \put(-10,0){\linethickness{0.15mm}
          \put(10,20){\line(1,0){20}}
        }
      }
      \put(20,0) {
        \linethickness{0.4mm} \thicklines
        \qbezier(0,0)(0,20)(0,40)
        \put(0,40){\vector(0,1){0}}
        \put(-10,10){\linethickness{0.15mm}
          \put(10,20){\line(1,0){20}}
        }
        \put(-10,-10){\linethickness{0.15mm}
          \put(10,20){\line(1,0){20}}
        }
      }
      \put(40,0) {
        \linethickness{0.4mm} \thicklines
        \qbezier(0,0)(0,20)(0,40)
        \put(0,40){\vector(0,1){0}}
      }
      \put(50,17){$-$}
      \put(65,0) {
        \linethickness{0.4mm} \thicklines
        \put(0,0) {
          \linethickness{0.4mm} \thicklines
          \qbezier(0,0)(0,20)(0,40)
          \put(0,40){\vector(0,1){0}}
        }
        \put(20,0) {
          \linethickness{0.4mm} \thicklines
          \qbezier(0,0)(0,20)(0,40)
          \put(0,40){\vector(0,1){0}}
          \put(-10,0){\linethickness{0.15mm}
            \put(10,20){\line(1,0){20}}
          }
        }
        \put(40,0) {
          \linethickness{0.4mm} \thicklines
          \qbezier(0,0)(0,20)(0,40)
          \put(0,40){\vector(0,1){0}}
        }
      }
    }
  \end{picture}
\\ \nn
& <(M^i_\rho M^\rho_j) (M^k_\sigma M^\sigma_\eta M^\eta_l) M^m_n F^{j l n}_{i k m}(M)>
- \delta^m_n <M^i_j M^k_l F^{j l n}_{i k m}(M)>
\begin{picture}(0,0)
  \put(-270,12){\put(0,0){\line(1,0){35}} \put(0,0){\line(0,-1){3}} \put(35,0){\line(0,-1){3}}}
  \put(-255,18){\put(0,0){\line(1,0){55}} \put(0,0){\line(0,-1){8}} \put(55,0){\line(0,-1){8}}}
  \put(-215,12){\put(0,0){\line(1,0){35}} \put(0,0){\line(0,-1){3}} \put(35,0){\line(0,-1){3}}}
  \put(-77,12){\put(0,0){\line(1,0){15}}\put(0,0){\line(0,-1){3}} \put(15,0){\line(0,-1){3}}}
\end{picture}
\\ \nn
\\ \nn
& =\ <M^i_j (M^k_\sigma M^\sigma_\eta M^\eta_l) (M^m_\rho M^\rho_n) F^{j l n}_{i k m}(M)>
- \delta^i_j <M^k_l M^m_n F^{j l n}_{i k m}(M)>
\begin{picture}(0,0)
  \put(-275,12){\put(0,0){\line(1,0){35}} \put(0,0){\line(0,-1){3}} \put(35,0){\line(0,-1){3}}}
  \put(-255,18){\put(0,0){\line(1,0){55}} \put(0,0){\line(0,-1){8}} \put(55,0){\line(0,-1){8}}}
  \put(-225,12){\put(0,0){\line(1,0){40}} \put(0,0){\line(0,-1){3}} \put(40,0){\line(0,-1){3}}}
  \put(-80,12){\put(0,0){\line(1,0){15}}\put(0,0){\line(0,-1){3}} \put(15,0){\line(0,-1){3}}}
\end{picture}
\ee

\noindent $\bullet$ \textbf{flip-rule} -- we can change the number of vertices of the fat graph:

\be \label{eq:flip-rule}
& \begin{picture}(300,60)(-100,-20)
    \linethickness{0.4mm} \thicklines
    \put(0,0) {
      \linethickness{0.4mm} \thicklines
      \qbezier(0,0)(20,20)(0,40)
      \put(0,40){\vector(-1,1){0}}
      \put(40,0){\qbezier(0,0)(-20,20)(0,40)}
      \put(40,0){\vector(1,-1){0}}
      \put(0,0){\linethickness{0.15mm}
        \put(10,20){\line(1,0){20}}
      }
    }
    \put(50,17){$\simeq\ \ -$}
    \put(75,0) {
      \linethickness{0.4mm} \thicklines
      \qbezier(0,0)(20,20)(40,0)
      \put(0,40){\vector(-1,1){0}}
      \put(40,40){\qbezier(0,0)(-20,-20)(-40,0)}
      \put(40,0){\vector(1,-1){0}}
      \put(0,0){\linethickness{0.15mm}
        \put(20,10){\line(0,1){20}}
      }
    }
  \end{picture}
\\ \nn
& <M^i_j M^k_l F^{j l}_{i k}(M)>\ =\ - <M^i_l M^k_j F^{j l}_{i k}(M)>
\begin{picture}(0,0)
  \put(-182.5,12){\put(0,0){\line(1,0){15}}\put(0,0){\line(0,-1){3}} \put(15,0){\line(0,-1){3}}}
\end{picture}
\begin{picture}(0,0)
  \put(-72.5,12){\put(0,0){\line(1,0){15}}\put(0,0){\line(0,-1){3}} \put(15,0){\line(0,-1){3}}}
\end{picture}
\ee

One can notice that MM-average analogues for transformations
\eqref{eq:2-rule}, \eqref{eq:1-rule} and \eqref{eq:flip-rule}
are somewhat different from ones for \eqref{eq:n-1-rule}, \eqref{eq:n-2-rule} and \eqref{eq:n-3-rule}.
Though they are equally easy to derive at $q = 1$ from explicit form of propagator
\eqref{propknot}, they are not so easy to express in terms of normal orderings.
This is because normal ordering sign is very good at saying which
matrices \textit{should not} pair, but can't be used to say which matrices \textit{should} pair.
And in these rules it is precisely this ``should pair'' relation, that we need
to express algebraically -- hence we resort to explicit pairing sign.

As we already emphasized in \cite{MMP1}, the flip-rule \eqref{eq:flip-rule} is different
from all the other rules in this section. As the next section demonstrates,
it is not required ensure Reidemeister invariance (under both usual and virtual moves!).
So, in principle, it should be possible to define dimensions
$D_\gamma$ differently, such that all other rules of this section are satisfied,
but the flip-rule is not, and result of hypercube calculation \eqref{HMMPmod} will still
be topological invariant. Investigation of this possibility is, however, out of scope
of the present paper.

\section{Reidemeister invariance justifying fat graph recursion rules
\label{sec:reid}} {
  \newtheorem{teo}{Theorem}[section]

  In this section we prove the following

  \begin{teo}
    The result of hypercube calculation \eqref{eq:hypercube-sum-over-fatgraphs} is a topological invariant,
    provided local fat graph transformations \eqref{eq:n-1-rule}- \eqref{eq:1-rule},
    i.e. all the rules from the section \ref{sec:n-rules},
    except for the flip-rule \eqref{eq:flip-rule},
    and factorization properties \eqref{fact}
 and \eqref{fact-reduced} hold.
  \end{teo}

  To prove this, it is sufficient to prove that the answer is invariant w.r.t
  all the Reidemeister moves (usual as well as virtual). In the following subsections
  we analyze invariance of \eqref{eq:hypercube-sum-over-fatgraphs} under each of them in turn.
  This analysis requires choice of types of intersections (black or white) and
  orientation of strands.
  To save space, we consider explicitly only essentially different cases.

  \subsection{1st Reidemeister}

  1st Reidemeister move is (without loss of generality,
  we choose black type of intersection -- analysis of the white one is the same)

  \begin{equation}
    \label{eq:1st-reidemeister}
    \noindent
    \begin{picture}(300,55)(-200,-25)
      \thicklines
      \put(-80,0){
        \put(0,10){\vector(0,1){10}}
        \put(0,-20){\line(0,1){10}}
        \qbezier(0,-10)(0,10)(20,10)
        \qbezier(20,10)(30,10)(30,0) \qbezier(30,0)(30,-10)(20,-10)
        \qbezier(20,-10)(10,-10)(5,0) \qbezier(2,5)(0,8)(0,10)
      }
      \put(-40,-2){\mbox{$=$}}
      \put(-20,0){\put(0,-20){\vector(0,1){40}}}
    \end{picture}
  \end{equation}

  \noindent Now let's imagine we do hypercube calculation for some link diagram ${\cal L}$
  with a loop like left hand side of \eqref{eq:1st-reidemeister}.
  Then sum \eqref{HMMPmod} over all subgraphs $\gamma$
  splits into two big subsums: where the edge corresponding
  to the intersection of this loop is kept, and where it is amputated
  (here we implicitly assume summation over keeping/deleting all other edges of the fat graph
  and don't draw it explicitly)

  \begin{equation}
    \label{eq:1st-reid-cube-calc}
    \noindent
    \begin{picture}(300,55)(-160,-25)
      \put(-100,-2){$\mathcal{H}_{lhs}\ = \left(-q \right)\cdot$}
      \put(10,0) {
        \thicklines
        \put(-40,0){
          \put(0,-20){\vector(0,1){40}}
          \put(20,0){\circle{20}}
          \put(25,10){\vector(1,0){0}}
          \qbezier[5](0,0)(5,0)(10,0)
        }
        \put(0,-2){\mbox{$+$}}
        \put(20,0){
          \put(0,-20){\vector(0,1){40}}
          \put(20,0){\circle{20}}
          \put(25,10){\vector(1,0){0}}
        }
      }
    \end{picture}
  \end{equation}

  \noindent Now, applying $[N-1]$-rule \eqref{eq:n-1-rule} and $[N]$-rule \eqref{eq:n-rule}
  we see, that it's equal
  \begin{equation}
  \label{1Rblackfact}
    \noindent
    \begin{picture}(300,55)(-150,-25)
      \put(-100,-2){$\mathcal{H}_{lhs}\ = \Big(\left(-q \right)\cdot[N-1] + [N]\Big)\cdot $}
      \put(15,0) {
        \thicklines
        \put(30,0){
          \put(0,-20){\vector(0,1){40}}
        }
        \put(45,-2){$= q^{1-N}\cdot$}
        \put(85,0){
          \put(0,-20){\vector(0,1){40}}
        }
      }
    \end{picture}
  \end{equation}

  \noindent At first sight, the \eqref{HMMPmod} sum for the right hand side
  has just one group of summands and with a wrong factor
  \begin{equation}
    \noindent
    \begin{picture}(300,55)(-220,-25)
      \put(-100,-2){$\mathcal{H}_{rhs}\ =$}
      \put(-50,0){
        \thicklines \put(0,-20){\vector(0,1){40}}
      }
    \end{picture}
  \end{equation}

  \noindent but one also needs to take into account
  that the total $q$-charge of the l.h.s. hypercube ($n_\bullet - n_\circ$)
  is bigger by one, because of an extra black edge, giving an extra factor $q^{N-1}$,
  so in fact we have
  \begin{equation}
  \label{1Rblackequ}
    \mathcal{H}_{lhs} = \mathcal{H}_{rhs}
  \end{equation}

\bigskip

If original vertex in (\ref{eq:1st-reidemeister}) was white, the horizontal edge in the first item
of (\ref{eq:1st-reid-cube-calc})
is straight, thus it enters with the coefficient $(-q)^{-1}$ and the factor in (\ref{1Rblackfact})
is $(-q)^{-1}\cdot[N-1]+[N] = q^{N-1}$.
At the same time the total $q$-charge of the l.h.s. hypercube ($n_{..} - n_-$)
  is now {\it smaller} by one and the prefactor is $q^{1-N}$, thus (\ref{1Rblackequ}) is
  again true.

  \subsection{2nd Reidemeister with parallel strands}

  Second Reidemeister move with parallel orientation of strands
  is

  \begin{equation}
    \label{eq:2nd-parallel-reidemeister}
    \noindent
    \begin{picture}(300,55)(-130,-25)
      \thicklines
      \put(-80,0){
        \qbezier(0,-20)(5,-15)(10,-10)
        \qbezier(10,-10)(20,0)(10,10)
        \put(10,10){\vector(-1,1){10}}
        \qbezier(20,-20)(15,-15)(12,-12)
        \qbezier(20,20)(15,15)(12,12)
        \put(20,20){\vector(1,1){0}}
        \qbezier(8,-8)(0,0)(8,8)
      }
      \put(-40,-2){\mbox{$=$}}
      \put(-20,0){\put(0,-20){\vector(0,1){40}} \put(20,-20){\vector(0,1){40}}}
    \end{picture}
  \end{equation}

  \noindent It is important, that one of the vertices of the planar diagram is black and the other is white,
  so fat graphs are essentially bi-colored.
  As everywhere in this paper we denote black edges as dotted straight lines
  and white edges as straight dotted lines.
  We need to draw them differently to easily track the correct $q$-prefactors.
  But let's remind once more, that quantum dimensions do not depend on colors of edges of the fat graph.
  Hence
  all the fat-graph equivalences of section \ref{sec:n-rules} apply
  regardless of the color of the edges.

  This time hypercube sum for the left hand side has four groups of summands.

  \begin{equation}
    \noindent
    \begin{picture}(300,55)(-130,-25)
      \put(-100,-2){$\mathcal{H}_{lhs}\ = $}
      \thicklines
      \put(10,0) {
        \put(-40,0){
          \put(-15,-2){$1\ \cdot$}
          \put(0,-20){\vector(0,1){40}}
          \put(20,-20){\vector(0,1){40}}
          \thinlines \linethickness{0.1mm}
          \put(0,-10) {
            \thinlines \linethickness{0.25mm}
            \qbezier[10](0,0)(10,0)(20,0)
          }
          \put(0,10){\line(1,0){20}}
        }
        \put(30,0){
          \put(-40,-2){$+\ \left(-q\right)$}
          \put(0,-20){\vector(0,1){40}}
          \put(20,-20){\vector(0,1){40}}
          \put(0,-10) {
            \thinlines \linethickness{0.25mm}
            \qbezier[10](0,0)(10,0)(20,0)
          }
        }
        \put(100,0){
          \put(-40,-2){$+\ \left(-\frac{1}{q}\right)$}
          \put(0,-20){\vector(0,1){40}}
          \put(20,-20){\vector(0,1){40}}
          \thinlines \linethickness{0.1mm}
          \put(0,10){\line(1,0){20}}
        }
        \put(170,0){
          \put(-30,-2){$+\ 1\ \cdot$}
          \put(0,-20){\vector(0,1){40}}
          \put(20,-20){\vector(0,1){40}}
        }
      }
    \end{picture}
  \end{equation}

  \noindent while hypercube sum for the right hand side still has only one group
  (an extra factor due to the difference in the total fat graph charges is now 1,
  because we have one extra black and one extra white edges):

  \begin{equation}
    \noindent
    \begin{picture}(300,55)(-130,-25)
      \put(-100,-2){$\mathcal{H}_{rhs}\ = $}
      \thicklines
      \put(10,0) {
        \put(-40,0){
          \put(-15,-2){$1\ \cdot$}
          \put(0,-20){\vector(0,1){40}}
          \put(20,-20){\vector(0,1){40}}
        }
      }
    \end{picture}
  \end{equation}

  \noindent Applying $[2]$-rule \eqref{eq:2-rule}
  to the first group in $\mathcal{H}_{lhs}$ we see that first three
  groups mutually cancel, and hence in this case lhs also coincides with rhs.

  \subsection{2nd Reidemeister with antiparallel strands}

  Second Reidemeister move with antiparallel orientation of strands
  is

  \begin{equation}
    \label{eq:2nd-antiparallel-reidemeister}
    \noindent
    \begin{picture}(300,55)(-130,-25)
      \thicklines
      \put(-80,0){
        \qbezier(0,-20)(5,-15)(10,-10)
        \qbezier(10,-10)(20,0)(10,10)
        \put(10,10){\vector(-1,1){10}}
        \qbezier(20,-20)(15,-15)(12,-12)
        \qbezier(20,20)(15,15)(12,12)
        \put(20,-20){\vector(1,-1){0}}
        \qbezier(8,-8)(0,0)(8,8)
      }
      \put(-40,-2){\mbox{$=$}}
      \put(-20,0){\put(0,-20){\vector(0,1){40}} \put(20,20){\vector(0,-1){40}}}
    \end{picture}
  \end{equation}

  \noindent This time hypercube sum for the left hand side again has four groups of summands

  \begin{equation}
    \indent
    \begin{picture}(300,55)(-80,-25)
      \put(-100,-2){$\mathcal{H}_{lhs}\ = $}
      \thicklines
      \put(10,0) {
        \put(-40,0){
          \put(-15,-2){$1\ \cdot$}
          \qbezier(20,20)(10,10)(0,20) \put(0,20){\vector(-1,1){0}}
          \qbezier(0,-20)(10,-10)(20,-20) \put(20,-20){\vector(1,-1){0}}
          \put(10,0){\put(0,0){\circle{15}} \put(-7.5,-5){\vector(0,-1){0}}}
          \thinlines \linethickness{0.1mm}
          \put(10,7.5){
            \thinlines \linethickness{0.25mm}
            \qbezier[6](0,0)(0,4)(0,8)
          }
          \put(10,-7.5) {
            \thinlines
            \qbezier(0,0)(0,-4)(0,-8)
          }
        }
        \put(30,0){
          \put(-40,-2){$+\ \left(-q\right)$}
          \qbezier(20,20)(10,10)(0,20) \put(0,20){\vector(-1,1){0}}
          \qbezier(0,-20)(10,-10)(20,-20) \put(20,-20){\vector(1,-1){0}}
          \put(10,0){\put(0,0){\circle{15}} \put(-7.5,-5){\vector(0,-1){0}}}
          \put(10,7.5){
            \thinlines \linethickness{0.25mm}
            \qbezier[6](0,0)(0,4)(0,8)
          }
        }
        \put(100,0){
          \put(-40,-2){$+\ \left(-\frac{1}{q}\right)$}
          \qbezier(20,20)(10,10)(0,20) \put(0,20){\vector(-1,1){0}}
          \qbezier(0,-20)(10,-10)(20,-20) \put(20,-20){\vector(1,-1){0}}
          \put(10,0){\put(0,0){\circle{15}} \put(-7.5,-5){\vector(0,-1){0}}}
          \put(10,-7.5) {
            \thinlines
            \qbezier(0,0)(0,-4)(0,-8)
          }
        }
        \put(170,0){
          \put(-30,-2){$+\ 1\ \cdot$}
          \qbezier(20,20)(10,10)(0,20) \put(0,20){\vector(-1,1){0}}
          \qbezier(0,-20)(10,-10)(20,-20) \put(20,-20){\vector(1,-1){0}}
          \put(10,0){\put(0,0){\circle{15}} \put(-7.5,-5){\vector(0,-1){0}}}
        }
      }
    \end{picture}
  \end{equation}

  \noindent while hypercube sum for the right hand side still has only one group
  (an extra factor due to the difference in the total fat graph charges is
  again absent)

  \begin{equation}
    \noindent
    \begin{picture}(300,55)(-130,-25)
      \put(-100,-2){$\mathcal{H}_{rhs}\ = $}
      \thicklines
      \put(10,0) {
        \put(-40,0){
          \put(-15,-2){$1\ \cdot$}
          \put(0,-20){\vector(0,1){40}}
          \put(20,-20){\vector(0,1){40}}
        }
      }
    \end{picture}
  \end{equation}

  \noindent Applying $[N-2]$-rule \eqref{eq:n-2-rule}
  to the first group in $\mathcal{H}_{lhs}$,
  $[N-1]$-rule \eqref{eq:n-1-rule} to the second and third groups
  and $[N]$-rule \eqref{eq:n-rule} to
  the fourth group and using simple identity for $q$-numbers
  \begin{equation}
    [2][N-1] = [N] + [N-2]
  \end{equation}
  we see that in this case also $\mathcal{H}_{lhs} = \mathcal{H}_{rhs}$.

  \subsection{``Parallel'' 3rd Reidemeister}

  As third Reidemeister move involves three strands, describing
  their mutual orientation as being ``parallel'' or ``antiparallel''
  is not exactly accurate.

  However, there are two essentially different ways to orient
  diagram of 3rd Reidemeister move: one contains extra Seifert cycle
  and the other does not. The check of invariance is different for these
  two ways, but literally the same for two orientation choices which belong
  to the same way. In this subsection we consider the way, which
  does not create extra Seifert cycle (and call it ``parallel'' for brevity)

  One of the typical representatives of this class of orientations, is

\begin{equation}
  \label{eq:3rd-parallel-reidemeister}
  \noindent
  \begin{picture}(300,55)(-130,-25)
    \linethickness{0.3mm} \thicklines
    \qbezier(0,-30)(-30,0)(0,30)
    \qbezier(-30,-10)(-20,-10)(-18,-10) \qbezier(-30,10)(-20,10)(-18,10)
    \qbezier(30,10)(20,-10)(-8,-9)
    \qbezier(30,-10)(28,-5)(25,-2) \qbezier(21,3)(10,10)(-8,9)
    \put(0,30){\vector(1,1){0}} \put(-5,-25){\vector(-1,1){0}} \put(-15,5){\vector(0,1){0}}
    \put(-20,-10){\vector(1,0){0}} \put(10,-8){\vector(1,0){0}} \put(30,10){\vector(2,3){0}}
    \put(-30,10){\vector(-1,0){0}} \put(0,9){\vector(-1,0){0}} \put(27,-5){\vector(-2,3){0}}
    \put(45,-2){$=$}
    \put(100,0) {
      \qbezier(0,30)(30,0)(0,-30)
      \qbezier(30,10)(20,10)(18,10) \qbezier(30,-10)(20,-10)(18,-10)
      \qbezier(-30,-10)(-20,10)(8,9)
      \qbezier(-30,10)(-28,5)(-25,2) \qbezier(-21,-3)(-10,-10)(8,-9)
      \put(0,30){\vector(-1,1){0}} \put(5,-25){\vector(1,1){0}} \put(15,5){\vector(0,1){0}}
      \put(25,-10){\vector(-1,0){0}} \put(-10,-8){\vector(-1,0){0}} \put(-30,10){\vector(-2,3){0}}
      \put(30,10){\vector(1,0){0}} \put(0,9){\vector(1,0){0}} \put(-25,-3){\vector(2,3){0}}
    }
  \end{picture}
\end{equation}

Total charges of hypercubes on the lhs and on the rhs are the same.
Let's explicitly write out different groups of contributions
(again, in all the pictures the rest of the fat graph, as well as
summation over removal of all its other edges is not drawn, but is, of course, assumed,
therefore each item in the following sums is actually a group of items
-- what is important, it is the same at both sides of Reidemeister identity)

\begin{equation}
  \noindent
  \begin{picture}(300,155)(-50,-125)
    \put(-100,-2){$\mathcal{H}_{lhs}\ = $}
    \thicklines
    \put(30,0) {
      \put(-40,0){
        \put(-50,-2){$\left(-q\right) \cdot$}
        \qbezier(-20,-15)(-5,0)(-20,15) \put(-20,15){\vector(-1,1){0}}
        \put(0,-30){\vector(0,1){60}}
        \qbezier(20,-15)(5,0)(20,15) \put(20,15){\vector(1,1){0}}
        \thinlines \linethickness{0.1mm}
        \put(-20,10) {
          \thinlines \linethickness{0.25mm}
          \qbezier[8](5,0)(10,0)(20,0)
        }
        \put(0,0) {
          \thinlines \linethickness{0.25mm}
          \qbezier[8](0,0)(6,0)(12,0)
        }
        \put(0,-10){\line(-1,0){15}}
      }
    }
    \put(25,-2){$+$}
    \put(120,0) {
      \put(-40,0){
        \put(-40,-2){$q^2 \cdot$}
        \qbezier(-20,-15)(-5,0)(-20,15) \put(-20,15){\vector(-1,1){0}}
        \put(0,-30){\vector(0,1){60}}
        \qbezier(20,-15)(5,0)(20,15) \put(20,15){\vector(1,1){0}}
        \thinlines \linethickness{0.1mm}
        \put(-20,10) {
          \thinlines \linethickness{0.25mm}
          \qbezier[8](5,0)(10,0)(20,0)
        }
        \put(0,0) {
          \thinlines \linethickness{0.25mm}
          \qbezier[8](0,0)(6,0)(12,0)
        }
      }
    }
    \put(115,-2){$+$}
    \put(200,0) {
      \put(-40,0){
        \put(-30,-2){$1 \cdot$}
        \qbezier(-20,-15)(-5,0)(-20,15) \put(-20,15){\vector(-1,1){0}}
        \put(0,-30){\vector(0,1){60}}
        \qbezier(20,-15)(5,0)(20,15) \put(20,15){\vector(1,1){0}}
        \thinlines \linethickness{0.1mm}
        \put(-20,10) {
          \thinlines \linethickness{0.25mm}
          \qbezier[8](5,0)(10,0)(20,0)
        }
        \put(0,-10){\line(-1,0){15}}
      }
    }
    \put(190,-2){$+$}
    \put(275,0) {
      \put(-40,0){
        \put(-30,-2){$1 \cdot$}
        \qbezier(-20,-15)(-5,0)(-20,15) \put(-20,15){\vector(-1,1){0}}
        \put(0,-30){\vector(0,1){60}}
        \qbezier(20,-15)(5,0)(20,15) \put(20,15){\vector(1,1){0}}
        \thinlines \linethickness{0.1mm}
        \put(0,0) {
          \thinlines \linethickness{0.25mm}
          \qbezier[8](0,0)(6,0)(12,0)
        }
        \put(0,-10){\line(-1,0){15}}
      }
    }
    \put(0,-80) {
      \put(30,0) {
        \put(-40,0){
          \put(-50,-2){$\left(-q\right) \cdot$}
          \qbezier(-20,-15)(-5,0)(-20,15) \put(-20,15){\vector(-1,1){0}}
          \put(0,-30){\vector(0,1){60}}
          \qbezier(20,-15)(5,0)(20,15) \put(20,15){\vector(1,1){0}}
          \thinlines \linethickness{0.1mm}
          \put(0,0) {
            \thinlines \linethickness{0.25mm}
            \qbezier[8](0,0)(6,0)(12,0)
          }
        }
      }
      \put(15,-2){$+$}
      \put(120,0) {
        \put(-40,0){
          \put(-50,-2){$\left(-q\right) \cdot$}
          \qbezier(-20,-15)(-5,0)(-20,15) \put(-20,15){\vector(-1,1){0}}
          \put(0,-30){\vector(0,1){60}}
          \qbezier(20,-15)(5,0)(20,15) \put(20,15){\vector(1,1){0}}
          \thinlines \linethickness{0.1mm}
          \put(-20,10) {
            \thinlines \linethickness{0.25mm}
            \qbezier[8](5,0)(10,0)(20,0)
          }
        }
      }
      \put(103,-2){$+$}
      \put(200,0) {
        \put(-40,0){
          \put(-50,-2){$\left (-\frac{\displaystyle 1}{\displaystyle q} \right ) \cdot$}
          \qbezier(-20,-15)(-5,0)(-20,15) \put(-20,15){\vector(-1,1){0}}
          \put(0,-30){\vector(0,1){60}}
          \qbezier(20,-15)(5,0)(20,15) \put(20,15){\vector(1,1){0}}
          \thinlines \linethickness{0.1mm}
          \put(0,-10){\line(-1,0){15}}
        }
      }
      \put(190,-2){$+$}
      \put(275,0) {
        \put(-40,0){
          \put(-30,-2){$1 \cdot$}
          \qbezier(-20,-15)(-5,0)(-20,15) \put(-20,15){\vector(-1,1){0}}
          \put(0,-30){\vector(0,1){60}}
          \qbezier(20,-15)(5,0)(20,15) \put(20,15){\vector(1,1){0}}
          \thinlines \linethickness{0.1mm}
        }
      }
    }
  \end{picture}
\end{equation}

\begin{equation}
  \notag \noindent
  \begin{picture}(300,155)(-50,-125)
    \put(-100,-2){$\mathcal{H}_{rhs}\ = $}
    \thicklines
    \put(30,0) {
      \put(-40,0){
        \put(-50,-2){$\left(-q \right) \cdot$}
        \qbezier(-20,-15)(-5,0)(-20,15) \put(-20,15){\vector(-1,1){0}}
        \put(0,-30){\vector(0,1){60}}
        \qbezier(20,-15)(5,0)(20,15) \put(20,15){\vector(1,1){0}}
        \thinlines \linethickness{0.1mm}
        \put(20,-10) {
          \thinlines \linethickness{0.25mm}
          \qbezier[8](-5,0)(-10,0)(-20,0)
        }
        \put(0,0) {
          \thinlines \linethickness{0.25mm}
          \qbezier[8](0,0)(-6,0)(-12,0)
        }
        \put(0,10){\line(1,0){15}}
      }
    }
    \put(25,-2){$+$}
    \put(120,0) {
      \put(-40,0){
        \put(-40,-2){$q^2 \cdot$}
        \qbezier(-20,-15)(-5,0)(-20,15) \put(-20,15){\vector(-1,1){0}}
        \put(0,-30){\vector(0,1){60}}
        \qbezier(20,-15)(5,0)(20,15) \put(20,15){\vector(1,1){0}}
        \thinlines \linethickness{0.1mm}
        \put(20,-10) {
          \thinlines \linethickness{0.25mm}
          \qbezier[8](-5,0)(-10,0)(-20,0)
        }
        \put(0,0) {
          \thinlines \linethickness{0.25mm}
          \qbezier[8](0,0)(-6,0)(-12,0)
        }
      }
    }
    \put(115,-2){$+$}
    \put(200,0) {
      \put(-40,0){
        \put(-30,-2){$1 \cdot$}
        \qbezier(-20,-15)(-5,0)(-20,15) \put(-20,15){\vector(-1,1){0}}
        \put(0,-30){\vector(0,1){60}}
        \qbezier(20,-15)(5,0)(20,15) \put(20,15){\vector(1,1){0}}
        \thinlines \linethickness{0.1mm}
        \put(20,-10) {
          \thinlines \linethickness{0.25mm}
          \qbezier[8](-5,0)(-10,0)(-20,0)
        }
        \put(0,10){\line(1,0){15}}
      }
    }
    \put(190,-2){$+$}
    \put(275,0) {
      \put(-40,0){
        \put(-30,-2){$1 \cdot$}
        \qbezier(-20,-15)(-5,0)(-20,15) \put(-20,15){\vector(-1,1){0}}
        \put(0,-30){\vector(0,1){60}}
        \qbezier(20,-15)(5,0)(20,15) \put(20,15){\vector(1,1){0}}
        \thinlines \linethickness{0.1mm}
        \put(0,0) {
          \thinlines \linethickness{0.25mm}
          \qbezier[8](0,0)(-6,0)(-12,0)
        }
        \put(0,10){\line(1,0){15}}
      }
    }
    \put(0,-80) {
      \put(30,0) {
        \put(-40,0){
          \put(-50,-2){$\left(-q\right) \cdot$}
          \qbezier(-20,-15)(-5,0)(-20,15) \put(-20,15){\vector(-1,1){0}}
          \put(0,-30){\vector(0,1){60}}
          \qbezier(20,-15)(5,0)(20,15) \put(20,15){\vector(1,1){0}}
          \thinlines \linethickness{0.1mm}
          \put(0,0) {
            \thinlines \linethickness{0.25mm}
            \qbezier[8](0,0)(-6,0)(-12,0)
          }
        }
      }
      \put(15,-2){$+$}
      \put(120,0) {
        \put(-40,0){
          \put(-50,-2){$\left(-q \right) \cdot$}
          \qbezier(-20,-15)(-5,0)(-20,15) \put(-20,15){\vector(-1,1){0}}
          \put(0,-30){\vector(0,1){60}}
          \qbezier(20,-15)(5,0)(20,15) \put(20,15){\vector(1,1){0}}
          \thinlines \linethickness{0.1mm}
          \put(20,-10) {
            \thinlines \linethickness{0.25mm}
            \qbezier[8](-5,0)(-10,0)(-20,0)
          }
        }
      }
      \put(103,-2){$+$}
      \put(200,0) {
        \put(-40,0){
          \put(-50,-2){$\left(-\frac{\displaystyle 1}{\displaystyle q}\right) \cdot$}
          \qbezier(-20,-15)(-5,0)(-20,15) \put(-20,15){\vector(-1,1){0}}
          \put(0,-30){\vector(0,1){60}}
          \qbezier(20,-15)(5,0)(20,15) \put(20,15){\vector(1,1){0}}
          \thinlines \linethickness{0.1mm}
          \put(20,-10) {
            \thinlines \linethickness{0.25mm}
            \qbezier[8](-5,0)(-10,0)(-20,0)
          }
        }
      }
      \put(190,-2){$+$}
      \put(275,0) {
        \put(-40,0){
          \put(-30,-2){$1 \cdot$}
          \qbezier(-20,-15)(-5,0)(-20,15) \put(-20,15){\vector(-1,1){0}}
          \put(0,-30){\vector(0,1){60}}
          \qbezier(20,-15)(5,0)(20,15) \put(20,15){\vector(1,1){0}}
          \thinlines \linethickness{0.1mm}
        }
      }
    }
  \end{picture}
\end{equation}

\noindent
We see, that if we apply $[2]$-rule \eqref{eq:2-rule}
to the third group of summands (on both sides), it
cancels with the sixth and seventh groups. Furthermore, the second, fourth and eighth groups
are the same on the left hand side and on the right hand side.
It may seem that fourth groups differ
(because on the l.h.s right edge is dotted, while on the r.h.s left edge is dotted),
but recall that the color of the edges actually does not matter for $q$-dimensions.
Finally, remaining groups are equal because of the $1$-rule \eqref{eq:1-rule}.

\subsection{``Antiparallel'' 3rd Reidemeister}

Typical representative of the ``antiparallel'' 3rd Reidemeister moves (i.e. ones that give extra
Seifert cycle) is

\begin{equation}
  \label{eq:3rd-antiparallel-reidemeister}
  \noindent
  \begin{picture}(300,55)(-130,-25)
    \linethickness{0.3mm} \thicklines
    \qbezier(0,-30)(-30,0)(0,30)
    \qbezier(-30,-10)(-20,-10)(-18,-10) \qbezier(-30,10)(-20,10)(-18,10)
    \qbezier(30,10)(20,-10)(-8,-9)
    \qbezier(30,-10)(28,-5)(25,-2) \qbezier(21,3)(10,10)(-8,9)
    \put(0,30){\vector(1,1){0}} \put(-5,-25){\vector(-1,1){0}} \put(-15,5){\vector(0,1){0}}
    \put(-30,-10){\vector(-1,0){0}} \put(0,-8){\vector(-1,0){0}} \put(27,4){\vector(-2,-3){0}}
    \put(-20,10){\vector(1,0){0}} \put(10,8){\vector(1,0){0}} \put(30,-10){\vector(2,-3){0}}
    \put(45,-2){$=$}
    \put(100,0) {
      \qbezier(0,30)(30,0)(0,-30)
      \qbezier(30,10)(20,10)(18,10) \qbezier(30,-10)(20,-10)(18,-10)
      \qbezier(-30,-10)(-20,10)(8,9)
      \qbezier(-30,10)(-28,5)(-25,2) \qbezier(-21,-3)(-10,-10)(8,-9)
      \put(0,30){\vector(-1,1){0}} \put(5,-25){\vector(1,1){0}} \put(15,5){\vector(0,1){0}}
      \put(30,-10){\vector(1,0){0}} \put(0,-8){\vector(1,0){0}} \put(-27,4){\vector(2,-3){0}}
      \put(20,10){\vector(-1,0){0}} \put(-10,8){\vector(-1,0){0}} \put(-30,-10){\vector(-2,-3){0}}
    }
  \end{picture}
\end{equation}

\noindent Again, the total charges of lhs- and rhs-hypercubes match. Explicitly
writing out all groups of summands, we get:

\begin{equation}
  \noindent
  \begin{picture}(300,155)(-50,-125)
    \put(-100,-2){$\mathcal{H}_{lhs}\ = $}
    \thicklines
    \put(30,0) {
      \put(-40,0){
        \put(-50,-2){$\left(-q\right) \cdot$}
        \qbezier(0,-30)(-5,-15)(-20,-15) \put(-20,-15){\vector(-1,0){0}}
        \qbezier(0,30)(-5,15)(-20,15) \put(0,30){\vector(1,2){0}}
        \qbezier(25,-15)(10,0)(25,15) \put(25,-15){\vector(1,-1){0}}
        \put(0,0){\circle{16}} \put(-8,5){\vector(0,1){0}}
        \put(0,0){\thinlines \linethickness{0.15mm} \put(-8,18){\line(1,-2){5}}}
        \put(8,0){\thinlines \linethickness{0.25mm} \qbezier[6](0,0)(5,0)(10,0)}
        \put(-8,-18){\thinlines \linethickness{0.25mm} \qbezier[6](0,0)(2.5,5)(5,10)}
      }
    }
    \put(25,-2){$+$}
    \put(120,0) {
      \put(-40,0){
        \put(-40,-2){$q^2 \cdot$}
        \qbezier(0,-30)(-5,-15)(-20,-15) \put(-20,-15){\vector(-1,0){0}}
        \qbezier(0,30)(-5,15)(-20,15) \put(0,30){\vector(1,2){0}}
        \qbezier(25,-15)(10,0)(25,15) \put(25,-15){\vector(1,-1){0}}
        \put(0,0){\circle{16}} \put(-8,5){\vector(0,1){0}}
        %
        \put(8,0){\thinlines \linethickness{0.25mm} \qbezier[6](0,0)(5,0)(10,0)}
        \put(-8,-18){\thinlines \linethickness{0.25mm} \qbezier[6](0,0)(2.5,5)(5,10)}
      }
    }
    \put(115,-2){$+$}
    \put(200,0) {
      \put(-40,0){
        \put(-30,-2){$1 \cdot$}
        \qbezier(0,-30)(-5,-15)(-20,-15) \put(-20,-15){\vector(-1,0){0}}
        \qbezier(0,30)(-5,15)(-20,15) \put(0,30){\vector(1,2){0}}
        \qbezier(25,-15)(10,0)(25,15) \put(25,-15){\vector(1,-1){0}}
        \put(0,0){\circle{16}} \put(-8,5){\vector(0,1){0}}
        \put(0,0){\thinlines \linethickness{0.15mm} \put(-8,18){\line(1,-2){5}}}
        \put(8,0){\thinlines \linethickness{0.25mm} \qbezier[6](0,0)(5,0)(10,0)}
      }
    }
    \put(190,-2){$+$}
    \put(275,0) {
      \put(-40,0){
        \put(-30,-2){$1 \cdot$}
        \qbezier(0,-30)(-5,-15)(-20,-15) \put(-20,-15){\vector(-1,0){0}}
        \qbezier(0,30)(-5,15)(-20,15) \put(0,30){\vector(1,2){0}}
        \qbezier(25,-15)(10,0)(25,15) \put(25,-15){\vector(1,-1){0}}
        \put(0,0){\circle{16}} \put(-8,5){\vector(0,1){0}}
        \put(0,0){\thinlines \linethickness{0.15mm} \put(-8,18){\line(1,-2){5}}}
        \put(-8,-18){\thinlines \linethickness{0.25mm} \qbezier[6](0,0)(2.5,5)(5,10)}
      }
    }
    \put(0,-80) {
      \put(30,0) {
        \put(-40,0){
          \put(-50,-2){$\left(-q\right) \cdot$}
          \qbezier(0,-30)(-5,-15)(-20,-15) \put(-20,-15){\vector(-1,0){0}}
          \qbezier(0,30)(-5,15)(-20,15) \put(0,30){\vector(1,2){0}}
          \qbezier(25,-15)(10,0)(25,15) \put(25,-15){\vector(1,-1){0}}
          \put(0,0){\circle{16}} \put(-8,5){\vector(0,1){0}}
          %
          \put(8,0){\thinlines \linethickness{0.25mm} \qbezier[6](0,0)(5,0)(10,0)}
        }
      }
      \put(15,-2){$+$}
      \put(120,0) {
        \put(-40,0){
          \put(-50,-2){$\left(-q\right) \cdot$}
          \qbezier(0,-30)(-5,-15)(-20,-15) \put(-20,-15){\vector(-1,0){0}}
          \qbezier(0,30)(-5,15)(-20,15) \put(0,30){\vector(1,2){0}}
          \qbezier(25,-15)(10,0)(25,15) \put(25,-15){\vector(1,-1){0}}
          \put(0,0){\circle{16}} \put(-8,5){\vector(0,1){0}}
          %
          \put(-8,-18){\thinlines \linethickness{0.25mm} \qbezier[6](0,0)(2.5,5)(5,10)}
        }
      }
      \put(105,-2){$+$}
      \put(210,0) {
        \put(-40,0){
          \put(-55,-2){$\left(-\frac{\displaystyle 1}{\displaystyle q}\right) \cdot$}
          \qbezier(0,-30)(-5,-15)(-20,-15) \put(-20,-15){\vector(-1,0){0}}
          \qbezier(0,30)(-5,15)(-20,15) \put(0,30){\vector(1,2){0}}
          \qbezier(25,-15)(10,0)(25,15) \put(25,-15){\vector(1,-1){0}}
          \put(0,0){\circle{16}} \put(-8,5){\vector(0,1){0}}
          \put(0,0){\thinlines \linethickness{0.15mm} \put(-8,18){\line(1,-2){5}}}
        }
      }
      \put(200,-2){$+$}
      \put(285,0) {
        \put(-40,0){
          \put(-30,-2){$1 \cdot$}
          \qbezier(0,-30)(-5,-15)(-20,-15) \put(-20,-15){\vector(-1,0){0}}
          \qbezier(0,30)(-5,15)(-20,15) \put(0,30){\vector(1,2){0}}
          \qbezier(25,-15)(10,0)(25,15) \put(25,-15){\vector(1,-1){0}}
          \put(0,0){\circle{16}} \put(-8,5){\vector(0,1){0}}
          %
        }
      }
    }
  \end{picture}
\end{equation}

\begin{equation}
  \notag \noindent
  \begin{picture}(300,155)(-50,-125)
    \put(-100,-2){$\mathcal{H}_{rhs}\ = $}
    \thicklines
    \put(30,0) {
      \put(-40,0){
        \put(-50,-2){$\left(-q\right) \cdot$}
        \qbezier(0,30)(5,15)(20,15) \put(20,-15){\vector(1,0){0}}
        \qbezier(0,-30)(5,-15)(20,-15) \put(0,30){\vector(-1,2){0}}
        \qbezier(-25,15)(-10,0)(-25,-15) \put(-25,-15){\vector(-1,-1){0}}
        \put(0,0){\circle{16}} \put(8,5){\vector(0,1){0}}
        \put(0,0){\thinlines \linethickness{0.15mm} \put(8,-18){\line(-1,2){5}}}
        \put(-8,0){\thinlines \linethickness{0.25mm} \qbezier[6](0,0)(-5,0)(-10,0)}
        \put(8,18){\thinlines \linethickness{0.25mm} \qbezier[6](0,0)(-2.5,-5)(-5,-10)}
      }
    }
    \put(25,-2){$+$}
    \put(120,0) {
      \put(-40,0){
        \put(-40,-2){$q^2 \cdot$}
        \qbezier(0,30)(5,15)(20,15) \put(20,-15){\vector(1,0){0}}
        \qbezier(0,-30)(5,-15)(20,-15) \put(0,30){\vector(-1,2){0}}
        \qbezier(-25,15)(-10,0)(-25,-15) \put(-25,-15){\vector(-1,-1){0}}
        \put(0,0){\circle{16}} \put(8,5){\vector(0,1){0}}
        %
        \put(-8,0){\thinlines \linethickness{0.25mm} \qbezier[6](0,0)(-5,0)(-10,0)}
        \put(8,18){\thinlines \linethickness{0.25mm} \qbezier[6](0,0)(-2.5,-5)(-5,-10)}
      }
    }
    \put(115,-2){$+$}
    \put(200,0) {
      \put(-40,0){
        \put(-30,-2){$1 \cdot$}
        \qbezier(0,30)(5,15)(20,15) \put(20,-15){\vector(1,0){0}}
        \qbezier(0,-30)(5,-15)(20,-15) \put(0,30){\vector(-1,2){0}}
        \qbezier(-25,15)(-10,0)(-25,-15) \put(-25,-15){\vector(-1,-1){0}}
        \put(0,0){\circle{16}} \put(8,5){\vector(0,1){0}}
        \put(0,0){\thinlines \linethickness{0.15mm} \put(8,-18){\line(-1,2){5}}}
        \put(8,18){\thinlines \linethickness{0.25mm} \qbezier[6](0,0)(-2.5,-5)(-5,-10)}
      }
    }
    \put(190,-2){$+$}
    \put(275,0) {
      \put(-40,0){
        \put(-30,-2){$1 \cdot$}
        \qbezier(0,30)(5,15)(20,15) \put(20,-15){\vector(1,0){0}}
        \qbezier(0,-30)(5,-15)(20,-15) \put(0,30){\vector(-1,2){0}}
        \qbezier(-25,15)(-10,0)(-25,-15) \put(-25,-15){\vector(-1,-1){0}}
        \put(0,0){\circle{16}} \put(8,5){\vector(0,1){0}}
        \put(0,0){\thinlines \linethickness{0.15mm} \put(8,-18){\line(-1,2){5}}}
        \put(-8,0){\thinlines \linethickness{0.25mm} \qbezier[6](0,0)(-5,0)(-10,0)}
      }
    }
    \put(0,-80) {
      \put(30,0) {
        \put(-40,0){
          \put(-50,-2){$\left(-q \right) \cdot$}
          \qbezier(0,30)(5,15)(20,15) \put(20,-15){\vector(1,0){0}}
          \qbezier(0,-30)(5,-15)(20,-15) \put(0,30){\vector(-1,2){0}}
          \qbezier(-25,15)(-10,0)(-25,-15) \put(-25,-15){\vector(-1,-1){0}}
          \put(0,0){\circle{16}} \put(8,5){\vector(0,1){0}}
          %
          \put(8,18){\thinlines \linethickness{0.25mm} \qbezier[6](0,0)(-2.5,-5)(-5,-10)}
        }
      }
      \put(15,-2){$+$}
      \put(120,0) {
        \put(-40,0){
          \put(-50,-2){$\left(-q \right) \cdot$}
          \qbezier(0,30)(5,15)(20,15) \put(20,-15){\vector(1,0){0}}
          \qbezier(0,-30)(5,-15)(20,-15) \put(0,30){\vector(-1,2){0}}
          \qbezier(-25,15)(-10,0)(-25,-15) \put(-25,-15){\vector(-1,-1){0}}
          \put(0,0){\circle{16}} \put(8,5){\vector(0,1){0}}
          %
          \put(-8,0){\thinlines \linethickness{0.25mm} \qbezier[6](0,0)(-5,0)(-10,0)}
        }
      }
      \put(105,-2){$+$}
      \put(210,0) {
        \put(-40,0){
          \put(-55,-2){$\left(-\frac{\displaystyle 1}{\displaystyle q}\right ) \cdot$}
          \qbezier(0,30)(5,15)(20,15) \put(20,-15){\vector(1,0){0}}
          \qbezier(0,-30)(5,-15)(20,-15) \put(0,30){\vector(-1,2){0}}
          \qbezier(-25,15)(-10,0)(-25,-15) \put(-25,-15){\vector(-1,-1){0}}
          \put(0,0){\circle{16}} \put(8,5){\vector(0,1){0}}
          \put(0,0){\thinlines \linethickness{0.15mm} \put(8,-18){\line(-1,2){5}}}
        }
      }
      \put(200,-2){$+$}
      \put(285,0) {
        \put(-40,0){
          \put(-30,-2){$1 \cdot$}
          \qbezier(0,30)(5,15)(20,15) \put(20,-15){\vector(1,0){0}}
          \qbezier(0,-30)(5,-15)(20,-15) \put(0,30){\vector(-1,2){0}}
          \qbezier(-25,15)(-10,0)(-25,-15) \put(-25,-15){\vector(-1,-1){0}}
          \put(0,0){\circle{16}} \put(8,5){\vector(0,1){0}}
          %
        }
      }
    }
  \end{picture}
\end{equation}

The (relatively) non-trivial part of the check is that applying $[N-2]$-rule \eqref{eq:n-2-rule}
to second, third and fourth groups, $[N-1]$-rule \eqref{eq:n-1-rule} to fifth, sixth and seventh groups
and $[N]$-rule \eqref{eq:n-rule} to the last group
, we get
the following coefficient in front of
$\begin{picture}(20,15)
  \put(20,0){
    \qbezier(0,0)(-5,5)(0,10)
    \qbezier(-10,15)(-10,10)(-15,10)
    \qbezier(-10,-5)(-10,0)(-15,0)
  }
\end{picture}$ and\ \ $\begin{picture}(20,15)
\qbezier(0,0)(5,5)(0,10)
\qbezier(10,15)(10,10)(15,10)
\qbezier(10,-5)(10,0)(15,0)
\end{picture}$
, respectively:
\begin{equation}
  \left(q^2 + 2\right) [N-2] - \left(\frac{1}{q} + 2 q \right) [N-1] + [N] \ = \ q[N-3]
\end{equation}
Now it is the $[N-3]$-rule that ensures that the lhs equals rhs also in this case.

\subsection{All the virtual Reidemeister moves}

  Invariance w.r.t. four virtual Reidemeister moves

  \begin{equation}
    \label{eq:virtual-reidemeister-moves}
    \noindent
    \begin{picture}(300,60)(-0,-30)
      \thicklines
      \put(-80,0){
        \qbezier(0,-20)(0,-10)(10,0) \qbezier(10,0)(20,10)(20,0)
        \qbezier(0,20)(0,10)(10,0) \qbezier(10,0)(20,-10)(20,0)
        \put(10,0){\put(-4,-3){\mbox{$\Box$}}}
      }
      \put(-50,-2){\mbox{$=$}}
      \put(-30,0){\put(0,-20){\line(0,1){40}}}
      \put(-20,-2){$;$}
      \put(0,0) {
        \qbezier(0,-20)(30,0)(0,20) \put(20,0) {\qbezier(0,-20)(-30,0)(0,20)}
         \put(10,0){\put(0,-12){\put(-4,-3){\mbox{$\Box$}}} \put(0,12){\put(-4,-3){\mbox{$\Box$}}}}
      }
      \put(30,-2){\mbox{$=$}}
      \put(50,0){\put(0,-20){\line(0,1){40}} \put(20,-20){\line(0,1){40}}}
      \put(80,-2){$;$}
      \put(120,0) {
        \put(0,0) {
          \qbezier(0,-20)(-20,0)(0,20)
          \qbezier(-20,10)(10,10)(20,-10) \qbezier(-20,-10)(10,-10)(20,10)
          \put(-8,9){\put(-4,-3){\mbox{$\Box$}}} \put(-8,-9){\put(-4,-3){\mbox{$\Box$}}}
          \put(12,0){\put(-4,-3){\mbox{$\Box$}}}
        }
        \put(30,-2){\mbox{$=$}}
        \put(70,0) {
          \qbezier(0,-20)(20,0)(0,20)
          \qbezier(20,10)(-10,10)(-20,-10) \qbezier(20,-10)(-10,-10)(-20,10)
           \put(8,9){\put(-4,-3){\mbox{$\Box$}}} \put(8,-9){\put(-4,-3){\mbox{$\Box$}}}
           \put(-12,0){\put(-4,-3){\mbox{$\Box$}}}
        }
      }
      \put(220,-2){$;$}
      %
      %
 \put(260,0){
        \put(0,0){
          \qbezier(0,-20)(-20,0)(0,20)
          \qbezier(-20,10)(10,10)(20,-10) \qbezier(-20,-10)(5,-8)(10,-2) \qbezier(14,2)(14,2)(20,10)
          \put(-8,9){\put(-4,-3){\mbox{$\Box$}}} \put(-8,-9){\put(-4,-3){\mbox{$\Box$}}}
        }
        \put(30,-2){\mbox{$=$}}
        \put(70,0) {
          \qbezier(0,-20)(20,0)(0,20)
          \qbezier(20,10)(-10,10)(-20,-10) \qbezier(20,-10)(-5,-8)(-10,-2) \qbezier(-14,2)(-14,2)(-20,10)
          \put(8,9){\put(-4,-3){\mbox{$\Box$}}} \put(8,-9){\put(-4,-3){\mbox{$\Box$}}}
        }
      }
      \end{picture}
  \end{equation}

  \noindent is, in fact, trivial.
  We just observe, that (regardless of the choice of orientations)
  the ``master'' bi-colored fat graph $\GLC$ is not changed
  at all when these transformations are applied to the link diagram ${\cal L}$.

  In particular, for the ``mixed'' move (involving usual and virtual crossings) even though
  pattern of Seifert cycles can take two forms, depending on the choice of orientation of strands

  \begin{equation}
    \label{eq:seifert-cycles-for-mixed-virtual-move}
    \noindent
    \begin{picture}(300,55)(0,-25)
      \thicklines
      \put(0,0) {
        \put(0,0) {
          \qbezier(0,-20)(-20,0)(0,20)
          \qbezier(-20,10)(10,10)(20,10) \qbezier(-20,-10)(10,-10)(20,-10)
          \put(0,0){\thinlines \put(10,-10){\line(0,1){20}}}
          \put(-8,9){\put(-4,-3){\mbox{$\Box$}}} \put(-8,-9){\put(-4,-4){\mbox{$\Box$}}}
        }
        \put(30,-2){\mbox{$=$}}
        \put(70,0) {
          \qbezier(0,-20)(20,0)(0,20)
          \qbezier(20,10)(-10,10)(-20,10) \qbezier(20,-10)(-10,-10)(-20,-10)
          \put(0,0){\thinlines \put(-10,-10){\line(0,1){20}}}
          \put(8,9){\put(-4,-3){\mbox{$\Box$}}} \put(8,-9){\put(-4,-4){\mbox{$\Box$}}}
        }
      }
      \put(120,-2){or}
      \put(180,0) {
        \put(0,0) {
          \qbezier(0,-20)(-20,0)(0,20)
          \qbezier(-20,10)(0,10)(0,0) \qbezier(-20,-10)(0,-10)(0,0)
          \qbezier(20,10)(10,10)(10,0) \qbezier(20,-10)(10,-10)(10,0)
          \put(0,0){\thinlines \put(0,0){\line(1,0){10}}}
          \put(-8,9){\put(-4,-3){\mbox{$\Box$}}} \put(-8,-9){\put(-4,-3){\mbox{$\Box$}}}
        }
        \put(30,-2){\mbox{$=$}}
        \put(70,0) {
          \qbezier(0,-20)(20,0)(0,20)
          \qbezier(20,10)(0,10)(0,0) \qbezier(20,-10)(0,-10)(0,0)
          \qbezier(-20,10)(-10,10)(-10,0) \qbezier(-20,-10)(-10,-10)(-10,0)
          \put(0,0){\thinlines \put(0,0){\line(-1,0){10}}}
         \put(8,9){\put(-4,-3){\mbox{$\Box$}}} \put(8,-9){\put(-4,-3){\mbox{$\Box$}}}
        }
      }

    \end{picture}
  \end{equation}

  \noindent in both cases pattern on the left is the same as the one on the right.

\Theremark {
  In contrast with these valid virtual Reidemeister moves, the one, which is invalid
  (explicitly forbidden in original Kauffman's construction)

  \begin{equation}
    \label{eq:forbidden-virtual-move}
    \noindent
    \begin{picture}(300,55)(-70,-25)
      \thicklines
      \put(0,0){
          \qbezier(0,-20)(0,-20)(-8,-12)
          \qbezier(0,20)(0,20)(-8,12)
          \qbezier(-11,-6)(-15,0)(-11,6)
          \qbezier(-20,10)(10,10)(20,-10) \qbezier(-20,-10)(10,-10)(20,10)
           \put(12,0){\put(-4,-3){\mbox{$\Box$}}}
        }
        \put(30,-2){\mbox{$\neq$}}
        \put(70,0){
          \qbezier(0,-20)(0,-20)(8,-12)
          \qbezier(0,20)(0,20)(8,12)
          \qbezier(11,-6)(15,0)(11,6)
          \qbezier(20,10)(-10,10)(-20,-10) \qbezier(20,-10)(-10,-10)(-20,10)
           \put(-12,0){\put(-4,-3){\mbox{$\Box$}}}
        }
    \end{picture}
  \end{equation}

  \noindent \underline{does} change the structure of Seifert cycles. For example, for
particular choice of strand orientations the l.h.s. and the r.h.s. are respectively

  \begin{equation}
    \label{eq:seifert-for-forbidden-virtual-move}
    \noindent
    \begin{picture}(300,55)(0,-25)
      \thicklines
      \put(0,0){
         \qbezier(0,-20)(0,-20)(-8,-12)
          \qbezier(0,20)(0,20)(-8,12)
          \qbezier(-11,-6)(-15,0)(-11,6)
          \qbezier(-20,10)(10,10)(20,-10) \qbezier(-20,-10)(10,-10)(20,10)
           \put(12,0){\put(-4,-3){\mbox{$\Box$}}}
          \put(-13,-5){\vector(0,-1){0}} \put(-5,15){\vector(-1,-1){0}} \put(0,-20){\vector(1,-1){0}}
          \put(-12.5,10){\vector(1,0){0}} \put(5,5){\vector(2,-1){0}} \put(20,-10){\vector(1,-2){0}}
          \put(18,5){\vector(-1,-2){0}} \put(-5,-9){\vector(-2,-1){0}} \put(-20,-10){\vector(-1,0){0}}
        }
        \put(30,-2){\mbox{$\rightarrow$}}
        \put(70,0){
          \qbezier(0,20)(0,0)(20,-10) \qbezier(0,-20)(0,0)(20,10)
           \put(7,0){\put(-4,-3){\mbox{$\Box$}}}
          \qbezier(-20,10)(-10,10)(-10,0) \qbezier(-20,-10)(-10,-10)(-10,0)
          \put(0,0){\thinlines
            \put(-10,5){\line(2,1){11}} \put(-10,-5){\line(2,-1){11}} }
          }
        \put(120,-2){and}
        \put(200,0){
          \qbezier(0,-20)(0,-20)(8,-12)
          \qbezier(0,20)(0,20)(8,12)
          \qbezier(11,-6)(15,0)(11,6)
          \qbezier(20,10)(-10,10)(-20,-10) \qbezier(20,-10)(-10,-10)(-20,10)
           \put(-12,0){\put(-4,-3){\mbox{$\Box$}}}
          \put(13,-5){\vector(0,-1){0}} \put(5,15){\vector(1,-1){0}} \put(0,-20){\vector(-1,-1){0}}
          \put(12.5,10){\vector(-1,0){0}} \put(-5,5){\vector(-2,-1){0}} \put(-20,-10){\vector(-1,-2){0}}
          \put(-18,5){\vector(1,-2){0}} \put(5,-9){\vector(2,-1){0}} \put(20,-10){\vector(1,0){0}}
        }
        \put(230,-2){\mbox{$\rightarrow$}}
        \put(270,0){
          \qbezier(0,20)(0,0)(-20,-10) \qbezier(0,-20)(0,0)(-20,10)
           \put(-7,0){\put(-4,-3){\mbox{$\Box$}}}
          \qbezier(20,10)(10,10)(10,0) \qbezier(20,-10)(10,-10)(10,0)
          \put(0,0){\thinlines
            \put(10,5){\line(-2,1){11}} \put(10,-5){\line(-2,-1){11}} }
          }
    \end{picture}
  \end{equation}
\noindent
i.e. are totally different.
}

\bigskip

Returning to valid virtual Reidemeister moves, we see that neither the r.h.s.
nor the l.h.s of \eqref{HMMPmod} changes under them, which completes the proof of the theorem.

\section{Computer program \label{sec:computer-program}}

Looking at local fatgraph transformation rules of section \ref{sec:n-rules}
one can't help but wonder, how do they allow to compute dimensions at all,
as some of them -- \eqref{eq:n-3-rule}, \eqref{eq:1-rule} and \eqref{eq:flip-rule} --
 do not simplify the fatgraph.
In this section we describe, how our program \cite{cl-vknots} works (more precisely,
how do versions found under git-tags ``master-dmdims'' and ``flipless-dmdims'' work).

\subsection{Flip-less version}

Let's first describe flip-less version (i.e. the one which does not use
the flip-rule \eqref{eq:flip-rule}), since it's simpler.

After mapping link diagram ${\cal L}$ to fat graph $\GLC$ the program
loops over all ways to delete edges from this graph, adding
together contributions of all subgraphs $\gamma$.
This is rather trivial.
What is non-trivial, is how the program calculates dimensions $D_\gamma$.

What it does can be summarized like this:
\begin{enumerate}
\item First, it looks, whether simplifying rules
-- the $[N-1]$-rule \eqref{eq:n-1-rule},
the $[2]$-rule \eqref{eq:2-rule} or the $[N-2]$-rule \eqref{eq:n-2-rule} --
can be applied to $\gamma$, i.e. whether some part of $\gamma$ looks like the left hand side of those formulas.
If it finds a site where simplifying rule can be applied, it immediately applies it.
\item If first step failed, it searches for all sites, where $1$-rule \eqref{eq:1-rule} could,
in principle, be applied. Then it performs full search over all ways to apply $1$-rule
at these sites (also including new sites, that appear only after some $1$-rules were applied).
For every such way it checks, whether resulting graph satisfies requirements of step 1
(i.e. is ``simplifiable''). If such a sequence of applications of $1$-rule, leading to
simplifiable graph, is found, it is applied.
As one may expect, this is a rather computationally-expensive procedure.
\item If the second step fails
(i.e. none of ways to apply $1$-rule lead to simplifiable graph),
 it signals an error, saying, that it can't calculate HOMFLY for
this particular link diagram.
\end{enumerate}

Though there is no guarantee whatsoever, that such naive algorithm would always work,
somehow it does calculate HOMFLY polynomials from braid representations for all knots in Rolfsen table
(though it takes twice as much time as the master version).
It's not clear, whether it got possible just because we restricted consideration to no more
than 10 crossings,
and/or looked only at braid representations.
It is certainly possible (though programmatically more tedious)
to retrieve other representations of
knots in Rolfsen table from {\ttfamily KnotTheory} Mathematica package
and this is a direction for future exercise.
Also, the $[N-3]$-rule \eqref{eq:n-3-rule} is not used in the program.
There are examples of virtual knots, that can't be calculated without it,
but it's an open question, whether we can do without it for non-virtual knots.

\subsection{Master version}

Master version is very similar in spirit to flip-less version (as it also
involves full searches), but it uses some of advantages, provided by flip-rule.

\begin{enumerate}
\item First, graph is question is checked to be ``easily simplifiable''.
Graph is ``easily simplifiable'' if $[N-1]$-rule, $[2]$-rule or
composition of flip-rule with $[N-1]$-rule (contraction of loop-edge)
can be applied to it.
\item Second, resulting ``non easily simplifiable'' graph
is converted to horde-diagram form using flip-rule \eqref{eq:flip-rule}
-- the form, where it has only one vertex. This form is not unique,
but already this simple optimization allows to significantly reduce
number of distinct graphs we need to calculate.
\item Third, it is checked, whether resulting horde diagram is ``simplifiable''
-- similarly to flip-less case, whether $[N-1]$-, $[2]-$ and $[N-2]$- rules,
plus their compositions with flip-rule, can be applied to graph
\item If the third step fails, really extensive full search starts.
Namely, first we generate \textit{all} horde-diagram forms of a given graph.
Then, for every such form we check if there is a sequence of applications of $1$-rule,
that results in ``simplifiable'' graph.
\item If all above steps fail, it signals an error.
\end{enumerate}

Thus, in the master version, even graphs, that could be calculated using only
graphs coming from non-virtual knots, are typically calculated through graphs
of virtual knots.

\bigskip

It is fascinating to observe, how different rules from section \ref{sec:n-rules}
play together to produce quantum dimension -- especially, when there are different
ways to apply these rules to the same graph. Sometimes answers seem different
at first glance, but they always coincide due to some $q$-number identities
-- hinting at some hidden structure, that orchestrates all these coincidences.
In the next section we describe our attempts to understand this hidden structure.
With certain reservations they can be considered as an attempt to lift the
restricted set of Reidemeister identities to the full
(knot-independent) set of quantum Ward identities,
which would lead to full   AMM/EO-like
topological recursion -- without explicit knowledge of the underlying matrix model
at $q\neq 1$, what is exactly in the spirit of the AMM/EO approach \cite{AMMEO}.
Its relation to particular explicit knot-dependent matrix-models {\it a la} \cite{Brini}
still remains to be revealed.


\section{Towards really recursive relations}
\label{sec:really-recursive-relations}

After exposition of the previous section it is clear, that
having relations that do not reduce complexity of the graph is not optimal.
In this section we try to cure this by trying to factor our objects by these
non-simplifying relations, to obtain smaller moduli space of all essentially different
$q$-dimensions, on which relations from section \ref{sec:n-rules} become proper recursions.

\subsection{Factoring by flip covariance}

Flip-covariance \eqref{eq:flip-rule}

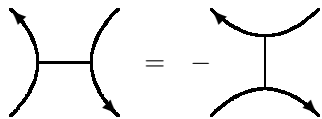
\noindent
\begin{picture}(300,60)(-130,-10)
  \linethickness{0.4mm} \thicklines
  \put(0,0) {
    \linethickness{0.4mm} \thicklines
    \qbezier(0,0)(20,20)(0,40)
    \put(0,40){\vector(-1,1){0}}
    \put(40,0){\qbezier(0,0)(-20,20)(0,40)}
    \put(40,0){\vector(1,-1){0}}
    \put(0,0){\linethickness{0.15mm}
      \put(10,20){\line(1,0){20}}
    }
  }
  \put(50,17){$=\ \ -$}
  \put(75,0) {
    \linethickness{0.4mm} \thicklines
    \qbezier(0,0)(20,20)(40,0)
    \put(0,40){\vector(-1,1){0}}
    \put(40,40){\qbezier(0,0)(-20,-20)(-40,0)}
    \put(40,0){\vector(1,-1){0}}
    \put(0,0){\linethickness{0.15mm}
      \put(20,10){\line(0,1){20}}
    }
  }
\end{picture}

\noindent
allows, in fact, to contract all the edges of the fat graph
to points. Then it becomes directed graph with (2,2)-type vertices
(i.e. each vertex has exactly 2 incoming and 2 outgoing edges).
The tradeoff, however, is that we lose information about the overall
sign of the corresponding $q$-dimension (we remind that it can easily
be negative for virtual knots, and our flips easily convert ordinary knots
into virtual and back).

Inverse transformation (from (2,2)-graph to a fat graph) is not
uniquely defined, however, we can canonically prescribe $q$-dimension to some (2,2)-graph $\Gamma$ by
\begin{align}
  D_\Gamma = (-1)^{\#(\text{connected components})}
  (-1)^{\#(\text{Seifert cycles})} D_{\Gamma_f}
\end{align}

where $\Gamma_f$ is any of resolutions of $\Gamma$ into a fat-graph.

\paragraph{Example} Fat graph for $4_1$ knot

\noindent
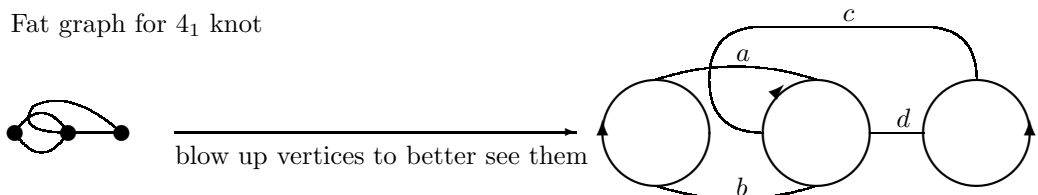
\begin{picture}(50,60)(-55,-25)
\put(0,0){\circle*{6}} \put(20,0){\circle*{6}} \put(40,0){\circle*{6}}
\qbezier(0,0)(10,15)(20,0)\qbezier(0,0)(10,-15)(20,0)
\qbezier(20,0)(30,0)(40,0)\qbezier(20,0)(0,0)(7,10)\qbezier(7,10)(21,17)(40,0)
\put(60,-12){\mbox{blow up vertices to better see them}}
\put(60,0){\vector(1,0){150}}
\put(240,0){
  \thicklines \linethickness{0.2mm}
  \put(0,0){\put(0,0){\circle{40}} \put(-20,5){\vector(0,1){0}}}
  \put(60,0){\put(0,0){\circle{40}} \put(-12,18){\vector(1,1){0}}}
  \put(120,0){\put(0,0){\circle{40}} \put(20,5){\vector(0,1){0}}}
  \thinlines \linethickness{0.15mm}
  \qbezier(0,20)(30,30)(60,20)
  \qbezier(0,-20)(30,-30)(60,-20)
  \qbezier(80,0)(90,0)(100,0)
  \qbezier(40,0)(20,0)(20,20)  \qbezier(20,20)(20,40)(40,40)
  \qbezier(40,40)(80,40)(100,40)
  \qbezier(100,40)(120,40)(120,20)
  \put(70,42){$c$}
  \put(90,2){$d$}
  \put(30,27){$a$}
  \put(30,-24){$b$}
}
\end{picture}

\noindent has the following (2,2)-graph reduction

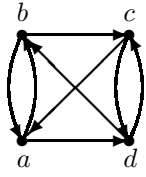
\begin{picture}(50,60)(-200,-15)
  \put(0,0){\circle*{4}} \put(40,0){\circle*{4}} \put(40,40){\circle*{4}} \put(0,40){\circle*{4}}
  \put(-2,-10){$a$} \put(40,0){\put(-2,-10){$d$}}
  \put(0,55){\put(-2,-10){$b$} \put(40,0){\put(-2,-10){$c$}}}
  \thicklines
  \qbezier(0,0)(-10,20)(0,40) \qbezier(0,0)(10,20)(0,40)
  \put(40,0){\qbezier(0,0)(-10,20)(0,40) \qbezier(0,0)(10,20)(0,40)}
  \put(0,0){\vector(1,0){40}}
  \put(0,40){\vector(1,0){40}}
  \put(40,40){\vector(-1,-1){38}}
  \put(40,0){\vector(-1,1){38}}
  \put(0,0){\vector(1,-2){0}} \put(0,40){\vector(-1,2){0}}
  \put(40,0){\put(0,0){\vector(1,-2){0}} \put(0,40){\vector(-1,2){0}}}
\end{picture}

\noindent Of course, it coincides, as a (2,2)-graph, with original link diagram
of $4_1$ knot, with important points to note:
\begin{itemize}
  \item to go from fat graph to (2,2)-graph we don't have to construct link diagram first
  \item there is no information about how exactly this (2,2)-graph should be laid on the plane to be
    interpreted as link diagram of some knot/link
  \item in particular, there is no cyclic order of incoming/outgoing edges
\end{itemize}

Since number of Seifert cycles in $4_1$ dessin is 3 and number of connected components is 1,
the canonical dimension of (2,2)-graph equals
\begin{align}
  D_{\Gamma_{4_1}} = [N][N-1]^2 + [N-2][2][N][N-1]
\end{align}

\subsection{Caveats: recursion relations}

Though we understand now, that dimensions, essentially, depend only on (2,2)-graph structure
and it's tempting just to work with (2,2)-graphs forgetting fat graphs entirely,
for now it is not possible.

The reason is, fat graphs, occurring in recursion relations, in principle, can have
number of connected components, different from each other -- hence, when projecting
relation to (2,2)-graphs, signs do appear, which depend on which graphs we consider.

\paragraph{Example}
Consider antiparallel edge elimination (combination of application of two flip-rules and $[N-2]$-rule)

\noindent
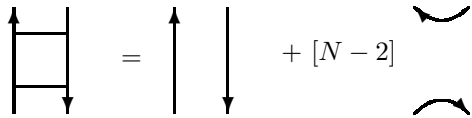
\begin{picture}(50,60)(-200,-25)
  \thicklines
  \put(0,-20){\vector(0,1){40}}
  \put(20,20){\vector(0,-1){40}}
      { \thinlines \put(0,10){\line(1,0){20}} \put(0,-10){\line(1,0){20}} }
      \put(40,-2){$=$}
      \put(60,0){
        \put(0,-20){\vector(0,1){40}}
        \put(20,20){\vector(0,-1){40}}
      }
      \put(100,0){$+\ [N-2]$}
      \put(150,0){
        \qbezier(0,20)(10,10)(20,20) \put(0,20){\vector(-1,1){0}}
        \qbezier(0,-20)(10,-10)(20,-20) \put(20,-20){\vector(1,-1){0}}
      }
\end{picture}

\noindent When applied to the horde diagram with two non-intersecting strands

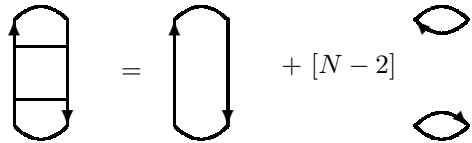
\begin{picture}(50,70)(-200,-35)
  \thicklines
  \put(0,-20){\vector(0,1){40}}
  \put(20,20){\vector(0,-1){40}}
  \qbezier(0,20)(10,30)(20,20) \qbezier(0,-20)(10,-30)(20,-20)
      { \thinlines \put(0,10){\line(1,0){20}} \put(0,-10){\line(1,0){20}} }
      \put(40,-2){$=$}
      \put(60,0){
        \put(0,-20){\vector(0,1){40}}
        \put(20,20){\vector(0,-1){40}}
        \qbezier(0,20)(10,30)(20,20) \qbezier(0,-20)(10,-30)(20,-20)
      }
      \put(100,0){$+\ [N-2]$}
      \put(150,0){
        \qbezier(0,20)(10,10)(20,20) \put(0,20){\vector(-1,1){0}}
        \qbezier(0,-20)(10,-10)(20,-20) \put(20,-20){\vector(1,-1){0}}
        \qbezier(0,20)(10,30)(20,20) \qbezier(0,-20)(10,-30)(20,-20)
      }
\end{picture}

\noindent it leads to relation on the level of (2,2)-graphs, which does not have any new signs
(because change in number of Seifert cycles is compensated by the change of the number of
connected components)

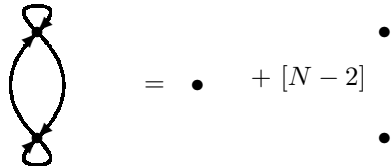
\begin{picture}(50,90)(-200,-55)
  \thicklines
  \put(0,-20){\circle*{4}} \put(0,20){\circle*{4}}
  \qbezier(0,-20)(20,0)(0,20) \qbezier(0,-20)(-20,0)(0,20)
  \qbezier(0,-20)(10,-30)(0,-30) \qbezier(0,-20)(-10,-30)(0,-30)
  \qbezier(0,20)(10,30)(0,30) \qbezier(0,20)(-10,30)(0,30)
  \put(0,-20){\vector(-1,-1){0}} \put(0,-20){\vector(1,1){0}}
  \put(0,20){\vector(1,1){0}} \put(0,20){\vector(-1,-1){0}}
  \put(40,-2){$=$}
  \put(60,0){
    \put(0,0){\circle*{4}}
  }
  \put(80,0){$+\ [N-2]$}
  \put(130,0){
    \put(0,20){\circle*{4}} \put(0,-20){\circle*{4}}
  }
  \put(-25,-50){$[N](-)^2[N-1]^2 = [N] + [N-2][N]^2$}
\end{picture}

\noindent However, if we add one more edge, such that all fat graphs in recursion stay connected

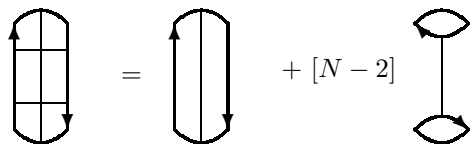
\begin{picture}(50,70)(-200,-35)
  \thicklines
  \put(0,-20){\vector(0,1){40}}
  \put(20,20){\vector(0,-1){40}}
  \qbezier(0,20)(10,30)(20,20) \qbezier(0,-20)(10,-30)(20,-20)
  \put(0,0) {
    \thinlines
    \put(0,10){\line(1,0){20}}
    \put(0,-10){\line(1,0){20}}
    \put(10,-25){\line(0,1){50}}
  }
      \put(40,-2){$=$}
      \put(60,0){
        \put(0,-20){\vector(0,1){40}}
        \put(20,20){\vector(0,-1){40}}
        \qbezier(0,20)(10,30)(20,20) \qbezier(0,-20)(10,-30)(20,-20)
        \put(0,0){ \thinlines \put(10,-25){\line(0,1){50}}}
      }
      \put(100,0){$+\ [N-2]$}
      \put(150,0){
        \qbezier(0,20)(10,10)(20,20) \put(0,20){\vector(-1,1){0}}
        \qbezier(0,-20)(10,-10)(20,-20) \put(20,-20){\vector(1,-1){0}}
        \qbezier(0,20)(10,30)(20,20) \qbezier(0,-20)(10,-30)(20,-20)
        \put(0,0){ \thinlines \put(10,-15){\line(0,1){30}}}
      }
\end{picture}

\noindent then extra sign does appear on the level of (2,2)-graph relation

\noindent
\begin{picture}(50,90)(-200,-55)
  \thicklines
  \put(-20,-20){\circle*{4}} \put(20,-20){\circle*{4}} \put(0,20){\circle*{4}}
  \put(-20,-20){
    \qbezier(0,0)(20,10)(40,0) \qbezier(0,0)(20,-10)(40,0)
    \qbezier(0,0)(0,20)(20,40) \qbezier(0,0)(20,20)(20,40)
  }
  \put(20,-20){
    \qbezier(0,0)(0,20)(-20,40) \qbezier(0,0)(-20,20)(-20,40)
  }
  \put(-20,-20){\vector(0,-1){0}} \put(-18,-20){\vector(-2,-1){0}}
  \put(0,15){\vector(1,3){0}} \put(0,20){\vector(-1,1){0}}
  \put(20,-20){\vector(2,1){0}} \put(18,-18){\vector(1,-1){0}}
  \put(40,-2){$=$}
  \put(60,0){
    \put(0,0){\circle*{4}}
    \qbezier(0,0)(10,-10)(0,-10) \qbezier(0,0)(-10,-10)(0,-10)
    \qbezier(0,0)(10,10)(0,10) \qbezier(0,0)(-10,10)(0,10)
    \put(0,0){\vector(-1,-1){0}} \put(0,0){\vector(1,1){0}}
  }
  \put(80,0){$-\ [N-2]$}
  \put(130,0){
    \put(0,0){\circle*{4}}
    \qbezier(0,0)(10,-10)(0,-10) \qbezier(0,0)(-10,-10)(0,-10)
    \qbezier(0,0)(10,10)(0,10) \qbezier(0,0)(-10,10)(0,10)
    \put(0,0){\vector(-1,-1){0}} \put(0,0){\vector(1,1){0}}
  }
\end{picture}

Thus, though for cataloguing purposes one can use (2,2)-graphs, when
actually computing dimensions one always needs to return to fat graph picture.

\subsection{Reduction of relations to (2,2)-graphs: sign freedoms}
\label{seq:reduction-to-2.2}

However, one can go with (2,2)-graphs a little bit further, and formulate recursions
in terms of them, but with unknown auxiliary variables $\epsilon_i$, which take
value $\pm1$. The values of these auxiliary variables can be later fixed by
the requirement to have correct $q \rightarrow 1$ limit.
But it is not at all clear, whether classical limit always fixes this freedom, no matter how
complicated graph in question is. It is definitely one of directions of further research.

We list here these projected rules with ``sign freedoms'' $\epsilon_i$

$[N-1]$-rule \eqref{eq:n-1-rule} becomes
\begin{equation}
    \begin{picture}(300,55)(-130,-25)
      \thicklines \linethickness{0.2mm}
      \put(0,0){\circle*{4}}
      \put(-10,-20){\vector(1,2){10}}
      \put(0,0){\vector(-1,2){10}}
      \qbezier(0,0)(20,20)(20,0) \qbezier(0,0)(20,-20)(20,0) \put(20,0){\vector(0,-1){0}}
      \put(30,-2){$=\ \epsilon\ [N-1]$}
      \put(90,-20){\vector(0,1){40}}
    \end{picture}
\end{equation}

$[2]$-rule \eqref{eq:2-rule} becomes
\begin{equation}
    \begin{picture}(300,80)(-100,-25)
      \thicklines \linethickness{0.2mm}
      \put(0,0){\circle*{4}} \put(0,30){\circle*{4}}
      \put(-10,-20){\vector(1,2){10}} \put(10,-20){\vector(-1,2){10}}
      \put(0,30){\put(0,0){\vector(-1,2){10}} \put(0,0){\vector(1,2){10}}}
      \qbezier(0,0)(15,15)(0,30) \qbezier(0,0)(-15,15)(0,30)
      \put(-7,20){\vector(0,1){0}} \put(7,20){\vector(0,1){0}}
      \put(0,15){\put(30,-2){$=\ \epsilon\ [2]$}}
      \put(80,15){
        \put(0,0){\circle*{4}}
        \put(-10,-30){\vector(1,3){9}} \put(10,-30){\vector(-1,3){9}}
        \put(0,0){\vector(1,3){10}} \put(0,0){\vector(-1,3){10}}
      }
    \end{picture}
\end{equation}

$[N-2]$-rule \eqref{eq:n-2-rule} becomes
\begin{equation}
  \begin{picture}(300,80)(-100,-25)
    \thicklines \linethickness{0.2mm}
    \put(0,0){\circle*{4}} \put(40,0){\circle*{4}}
    \put(-20,-20){\vector(1,1){20}} \put(0,0){\vector(-1,1){20}}
    \qbezier(0,0)(20,20)(40,0)     \qbezier(0,0)(20,-20)(40,0)
    \put(25,10){\vector(1,0){0}} \put(15,-10){\vector(-1,0){0}}
    \put(40,0){\put(0,0){\vector(1,-1){20}} \put(20,20){\vector(-1,-1){20}}}
    \put(70,-2){$=\ \epsilon_1\ [N-2]$}
    \put(130,0){
      \qbezier(0,-20)(20,0)(0,20) \qbezier(40,-20)(20,0)(40,20)
      \put(0,20){\vector(-1,1){0}} \put(40,-20){\vector(1,-1){0}}
    }
    \put(180,-2){$+\ \epsilon_2$}
    \put(200,0){
      \qbezier(40,20)(20,0)(0,20) \qbezier(40,-20)(20,0)(0,-20)
      \put(0,20){\vector(-1,1){0}} \put(40,-20){\vector(1,-1){0}}
    }
  \end{picture}
\end{equation}

$1$-rule \eqref{eq:1-rule} becomes
\begin{equation}
  \begin{picture}(300,80)(-50,-25)
    \thicklines \linethickness{0.2mm}
    \put(0,0){\circle*{4}} \put(0,30){\circle*{4}}
    \put(-10,-20){\vector(1,2){10}} \put(10,-20){\vector(-1,2){10}}
    \put(0,30){\put(0,0){\vector(-1,2){10}} \put(0,0){\vector(1,2){10}}}
    \qbezier(0,0)(-15,15)(0,30)
    \put(20,15){
      \put(0,0){\circle*{4}}
      \put(10,-35){\vector(-1,3){11}}
      \put(0,0){\vector(1,3){11}} 
    }
    \qbezier(0,0)(0,0)(20,15) \qbezier(0,30)(0,30)(20,15)
    \put(-7,20){\vector(0,1){0}} \put(15,12){\vector(1,1){0}}
    \put(5,26){\vector(-1,1){0}}
    \put(10,15){\put(30,-2){$-\ \epsilon_1$}}
    \put(80,15){
      \put(0,0){\circle*{4}}
      \put(-12,-35){\vector(1,3){11}} \put(12,-35){\vector(-1,3){11}}
      \put(0,0){\vector(1,3){11}} \put(0,0){\vector(-1,3){11}}
      \put(24,-35){\vector(0,1){68}}
    }
    \put(90,15){\put(30,-2){$=\ \epsilon_2$}}

    \put(180,0) {
      \put(0,0){\circle*{4}} \put(0,30){\circle*{4}}
      \put(-10,-20){\vector(1,2){10}} \put(10,-20){\vector(-1,2){10}}
      \put(0,30){\put(0,0){\vector(-1,2){10}} \put(0,0){\vector(1,2){10}}}
      \qbezier(0,0)(15,15)(0,30)
      \qbezier(0,0)(0,0)(-20,15)
      \qbezier(0,30)(0,30)(-20,15)
      \put(7,20){\vector(0,1){0}}
      \put(-15,12){\vector(-1,1){0}}
      \put(-5,26){\vector(1,1){0}}
      \put(-20,15){
        \put(0,0){\circle*{4}}
        \put(-12,-35){\vector(1,3){11}}
        \put(0,0){\vector(-1,3){11}}
      }
    }
    \put(200,15){\put(0,-2){$-\ \epsilon_3$}}
    \put(255,15){
      \put(0,0){\circle*{4}}
      \put(-12,-35){\vector(1,3){11}} \put(12,-35){\vector(-1,3){11}}
      \put(0,0){\vector(1,3){11}} \put(0,0){\vector(-1,3){11}}
      \put(-24,-35){\vector(0,1){68}}
    }
  \end{picture}
\end{equation}

$[N-3]$-rule \eqref{eq:n-3-rule} becomes
\begin{equation}
  \begin{picture}(300,160)(50,-105)
    \thicklines \linethickness{0.2mm}
    \put(0,0) {
      \put(0,5) {
        \put(0,-20){\circle*{4}} \put(-20,10){\circle*{4}} \put(20,10){\circle*{4}}
        \qbezier(-40,0)(-40,0)(-20,10) \qbezier(-20,10)(-20,10)(-20,30)
        \qbezier(40,0)(40,0)(20,10) \qbezier(20,10)(20,10)(20,30)
        \qbezier(-20,-30)(-20,-30)(0,-20) \qbezier(20,-30)(20,-30)(0,-20)
        \qbezier(-20,10)(0,20)(20,10) \qbezier(-20,10)(-20,-10)(0,-20) \qbezier(20,10)(20,-10)(0,-20)
        \put(5,15){\vector(1,0){0}} \put(12,-12){\vector(-1,-2){0}} \put(-16,-6){\vector(-1,2){0}}
        \put(-20,30){\vector(0,1){0}}  \put(40,0){\vector(2,-1){0}} \put(-20,-30){\vector(-2,-1){0}}
        \put(-30,5){\vector(2,1){0}} \put(10,-25){\vector(-2,1){0}} \put(20,20){\vector(0,-1){0}}
      }
    }
    \put(0,5){\put(50,-2){$-\epsilon_1\ [N-3]$}}
    \thicklines \linethickness{0.2mm}
    \put(150,0) {
      \put(0,5) {
        \qbezier(-20,-30)(0,-10)(20,-30) \put(-5,-20){\vector(-1,0){0}}
        \qbezier(-40,0)(-10,2.5)(-20,30) \put(-17,12){\vector(1,1){0}}
        \qbezier(40,0)(10,2.5)(20,30) \put(24,5){\vector(1,-1){0}}
      }
    }
    \put(-120,-80){
      \put(200,3){$=\ \epsilon_2$}
      \put(270,0) {
        \put(0,5) {
          \put(0,20){\circle*{4}} \put(-20,-10){\circle*{4}} \put(20,-10){\circle*{4}}
          \qbezier(-40,0)(-40,0)(-20,-10) \qbezier(-20,-10)(-20,-10)(-20,-30)
          \qbezier(40,0)(40,0)(20,-10) \qbezier(20,-10)(20,-10)(20,-30)
          \qbezier(-20,30)(-20,30)(0,20) \qbezier(20,30)(20,30)(0,20)
          \qbezier(-20,-10)(0,-20)(20,-10) \qbezier(-20,-10)(-20,10)(0,20) \qbezier(20,-10)(20,10)(0,20)
          \put(5,-15){\vector(1,0){0}} \put(12,12){\vector(-1,2){0}} \put(-16,6){\vector(-1,-2){0}}
          \put(-20,-30){\vector(0,-1){0}}  \put(40,0){\vector(2,1){0}} \put(-20,30){\vector(-2,1){0}}
          \put(-30,-5){\vector(2,-1){0}} \put(10,25){\vector(-2,-1){0}} \put(20,-20){\vector(0,1){0}}
        }
      }
      \put(0,5){\put(320,-2){$-\epsilon_3\ [N-3]$}}
      \put(420,0) {
        \put(0,5) {
          \qbezier(-20,30)(0,10)(20,30) \put(-5,20){\vector(-1,0){0}}
          \qbezier(-40,0)(-10,-2.5)(-20,-30) \put(-17,-12){\vector(1,-1){0}}
          \qbezier(40,0)(10,-2.5)(20,-30) \put(24,-5){\vector(1,1){0}}
        }
      }
    }
  \end{picture}
\end{equation}

\noindent Needless to say that in practice one does not have to keep track of all the different
$\epsilon$'s that arise at each decomposition step. It is sufficient to
have new unknown phase $\epsilon_i$ each time recursion ``hits'' empty fat graph.

\subsection{Factoring by $1$-rule: higher-valency vertices}

Still, after reduction to (2,2)-graphs we have not reached our goal yet:
projections of $1$-rule and $[N-3]$-rule do not simplify graphs.

We can try to bend this for $1$-rule, by declaring its left and right hand sides to be equal to
vertex of higher valency - (3,3)-vertex
(from now on we omit $\epsilon$ sign factors, but they are there in each and every graph transformation
formula)

\begin{equation}
  \begin{picture}(300,80)(-50,-25)
    \thicklines \linethickness{0.2mm}
    \put(-80,0) {
      \put(0,15){
        \put(0,){\circle*{4}}
        \put(0,-35){\vector(0,1){70}}
        \put(-17.5,-35){\vector(1,2){35}}
        \put(17.5,-35){\vector(-1,2){35}}
        \put(40,-2){$=$}
      }
    }
    \put(0,0){\circle*{4}} \put(0,30){\circle*{4}}
    \put(-10,-20){\vector(1,2){10}} \put(10,-20){\vector(-1,2){10}}
    \put(0,30){\put(0,0){\vector(-1,2){10}} \put(0,0){\vector(1,2){10}}}
    \qbezier(0,0)(-15,15)(0,30)
    \put(20,15){
      \put(0,0){\circle*{4}}
      \put(10,-35){\vector(-1,3){11}}
      \put(0,0){\vector(1,3){11}} 
    }
    \qbezier(0,0)(0,0)(20,15) \qbezier(0,30)(0,30)(20,15)
    \put(-7,20){\vector(0,1){0}} \put(15,12){\vector(1,1){0}}
    \put(5,26){\vector(-1,1){0}}
    \put(10,15){\put(30,-2){$-$}}
    \put(80,15){
      \put(0,0){\circle*{4}}
      \put(-12,-35){\vector(1,3){11}} \put(12,-35){\vector(-1,3){11}}
      \put(0,0){\vector(1,3){11}} \put(0,0){\vector(-1,3){11}}
      \put(24,-35){\vector(0,1){68}}
    }
    \put(90,15){\put(30,-2){$=$}}

    \put(180,0) {
      \put(0,0){\circle*{4}} \put(0,30){\circle*{4}}
      \put(-10,-20){\vector(1,2){10}} \put(10,-20){\vector(-1,2){10}}
      \put(0,30){\put(0,0){\vector(-1,2){10}} \put(0,0){\vector(1,2){10}}}
      \qbezier(0,0)(15,15)(0,30)
      \qbezier(0,0)(0,0)(-20,15)
      \qbezier(0,30)(0,30)(-20,15)
      \put(7,20){\vector(0,1){0}}
      \put(-15,12){\vector(-1,1){0}}
      \put(-5,26){\vector(1,1){0}}
      \put(-20,15){
        \put(0,0){\circle*{4}}
        \put(-12,-35){\vector(1,3){11}}
        \put(0,0){\vector(-1,3){11}}
      }
    }
    \put(200,15){\put(0,-2){$-$}}
    \put(255,15){
      \put(0,0){\circle*{4}}
      \put(-12,-35){\vector(1,3){11}} \put(12,-35){\vector(-1,3){11}}
      \put(0,0){\vector(1,3){11}} \put(0,0){\vector(-1,3){11}}
      \put(-24,-35){\vector(0,1){68}}
    }
  \end{picture}
\end{equation}

\noindent from where it is only small step figure out, that all the projectors to
totally antisymmetric representations, in this picture, become vertices
of higher valencies
(see \cite{Cvitanovich}, chapter 6.2 for notation for antisymmetric projectors and
useful formulas at $q = 1$)

\begin{picture}(300,80)(-130,-40)
  \thicklines \linethickness{0.2mm}
  \put(-80,0){
    \thicklines
    \put(0,-20){\vector(0,1){40}}
    \put(20,-20){\vector(0,1){40}}
    \put(40,-20){\vector(0,1){40}}
    \put(60,-20){\vector(0,1){40}}
    \linethickness{2mm}
    \put(-2,0){\line(1,0){64}}
    \put(30,-25){p}
  }
  \put(0,-2){\mbox{$= \frac{\displaystyle 1}{\displaystyle [p]!}$}}
  \put(80,0) {
    \put(-40,-20){\vector(2,1){80}}
    \put(40,-20){\vector(-2,1){80}}
    \put(0,0){\circle*{4}}
    \put(-10,-20){\vector(1,2){20}}
    \put(10,-20){\vector(-1,2){20}}
    \put(-2.5,-25){p}
  }
\end{picture}

Relation between (n-1,n-1)-vertex and (n,n)-vertex
(projection of relation between projectors on totally antisymmetric representations)
looks most naturally (again, for $q = 1$ this is in \cite{Cvitanovich},
but to get formula, that generalizes to $q \neq 1$ one needs to substitute
 permutation by difference of identity and projector on [1,1]-representation)

\begin{equation}
  \label{eq:n-1-n-1-through-n-n}
  \begin{picture}(300,80)(-50,-25)
    \thicklines \linethickness{0.2mm}
    \put(0,0){\circle*{4}} \put(0,30){\circle*{4}}
    \put(-10,-20){\line(1,2){10}} \put(10,-20){\line(-1,2){10}} \put(0,-20){\line(0,1){20}}
    \put(-5,-10){\vector(1,2){0}} \put(5,-10){\vector(-1,2){0}} \put(0,-10){\vector(0,1){0}}
    \put(0,30){\put(0,0){\vector(-1,2){10}} \put(0,0){\vector(1,2){10}} \put(0,0){\vector(0,1){20}}}
    \put(-9,-27){$\displaystyle p + 1$}
    \put(-9,53){$\displaystyle p + 1$}
    \put(-15,15){$\textbf{p}$}
    \qbezier(0,0)(-15,15)(0,30)
    \put(20,15){
      \put(0,0){\circle*{4}}
      \put(10,-35){\vector(-1,3){11}}
      \put(0,0){\vector(1,3){11}} 
    }
    \qbezier(0,0)(0,0)(20,15) \qbezier(0,30)(0,30)(20,15)
    \put(-7,20){\vector(0,1){0}} \put(15,12){\vector(1,1){0}}
    \put(5,26){\vector(-1,1){0}}
    \put(10,15){\put(30,-2){$=$}}
    \put(80,0) {
      \put(0,15){
        \linethickness{0.2mm}
        \put(0,0){\circle*{4}}
        \qbezier(-20,-35)(0,0)(20,35) \qbezier(20,-35)(0,0)(-20,35)
        \qbezier(-10,-35)(0,0)(10,35) \qbezier(10,-35)(0,0)(-10,35)
        \put(-20,35){\vector(-1,2){0}} \put(-10,35){\vector(-1,3){0}}
        \put(20,35){\vector(1,2){0}} \put(10,35){\vector(1,3){0}}
        \put(-14,-25){\vector(1,2){0}} \put(-7,-25){\vector(1,3){0}}
        \put(14,-25){\vector(-1,2){0}} \put(7,-25){\vector(-1,3){0}}
        \put(-9,-42){$\displaystyle p + 2$}
        \put(-9,40){$\displaystyle p + 2$}
      }
    }
    \put(90,15){\put(30,-2){$+\ [p]$}}
    \put(160,0) {
      \put(0,15){
        \put(0,){\circle*{4}}
        \put(0,-35){\vector(0,1){70}}
        \put(-17.5,-35){\vector(1,2){35}}
        \put(17.5,-35){\vector(-1,2){35}}
        \put(30,-35){\vector(0,1){70}}
        \put(-9,-42){$\displaystyle p + 1$}
        \put(-9,40){$\displaystyle p + 1$}
      }
    }
  \end{picture}
\end{equation}

\noindent if one introduces normalized multi-edges (we denote multiplicity
of the edge by \textbf{bold}, so it does not get confused with anything else).
Sometimes, we also call multiplicities \textit{momenta}, as they are
conserved at every vertex

\begin{equation}
  \label{eq:multiedge-def}
  \begin{picture}(300,80)(-50,-25)
    \thicklines \linethickness{0.2mm}
    \put(0,-20){\circle*{4}} \put(0,20){\circle*{4}}
    \put(0,-20){\line(0,1){40}} \put(0,5){\vector(0,1){0}}
    \put(-10,0){$\textbf{p}$}
    \put(20,0){$=\ \frac{\displaystyle 1}{\displaystyle [p]!}$}
    \put(100,0) {
      \put(0,-20){\circle*{4}} \put(0,20){\circle*{4}}
      \put(-6,0){$\dots$} \put(-16,-10){$p\ \text{edges}$}
      \qbezier(0,-20)(-40,-20)(-40,0) \qbezier(0,20)(-40,20)(-40,0) \put(-40,5){\vector(0,1){0}}
      \qbezier(0,-20)(-30,-20)(-30,0) \qbezier(0,20)(-30,20)(-30,0) \put(-30,5){\vector(0,1){0}}
      \qbezier(0,-20)(40,-20)(40,0) \qbezier(0,20)(40,20)(40,0) \put(40,5){\vector(0,1){0}}
      \qbezier(0,-20)(30,-20)(30,0) \qbezier(0,20)(30,20)(30,0) \put(30,5){\vector(0,1){0}}
    }
  \end{picture}
\end{equation}

It is a useful convention to treat edge with zero momentum as absence of any edge,
and edges with negative momentum as zeroes -- dimension of graphs containing such
edges is automatically zero

\begin{equation} \label{eq:zero-convention}
  \begin{picture}(300,60)(-50,-25)
    \thicklines \linethickness{0.2mm}
    \put(0,-20){\circle*{4}} \put(0,20){\circle*{4}}
    \put(0,-20){\line(0,1){40}} \put(0,5){\vector(0,1){0}}
    \put(-10,0){$\textbf{0}$}
    \put(15,0){$=$}
    \put(40,0){\put(0,-20){\circle*{4}} \put(0,20){\circle*{4}}}
    \put(100,0) {
      \put(0,-20){\circle*{4}} \put(0,20){\circle*{4}}
      \put(0,-20){\line(0,1){40}} \put(0,5){\vector(0,1){0}}
      \put(-15,0){$\textbf{-k}$}
      \put(15,0){$=\ 0,\ k \in \mathbb{Z}_{>0}$}
    }
  \end{picture}
\end{equation}

\subsection{Absorption property and clustering of edges}

Absorption property of higher antisymmetric projectors (\cite{Cvitanovich} formula 6.16)

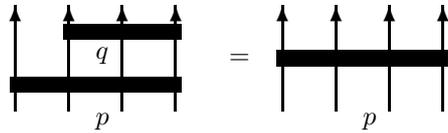
\begin{picture}(300,55)(-130,-25)
  \put(-80,0){
    \thicklines
    \put(0,-20){\vector(0,1){40}}
    \put(20,-20){\vector(0,1){40}}
    \put(40,-20){\vector(0,1){40}}
    \put(60,-20){\vector(0,1){40}}
    \linethickness{2mm}
    \put(-2,-10){\line(1,0){64}}
    \put(18,10){\line(1,0){44}}
    \put(30,-25){$p$}
    \put(30,0){$q$}
  }
  \put(0,-2){\mbox{$=$}}
  \put(20,0){
    \thicklines
    \put(0,-20){\vector(0,1){40}}
    \put(20,-20){\vector(0,1){40}}
    \put(40,-20){\vector(0,1){40}}
    \put(60,-20){\vector(0,1){40}}
    \linethickness{2mm}
    \put(-2,0){\line(1,0){64}}
    \put(30,-25){$p$}
  }
\end{picture}

\noindent also projects on directed graphs. Here it takes the following form:
any group of similarly directed arrows can be ``separated'' from any vertex,
by adding intermediate edge (the momentum of this intermediate edge is such
that momentum is conserved at each vertex)

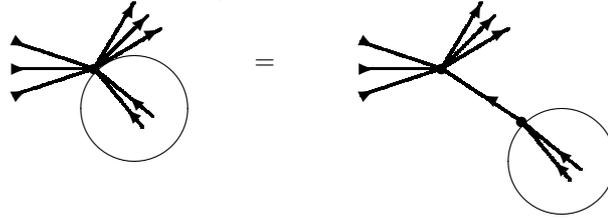
\begin{picture}(300,100)(-130,-75)
  \put(0,0){\circle*{4}}
  \thicklines
  \qbezier(0,0)(0,0)(-30,0) \put(-25,0){\vector(1,0){0}}
  \qbezier(0,0)(0,0)(-30,-10) \put(-25,-8){\vector(2,1){0}}
  \qbezier(0,0)(0,0)(-30,10) \put(-25,8){\vector(2,-1){0}}
  \qbezier(0,0)(0,0)(20,20) \put(20,20){\vector(1,1){0}}
  \qbezier(0,0)(0,0)(15,25) \put(15,25){\vector(1,2){0}}
  \qbezier(0,0)(0,0)(25,15) \put(25,15){\vector(2,1){0}}
  \thinlines
  \put(15,-15){\circle{40}}
  \thicklines
  \qbezier(0,0)(0,0)(18,-22) \put(12,-15){\vector(-1,1){0}}
  \qbezier(0,0)(0,0)(22,-18) \put(13,-10){\vector(-1,1){0}}
  \put(60,0){$=$}
  \put(130,0) {
    \put(0,0){\circle*{4}}
    \qbezier(0,0)(0,0)(-30,0) \put(-25,0){\vector(1,0){0}}
    \qbezier(0,0)(0,0)(-30,-10) \put(-25,-8){\vector(2,1){0}}
    \qbezier(0,0)(0,0)(-30,10) \put(-25,8){\vector(2,-1){0}}
    \qbezier(0,0)(0,0)(20,20) \put(20,20){\vector(1,1){0}}
    \qbezier(0,0)(0,0)(15,25) \put(15,25){\vector(1,2){0}}
    \qbezier(0,0)(0,0)(25,15) \put(25,15){\vector(2,1){0}}
    \qbezier(0,0)(0,0)(30,-20) \put(15,-10){\vector(-2,1){0}}
    \put(30,-20) {
      \put(0,0){\circle*{4}}
      \thinlines
      \put(15,-15){\circle{40}}
      \thicklines
      \qbezier(0,0)(0,0)(18,-22) \put(12,-15){\vector(-1,1){0}}
      \qbezier(0,0)(0,0)(22,-18) \put(13,-10){\vector(-1,1){0}}
    }
  }
\end{picture}

\noindent This has a very important implication:
\textbf{for any recursion relation we formulate, it suffices to
have just two (one incoming and one outgoing) external edges at each vertex.}
For example, relation \eqref{eq:n-1-n-1-through-n-n} can be written
as the following rule for elimination of a loop-free triangle

\begin{equation}
  \begin{picture}(300,80)(-10,-25)
    \thicklines \linethickness{0.2mm}
    \put(0,0){\circle*{4}} \put(0,30){\circle*{4}}
    \put(0,-20){\line(0,1){20}}
    \put(0,-10){\vector(0,1){0}}
    \put(0,30){
      \put(0,0){\vector(0,1){20}}
    }
    \put(-15,-27){$\displaystyle {\bf p + 1}$}
    \put(-15,53){$\displaystyle {\bf p + 1}$}
    \put(-15,15){$\textbf{p}$}
    \qbezier(0,0)(-15,15)(0,30)
    \put(20,15){
      \put(0,0){\circle*{4}}
      \put(10,-35){\vector(-1,3){11}}
      \put(0,0){\vector(1,3){11}} 
    }
    \qbezier(0,0)(0,0)(20,15) \qbezier(0,30)(0,30)(20,15)
    \put(-7,20){\vector(0,1){0}} \put(15,12){\vector(1,1){0}}
    \put(5,26){\vector(-1,1){0}}
    \put(10,15){\put(30,-2){$=$}}
    \put(80,0) {
      \put(0,0){\circle*{4}} \put(0,30){\circle*{4}}
      \put(-10,-20){\line(1,2){10}} \put(-5,-10){\vector(1,2){0}}
      \put(10,-20){\line(-1,2){10}} \put(5,-10){\vector(-1,2){0}}
      \put(0,30){
        \put(0,0){\vector(-1,2){10}} \put(0,0){\vector(1,2){10}}
      }
      \put(-25,-27){$\displaystyle {\bf p + 1}$}
      \put(-25,53){$\displaystyle {\bf p + 1}$}
      \put(5,15){$\textbf{p+2}$}
      \qbezier(0,0)(0,15)(0,30)
      \put(0,20){\vector(0,1){0}}
    }
    \put(100,15){\put(30,-2){$+\ [p]$}}
    \put(180,0) {
      \put(-25,-27){$\displaystyle {\bf p + 1}$}
      \put(-25,53){$\displaystyle {\bf p + 1}$}
      \put(-10,-20){\vector(0,1){70}}
      \put(10,-20){\vector(0,1){70}}
    }
  \end{picture}
\end{equation}

\noindent and the form \eqref{eq:n-1-n-1-through-n-n}, with arbitrary
number of incoming and outgoing legs, \textit{automatically follows}.
Hence, in what follows we formulate all the relations in this concise form.

Directed graphs with multivalent vertices and momentum on edges look beautiful.
What's more important, $1$-rule now takes form of reduction to (3,3)-valent vertex
-- we made one more step towards really recursive relations.
There is, however, again a tradeoff -- now we have expanded
space of objects we work with and so far we don't have enough relations for this space
 -- we must somehow deduce them (or guess additional ones)
 from the basic recursion relations on (2,2)-graphs
in subsection \ref{seq:reduction-to-2.2}.

Next subsection lists our progress in finding these effective relations,
but so far they do not yet form the complete set.

\subsection{Recursion relations for directed momentum-labeled graphs}

In this subsection we list relations on directed multivalent momentum-labeled
graphs (introduction of which was motivated in the previous subsection).
For ones which can be derived from relations on (2,2)-graphs, we give a hint,
how. For ones which can't (or we don't know, how) we clearly say so.

\newcommand\qbinom[2]{\genfrac{[}{]}{0pt}{}{#1}{#2}}

First, trace of an edge is (corollary of \eqref{eq:multiedge-def} and \eqref{eq:n-1-n-1-through-n-n})
\begin{equation}
  \begin{picture}(300,50)(-50,-25)
    \thicklines \linethickness{0.2mm}
    \put(0,0){\circle*{4}}
    \put(20,0){\circle{40}}
    \put(40,-5){\vector(0,-1){0}}
    \put(30,0){$\textbf{p}$}
    \put(50,-2){$= \displaystyle \qbinom{N}{p} = \frac{[N]!}{[N-p]![p]!}$}
  \end{picture}
\end{equation}

\noindent furthermore, same equations lead to formula for contraction of any 1-gon
\begin{equation}
  \begin{picture}(300,50)(-50,-25)
    \thicklines \linethickness{0.2mm}
    \put(0,0){\circle*{4}}
    \put(20,0){\circle{40}}
    \put(0,-20){\vector(0,1){40}}
    \put(0,-10){\vector(0,1){0}} 
    \put(40,-5){\vector(0,-1){0}}
    \put(30,0){$\textbf{p}$}
    \put(-10,-20){$\textbf{q}$} \put(-10,15){$\textbf{q}$}
    \put(50,-2){$= \displaystyle \qbinom{N-q}{p}$}
    \put(110,0) {
      \put(0,-20){\vector(0,1){40}}
      \put(0,-10){\vector(0,1){0}} 
      \put(0,0){\circle*{4}}
      \put(-10,-20){$\textbf{q}$} \put(-10,15){$\textbf{q}$}
    }
  \end{picture}
\end{equation}

On the language of directed graphs, $1/[N]$-decomposition rule
\eqref{fact-reduced}
becomes (it is not derivable from anywhere). Graphs $\gamma_1$ and $\gamma_2$
here are assumed to be connected only through two edges, which are drawn

\begin{equation}
  \begin{picture}(300,50)(-50,-25)
    \thicklines \linethickness{0.2mm}
    \put(0,0){\circle{30}}\put(-4,-2){\mbox{$\gamma_1$}}
    \put(60,0){\put(0,0){\circle{30}} \put(-4,-2){\mbox{$\gamma_2$}}}
    \put(13,10){\line(1,0){34}} \put(35,10){\vector(1,0){0}}
    \put(13,-10){\line(1,0){34}} \put(25,-10){\vector(-1,0){0}}
    \put(80,-2){$=\ \displaystyle \frac{1}{[N]}$}
    \put(130,0) {
      \put(0,0){\circle{30}}\put(-4,-2){\mbox{$\gamma_1$}}
      \put(60,0){\put(0,0){\circle{30}} \put(-4,-2){\mbox{$\gamma_2$}}}
      \qbezier(13,10)(25,10)(25,0) \qbezier(13,-10)(25,-10)(25,0)
      \qbezier(47,10)(35,10)(35,0) \qbezier(47,-10)(35,-10)(35,0)
      \put(25,-5){\vector(0,-1){0}} \put(35,5){\vector(0,1){0}}
    }
  \end{picture}
\end{equation}

\noindent and its momentum generalization is (again, $\gamma_1$
and $\gamma_2$ are connected only via these two multi-edges)

\begin{equation}
  \label{eq:real-fact-p}
  \begin{picture}(300,50)(-50,-25)
    \thicklines \linethickness{0.2mm}
    \put(0,0){\circle{30}}\put(-4,-2){\mbox{$\gamma_1$}}
    \put(60,0){\put(0,0){\circle{30}} \put(-4,-2){\mbox{$\gamma_2$}}}
    \put(13,10){\line(1,0){34}} \put(35,10){\vector(1,0){0}}
    \put(13,-10){\line(1,0){34}} \put(25,-10){\vector(-1,0){0}}
    \put(25,14){$\textbf{p}$} \put(25,-20){$\textbf{p}$}
    \put(80,-2){$=\ \displaystyle \qbinom{N}{p}^{-1}$}
    \put(140,0) {
      \put(0,0){\circle{30}}\put(-4,-2){\mbox{$\gamma_1$}}
      \put(60,0){\put(0,0){\circle{30}} \put(-4,-2){\mbox{$\gamma_2$}}}
      \qbezier(13,10)(25,10)(25,0) \qbezier(13,-10)(25,-10)(25,0)
      \qbezier(47,10)(35,10)(35,0) \qbezier(47,-10)(35,-10)(35,0)
      \put(25,-5){\vector(0,-1){0}} \put(35,5){\vector(0,1){0}}
      \put(15,-18){$\textbf{p}$} \put(40,-18){$\textbf{p}$}
    }
  \end{picture}
\end{equation}

\noindent This formula is interesting, because
at $p = 0$ it contains factorization property \eqref{fact}
and at $p = 1$  -- factorization property
\eqref{fact-reduced} (remember \eqref{eq:zero-convention}).

\subsubsection{Di-gon recursions}

Di-gon with similarly oriented sides can be always removed, regardless
of other inputs and outputs to its vertices (just by definition of multiedges)

\begin{equation}
  \label{eq:sim-digon}
  \begin{picture}(300,80)(-50,-40)
    \thicklines \linethickness{0.2mm}
    \put(0,-20){\circle*{4}} \put(0,20){\circle*{4}}
    \qbezier(0,-20)(-20,0)(0,20) \qbezier(0,-20)(20,0)(0,20)
    \put(0,20){\vector(-1,2){10}} \put(10,40){\vector(-1,-2){10}}
    \put(-15,30){$\textbf{a}$} \put(10,30){$\textbf{b}$}
    \put(-10,-40){\line(1,2){10}} \put(-5,-30){\vector(1,2){0}}
    \put(10,-40){\line(-1,2){10}} \put(10,-40){\vector(1,-2){0}}
    \put(-15,-30){$\textbf{c}$} \put(10,-30){$\textbf{d}$}
    \put(-10,5){\vector(0,1){0}} \put(10,5){\vector(0,1){0}}
    \put(-20,-2){\textbf{p}} \put(15,-2){\textbf{q}}
    \put(30,-2){$\displaystyle = \frac{[p + q]!}{[p]![q]!}$}
    \put(90,0) {
      \put(0,-20){\circle*{4}} \put(0,20){\circle*{4}}
      \put(0,-20){\vector(0,1){40}}
      \put(5,-2){\textbf{p+q}}
      \put(0,20){\vector(-1,2){10}} \put(10,40){\vector(-1,-2){10}}
      \put(-15,30){$\textbf{a}$} \put(10,30){$\textbf{b}$}
      \put(-10,-40){\line(1,2){10}} \put(-5,-30){\vector(1,2){0}}
      \put(10,-40){\line(-1,2){10}} \put(10,-40){\vector(1,-2){0}}
      \put(-15,-30){$\textbf{c}$} \put(10,-30){$\textbf{d}$}
    }
  \end{picture}
\end{equation}

Turning to di-gon with differently oriented edges (such that it introduces
a loop into the graph), from decomposition rule \eqref{eq:real-fact-p}
we derive

\begin{equation}
  \label{eq:anti-digon}
  \begin{picture}(300,80)(-50,-30)
    \thicklines \linethickness{0.2mm}
    \put(0,-20){\circle*{4}} \put(0,20){\circle*{4}}
    \qbezier(0,-20)(-20,0)(0,20) \qbezier(0,-20)(20,0)(0,20)
    \put(-10,5){\vector(0,1){0}} \put(10,-5){\vector(0,-1){0}}
    \put(-45,-2){\textbf{p + q}} \put(15,-2){\textbf{p}}
    \put(-10,27){\textbf{q}} \put(-10,-30){\textbf{q}}
    \put(0,20){\vector(0,1){20}} \put(0,-40){\vector(0,1){20}}
    \put(30,-2){$\displaystyle = \qbinom{N - q}{p}$}
    \put(90,0) {
      \put(0,-40){\vector(0,1){80}}
      \put(5,-2){\textbf{q}}
    }
  \end{picture}
\end{equation}



When there are inputs as well as outputs at each vertex of the digon,
but momentum of both edges of digon is equal to 1, decomposition formula
is a consequence of $[N-2]$-rule \eqref{eq:n-2-rule}

\begin{equation}
  \label{eq:anti-dirty-inout-simple-digon}
  \begin{picture}(300,100)(0,-50)
    \thicklines \linethickness{0.2mm}
    \put(-20,0){\circle*{4}} \put(20,0){\circle*{4}}
    \qbezier(-20,0)(0,-20)(20,0) \qbezier(-20,0)(0,20)(20,0)
    \put(5,-10){\vector(1,0){0}} \put(-5,10){\vector(-1,0){0}}
    \put(-20,0) {
      \qbezier(-20,20)(0,10)(0,0) \put(-10,15){\vector(2,-1){0}}
      \qbezier(-20,-20)(0,-10)(0,0) \put(-20,-20){\vector(-2,-1){0}}
      \put(-10,-20){${\bf q_1}$} \put(-10,20){${\bf q_1}$}
    }
    \put(20,0) {
      \qbezier(20,20)(0,10)(0,0) \put(10,-15){\vector(-2,1){0}}
      \qbezier(20,-20)(0,-10)(0,0) \put(20,20){\vector(2,1){0}}
      \put(0,-20){${\bf q_2}$} \put(0,20){${\bf q_2}$}
    }
    \put(50,-2){$\displaystyle = [N - q_1 - q_2]$}
    \put(160,0) {
      \put(-20,0){\circle*{4}} \put(20,0){\circle*{4}}
      \put(-20,0) {
        \qbezier(-20,20)(0,10)(0,0) \put(-10,15){\vector(2,-1){0}}
        \qbezier(-20,-20)(0,-10)(0,0) \put(-20,-20){\vector(-2,-1){0}}
        \put(-10,-20){${\bf q_1}$} \put(-10,20){${\bf q_1}$}
      }
      \put(20,0) {
        \qbezier(20,20)(0,10)(0,0) \put(10,-15){\vector(-2,1){0}}
        \qbezier(20,-20)(0,-10)(0,0) \put(20,20){\vector(2,1){0}}
        \put(0,-20){${\bf q_2}$} \put(0,20){${\bf q_2}$}
      }
    }
    \put(210,-2){$\displaystyle +$}
    \put(270,0) {
      \put(-20,0) {
        \put(0,20) {
          \put(0,0){\circle*{4}}
          \qbezier(-20,20)(0,10)(0,0) \put(-10,15){\vector(2,-1){0}}
          \put(-10,20){${\bf q_1}$}
        }
        \put(0,-20) {
          \put(0,0){\circle*{4}}
          \qbezier(-20,-20)(0,-10)(0,0) \put(-20,-20){\vector(-2,-1){0}}
          \put(-10,-20){${\bf q_1}$}
        }
      }
      \put(20,0) {
        \put(0,20) {
          \put(0,0){\circle*{4}}
          \qbezier(20,20)(0,10)(0,0)
          \put(20,20){\vector(2,1){0}}
          \put(0,20){${\bf q_2}$}
        }
        \put(0,-20) {
          \put(0,0){\circle*{4}}
          \put(10,-15){\vector(-2,1){0}}
          \qbezier(20,-20)(0,-10)(0,0)
          \put(0,-20){${\bf q_2}$}
        }
      }
      \qbezier(-20,20)(0,20)(20,20) \qbezier(20,20)(20,0)(20,-20)
      \put(-52,-2){${\bf q_1-1}$}
      \qbezier(-20,-20)(-20,0)(-20,20) \qbezier(-20,-20)(0,-20)(20,-20)
      \put(25,-2){${\bf q_2-1}$}
      \put(5,20){\vector(1,0){0}} \put(-5,-20){\vector(-1,0){0}}
      \put(20,5){\vector(0,1){0}} \put(-20,-5){\vector(0,-1){0}}
    }
  \end{picture}
\end{equation}

\noindent which, conjecturally, for multiedges inside loop takes form

\begin{equation}
  \label{eq:anti-dirty-inout-p-digon}
  \begin{picture}(300,100)(0,-50)
    \thicklines \linethickness{0.2mm}
    \put(-20,0){\circle*{4}} \put(20,0){\circle*{4}}
    \qbezier(-20,0)(0,-20)(20,0) \qbezier(-20,0)(0,20)(20,0)
    \put(5,-10){\vector(1,0){0}} \put(-5,10){\vector(-1,0){0}}
    \put(-5,15){\textbf{p}} \put(-5,-20){\textbf{p}}
    \put(-20,0) {
      \qbezier(-20,20)(0,10)(0,0) \put(-10,15){\vector(2,-1){0}}
      \qbezier(-20,-20)(0,-10)(0,0) \put(-20,-20){\vector(-2,-1){0}}
      \put(-10,-20){${\bf q_1}$} \put(-10,20){${\bf q_1}$}
    }
    \put(20,0) {
      \qbezier(20,20)(0,10)(0,0) \put(10,-15){\vector(-2,1){0}}
      \qbezier(20,-20)(0,-10)(0,0) \put(20,20){\vector(2,1){0}}
      \put(0,-20){${\bf q_2}$} \put(0,20){${\bf q_2}$}
    }
    \put(50,-2){$\displaystyle = \sum_{k = 0}^p \qbinom{N - q_1 - q_2}{p-k}$}
    \put(200,0) {
      \put(-20,0) {
        \put(0,20) {
          \put(0,0){\circle*{4}}
          \qbezier(-20,20)(0,10)(0,0) \put(-10,15){\vector(2,-1){0}}
          \put(-15,25){${\bf q_1}$}
        }
        \put(0,-20) {
          \put(0,0){\circle*{4}}
          \qbezier(-20,-20)(0,-10)(0,0) \put(-20,-20){\vector(-2,-1){0}}
          \put(-15,-30){${\bf q_1}$}
        }
      }
      \put(20,0) {
        \put(0,20) {
          \put(0,0){\circle*{4}}
          \qbezier(20,20)(0,10)(0,0)
          \put(20,20){\vector(2,1){0}}
          \put(5,25){${\bf q_2}$}
        }
        \put(0,-20) {
          \put(0,0){\circle*{4}}
          \put(10,-15){\vector(-2,1){0}}
          \qbezier(20,-20)(0,-10)(0,0)
          \put(5,-30){${\bf q_2}$}
        }
      }
      \qbezier(-20,20)(0,20)(20,20) \qbezier(20,20)(20,0)(20,-20)
      \qbezier(-20,-20)(-20,0)(-20,20) \qbezier(-20,-20)(0,-20)(20,-20)
      \put(5,20){\vector(1,0){0}} \put(-5,-20){\vector(-1,0){0}}
      \put(20,5){\vector(0,1){0}} \put(-20,-5){\vector(0,-1){0}}
      \put(-5,25){\textbf{k}} \put(-5,-30){\textbf{k}}
      \put(25,-2){$\textbf{q}_\textbf{1}-\textbf{k}$}
      \put(-55,-2){$\textbf{q}_\textbf{2}-\textbf{k}$}
    }
  \end{picture}
\end{equation}

\noindent and, again conjecturally, when there is a momentum transfer
between vertices of digon, it takes form

\begin{equation}
  \label{eq:anti-dirty-inout-p-m-digon}
  \begin{picture}(300,100)(0,-50)
    \thicklines \linethickness{0.2mm}
    \put(-20,0){\circle*{4}} \put(20,0){\circle*{4}}
    \qbezier(-20,0)(0,-20)(20,0) \qbezier(-20,0)(0,20)(20,0)
    \put(5,-10){\vector(1,0){0}} \put(-5,10){\vector(-1,0){0}}
    \put(-5,15){\textbf{p}} \put(-15,-20){\textbf{p+m}}
    \put(-20,0) {
      \qbezier(-20,20)(0,10)(0,0) \put(-10,15){\vector(2,-1){0}}
      \qbezier(-20,-20)(0,-10)(0,0) \put(-20,-20){\vector(-2,-1){0}}
      \put(-10,-20){$\textbf{q}_\textbf{1}$} \put(-10,20){$\textbf{q}_\textbf{1}$}
    }
    \put(20,0) {
      \qbezier(20,20)(0,10)(0,0) \put(10,-15){\vector(-2,1){0}}
      \qbezier(20,-20)(0,-10)(0,0) \put(20,20){\vector(2,1){0}}
      \put(0,-20){${\bf q_2}$} \put(0,20){${\bf q_2}$}
    }
    \put(50,-2){$\displaystyle = \sum_{k = 0}^p \qbinom{N - q_1 - q_2 - m}{p-k}$}
    \put(220,0) {
      \put(-20,0) {
        \put(0,20) {
          \put(0,0){\circle*{4}}
          \qbezier(-20,20)(0,10)(0,0) \put(-10,15){\vector(2,-1){0}}
          \put(-25,25){${\bf q_1+m}$}
        }
        \put(0,-20) {
          \put(0,0){\circle*{4}}
          \qbezier(-20,-20)(0,-10)(0,0) \put(-20,-20){\vector(-2,-1){0}}
          \put(-15,-30){${\bf q_1}$}
        }
      }
      \put(20,0) {
        \put(0,20) {
          \put(0,0){\circle*{4}}
          \qbezier(20,20)(0,10)(0,0)
          \put(20,20){\vector(2,1){0}}
          \put(-5,25){${\bf q_2+m}$}
        }
        \put(0,-20) {
          \put(0,0){\circle*{4}}
          \put(10,-15){\vector(-2,1){0}}
          \qbezier(20,-20)(0,-10)(0,0)
          \put(5,-30){${\bf q_2}$}
        }
      }
      \qbezier(-20,20)(0,20)(20,20) \qbezier(20,20)(20,0)(20,-20)
      \qbezier(-20,-20)(-20,0)(-20,20) \qbezier(-20,-20)(0,-20)(20,-20)
      \put(5,20){\vector(1,0){0}} \put(-5,-20){\vector(-1,0){0}}
      \put(20,5){\vector(0,1){0}} \put(-20,-5){\vector(0,-1){0}}
      \put(-15,25){\textbf{k+m}} \put(-5,-30){\textbf{k}}
      \put(25,-2){$\textbf{q}_\textbf{1}-\textbf{k}$}
      \put(-55,-2){$\textbf{q}_\textbf{2}-\textbf{k}$}
    }
  \end{picture}
\end{equation}


One may wonder, in which sense formulas \eqref{eq:anti-dirty-inout-p-digon}
and \eqref{eq:anti-dirty-inout-p-m-digon} are recursions, since diagrams on the right hand side
look more complicated. But, in fact, they either have strictly smaller total sum of momenta
(all $k \neq 0$ cases) or are structurally simpler due to $0$-momentum edges (case $k = 0$).

\subsubsection{Tri-gon (triangle) recursions}

Via repetitive application of relation between high-valency vertices \eqref{eq:n-1-n-1-through-n-n} it's
easy to derive recursion relation for elimination of loop-free trigon

\begin{equation}
  \label{eq:elimination-of-simple-loop-free-trigon}
  \begin{picture}(300,80)(-10,-25)
    \thicklines \linethickness{0.2mm}
    \put(0,0){\circle*{4}} \put(0,30){\circle*{4}}
    \put(0,-20){\line(0,1){20}}
    \put(0,-10){\vector(0,1){0}}
    \put(0,30){
      \put(0,0){\vector(0,1){20}}
    }
    \put(-15,-27){$\displaystyle {\bf p + q}$}
    \put(-15,53){$\displaystyle {\bf p + q}$}
    \put(-15,15){$\textbf{p}$}
    \put(10,25){$\displaystyle {\bf q}$}
    \put(10,0){$\displaystyle {\bf q}$}
    \qbezier(0,0)(-15,15)(0,30)
    \put(20,15){
      \put(0,0){\circle*{4}}
      \put(10,-35){\vector(-1,3){11}}
      \put(0,0){\vector(1,3){11}} 
    }
    \qbezier(0,0)(0,0)(20,15) \qbezier(0,30)(0,30)(20,15)
    \put(-7,20){\vector(0,1){0}} \put(15,12){\vector(1,1){0}}
    \put(5,26){\vector(-1,1){0}}
    \put(10,15){\put(30,-2){$= \frac{\displaystyle [p + q - 1]!}{\displaystyle [q-1]![p]!}$}}
    \put(120,0) {
      \put(0,0){\circle*{4}} \put(0,30){\circle*{4}}
      \put(-10,-20){\line(1,2){10}} \put(-5,-10){\vector(1,2){0}}
      \put(10,-20){\line(-1,2){10}} \put(5,-10){\vector(-1,2){0}}
      \put(0,30){
        \put(0,0){\vector(-1,2){10}} \put(0,0){\vector(1,2){10}}
      }
      \put(-25,-27){$\displaystyle {\bf p + q}$}
      \put(-25,53){$\displaystyle {\bf p + q}$}
      \put(5,15){$\textbf{p+q+1}$}
      \qbezier(0,0)(0,15)(0,30)
      \put(0,20){\vector(0,1){0}}
    }
    \put(150,15){\put(30,-2){$+\ \frac{\displaystyle [p+q-1]!}{\displaystyle [p-1]![q]!}$}}
    \put(270,0) {
      \put(-25,-27){$\displaystyle {\bf p + q}$}
      \put(-25,53){$\displaystyle {\bf p + q}$}
      \put(-10,-20){\vector(0,1){70}}
      \put(10,-20){\vector(0,1){70}}
    }
  \end{picture}
\end{equation}

More-or-less analogously to di-gon recursion formula, one can conjecture
a generalization to the case, when right edges carry momentum $m$
(it passes elementary trace test)

\begin{equation}
  \label{eq:elimination-of-m-loop-free-trigon}
  \begin{picture}(300,80)(-10,-25)
    \thicklines \linethickness{0.2mm}
    \put(0,0){\circle*{4}} \put(0,30){\circle*{4}}
    \put(0,-20){\line(0,1){20}}
    \put(0,-10){\vector(0,1){0}}
    \put(0,30){
      \put(0,0){\vector(0,1){20}}
    }
    \put(-15,-27){$\displaystyle {\bf p + q}$}
    \put(-15,53){$\displaystyle {\bf p + q}$}
    \put(-15,15){$\textbf{p}$}
    \put(10,25){$\displaystyle {\bf q}$}
    \put(10,0){$\displaystyle {\bf q}$}
    \put(25,-27){$\displaystyle {\bf m}$}
    \put(25,53){$\displaystyle {\bf m}$}
    \qbezier(0,0)(-15,15)(0,30)
    \put(20,15){
      \put(0,0){\circle*{4}}
      \put(10,-35){\vector(-1,3){11}}
      \put(0,0){\vector(1,3){11}} 
    }
    \qbezier(0,0)(0,0)(20,15) \qbezier(0,30)(0,30)(20,15)
    \put(-7,20){\vector(0,1){0}} \put(15,12){\vector(1,1){0}}
    \put(5,26){\vector(-1,1){0}}
    \put(10,15){\put(30,-2){$= \displaystyle \sum_{k = 0}^q \qbinom{p + q - m}{q - k}$}}
    \put(160,0) {
      \put(0,0){\circle*{4}} \put(0,30){\circle*{4}}
      \put(0,-20){\line(0,1){20}}
      \put(0,-10){\vector(0,1){0}}
      \put(0,30){
        \put(0,0){\vector(0,1){20}}
      }
      \put(-30,0) {
        \put(-15,-27){$\displaystyle {\bf p+q}$}
        \put(-15,53){$\displaystyle {\bf p+q}$}
        \put(25,-27){$\displaystyle {\bf m}$}
        \put(25,53){$\displaystyle {\bf m}$}
      }
      \put(15,15){$\textbf{m-k}$}
      \put(-15,25){$\displaystyle {\bf k}$}
      \put(-15,0){$\displaystyle {\bf k}$}
      \qbezier(0,0)(15,15)(0,30)
      \put(-20,15){
        \put(0,0){\circle*{4}}
        \put(-10,-35){\vector(1,3){11}}
        \put(0,0){\vector(-1,3){11}} 
      }
      \qbezier(0,0)(0,0)(-20,15) \qbezier(0,30)(0,30)(-20,15)
      \put(7,20){\vector(0,1){0}} \put(-15,12){\vector(-1,1){0}}
      \put(-5,26){\vector(1,1){0}}
    }
  \end{picture}
\end{equation}

\subsubsection{General observations and comments}
\begin{itemize}
\item Looks like when recursion is breaking the loop, then coefficients involve $N$,
otherwise -- not;
\item General pattern seems to be, that subgraph (with some incoming and outgoing arrows)
can be substituted by the some over \textit{all possible momentum transfers} between
these external legs, with some coefficients. However, what is the general formula
for these coefficients, is not clear.
\end{itemize}

This concludes exposition of our progress in understanding the full structure
of $q \neq 1$ recursions. We continue working on better understanding.

\bigskip

\appendix

\section{On the map \{link diagram\} $\rightarrow$ \{fat graph\}
         \label{sec:link->graph}}

\subsection{The map is not injective: Kishino ``unknotting''}
Here we demonstrate that two different link diagrams can
have the same associated fat graph. To make presentation
more dramatic, we consider celebrated example of Kishino knot.

Its link diagram and associated bi-colored fat graph are

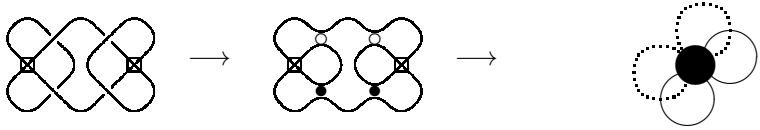
\newcommand\smallScross[0] {
  \qbezier(-5,-5)(0,0)(5,5) \qbezier(5,-5)(0,0)(-5,5)
}

\newcommand\smallVirtInt[0] {
  \put(0,0){
    \smallScross
    \qbezier(-2.5,-2.5)(-2.5,0)(-2.5,2.5)
    \qbezier(-2.5,2.5)(0,2.5)(2.5,2.5)
    \qbezier(2.5,2.5)(2.5,0)(2.5,-2.5)
    \qbezier(2.5,-2.5)(0,-2.5)(-2.5,-2.5)
  }
}
\newcommand\smallPosInt[0] {
  \put(0,0){
    \qbezier(-5,-5)(0,0)(5,5)
    \qbezier(5,-5)(2.5,-2.5)(1.5,-1.5)
    \qbezier(-5,5)(-2.5,2.5)(-1.5,1.5)
  }
}
\newcommand\smallNegInt[0] {
  \put(0,0){
    \qbezier(5,-5)(0,0)(-5,5)
    \qbezier(-5,-5)(-2.5,-2.5)(-1.5,-1.5)
    \qbezier(5,5)(2.5,2.5)(1.5,1.5)
  }
}

\newcommand\rarc[0]{
  \qbezier(0,-5)(5,0)(0,5)
}
\newcommand\larc[0]{
  \qbezier(0,-5)(-5,0)(0,5)
}
\newcommand\uarc[0]{
  \qbezier(-5,0)(0,5)(5,0)
}
\newcommand\darc[0]{
  \qbezier(-5,0)(0,-5)(5,0)
}

\newcommand\kishinoTemplate[0] {
  \put(0,0)\smallVirtInt \put(40,0)\smallVirtInt
  \put(15,0)\rarc \put(25,0)\larc
  \put(20,-15)\darc \put(20,15)\uarc
  \put(0,10){\put(-5,0)\larc \put(0,5)\uarc}
  \put(0,-10){\put(-5,0)\larc \put(0,-5)\darc}
  \put(40,10){\put(5,0)\rarc \put(0,5)\uarc}
  \put(40,-10){\put(5,0)\rarc \put(0,-5)\darc}
}

\newcommand\smallHBInt[0]{
  {\put(0,-5)\uarc \put(0,5)\darc \put(0,0){\circle*{4}}}
}

\newcommand\smallHWInt[0]{
  {\put(0,-5)\uarc \put(0,5)\darc \put(0,0){\circle{4}}}
}

\newcommand\Slash[0]{
  \qbezier(-5,-5)(0,0)(5,5)
}

\newcommand\Backslash[0]{
  \qbezier(5,-5)(0,0)(-5,5)
}

\begin{picture}(300,75)(-50,-40)
  \put(10,10)\smallPosInt \put(10,-10)\smallNegInt
  \put(30,10)\smallPosInt \put(30,-10)\smallNegInt
  \put(0,0)\kishinoTemplate

  \put(60,0){$\longrightarrow$}

  \put(100,0){
    \kishinoTemplate
    \put(10,10)\smallHWInt \put(10,-10)\smallHBInt
    \put(30,10)\smallHWInt \put(30,-10)\smallHBInt
  }

  \put(160,0){$\longrightarrow$}

  \put(250,0) {
    \put(0,0){\circle*{20}}
    \put(-3,-3){
      \put(0,-10){\circle{20}}
      \thicklines
      \qbezier[7](0,0)(0,-10)(-10,-10) \qbezier[7](-10,-10)(-20,-10)(-20,0)
      \qbezier[7](-20,0)(-20,10)(-10,10) \qbezier[7](-10,10)(0,10)(0,0)
    }
    \put(3,3){
      \put(10,0){\circle{20}}
      \thicklines
      \put(10,10){
        \qbezier[7](0,0)(0,-10)(-10,-10) \qbezier[7](-10,-10)(-20,-10)(-20,0)
        \qbezier[7](-20,0)(-20,10)(-10,10) \qbezier[7](-10,10)(0,10)(0,0)
      }
    }
  }
\end{picture}

Now notice that if we consider different composition of two Kishino's
virtual trefoil subcomponents the fat graph stays the same
(including the coloring of the graph)

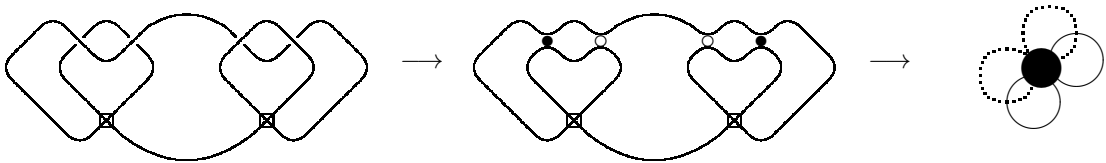
\newcommand\unknottedKishinoTemplate[0] {{
  \put(0,-20)\smallVirtInt
  \put(60,0){\put(0,-20)\smallVirtInt}
  \put(0,15)\uarc \put(0,5)\darc
  \put(15,0)\rarc \put(-15,0)\larc \put(10,-10)\Slash \put(-10,-10)\Backslash
  \put(-20,15)\uarc \put(-10,-25)\darc \put(-20,-20)\Backslash \put(-30,-10)\Backslash
  \put(-30,10)\Slash \put(-35,0)\larc
  \put(60,0) {
    \put(0,15)\uarc \put(0,5)\darc
    \put(-15,0)\larc \put(15,0)\rarc \put(-10,-10)\Backslash \put(10,-10)\Slash
    \put(20,15)\uarc \put(10,-25)\darc \put(20,-20)\Slash \put(30,-10)\Slash
    \put(30,10)\Backslash \put(35,0)\rarc
  }
  \qbezier(15,15)(30,25)(45,15)
  \qbezier(5,-25)(30,-45)(55,-25)
}}

\begin{picture}(300,75)(-50,-40)
\put(-10,10)\smallNegInt \put(10,10) \smallPosInt
\put(60,0){\put(-10,10)\smallPosInt \put(10,10) \smallNegInt}
\unknottedKishinoTemplate

\put(110,0){$\longrightarrow$}

\put(175,0) {
  \put(-10,10)\smallHBInt \put(10,10) \smallHWInt
  \put(60,0){\put(-10,10)\smallHWInt \put(10,10) \smallHBInt}
  \unknottedKishinoTemplate
}

\put(285,0){$\longrightarrow$}

\put(350,0) {
  \put(0,0){\circle*{20}}
  \put(-3,-3){
    \put(0,-10){\circle{20}}
    \thicklines
    \qbezier[7](0,0)(0,-10)(-10,-10) \qbezier[7](-10,-10)(-20,-10)(-20,0)
    \qbezier[7](-20,0)(-20,10)(-10,10) \qbezier[7](-10,10)(0,10)(0,0)
  }
  \put(3,3){
    \put(10,0){\circle{20}}
    \thicklines
    \put(10,10){
      \qbezier[7](0,0)(0,-10)(-10,-10) \qbezier[7](-10,-10)(-20,-10)(-20,0)
      \qbezier[7](-20,0)(-20,10)(-10,10) \qbezier[7](-10,10)(0,10)(0,0)
    }
  }
}
\end{picture}

However, two second Reidemeister moves followed by two first virtual Reidemeister moves
unknot this second composite, while Kishino is known to be nontrivial.

\subsection{The map is surjective: spirals-and-arcs algorithm}

Here we describe the idea of the algorithm, that allows to construct some preimage
link diagram for arbitrary fat graph.
This link diagram may not be (and usually is not) the simplest diagram with
this fat graph. We also not formulate the algorithm rigorously, instead
giving couple of examples of how it is supposed to work.
We believe that working out all the details to the level
of actually writing down explicit \textit{formulas}
for the graph of the link diagram is certainly doable (if a little bit tedious) exercise.
Hence, implementing a computer program, that, given a fat graph, outputs
LaTeX commands to draw link diagram, is straightforward.

The idea of the algorithm is the following:
\begin{itemize}
\item every fat graph vertex becomes a spiral, situated in the
annulus $2 i < r < 2 i + 1$ (for some arbitrary numbering of vertices),
with occasional ``spikes'' towards center of the circle;
\item every fat graph edge becomes an arc in the unit circle,
connecting points $r = 1, \phi = \pi \frac{2 j}{\#\text{edges}}$
and $r = 1, \phi = \pi \frac{2 j + 1}{\#\text{edges}}$
(again, for some arbitrary numbering of edges);
\item finally, every arc becomes intersection + virtual intersection
(as parts of Seifert cycles next to the arc are contra-oriented).
\end{itemize}

This algorithm, applied to virtual trefoil fat graph, gives
(we omit squares around virtual crossings, since there are too many of them)

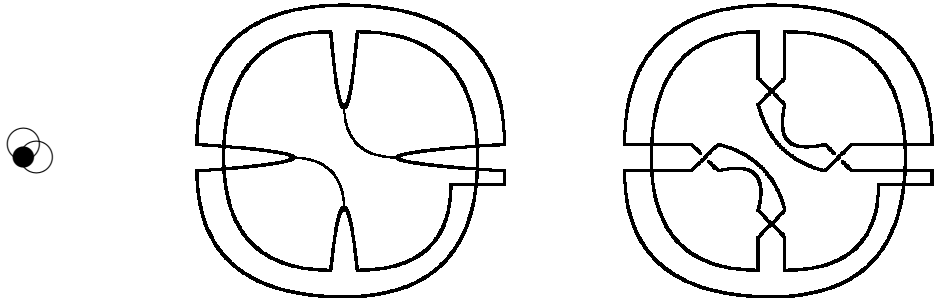
\newcommand\vtrefoilSpiralTemplate[0] {
  \put(0,0) {
    \thicklines
    \qbezier(60,5)(60,57.5)(0,57.5) \qbezier(0,57.5)(-55,57.5)(-55,5)
    \qbezier(-55,-5)(-55,-52.5)(0,-52.5) \qbezier(0,-52.5)(50,-52.5)(50,0)
    \qbezier(50,0)(50,47.5)(5,47.5) \qbezier(-5,47.5)(-45,47.5)(-45,0)
    \qbezier(-45,0)(-45,-42.5)(-5,-42.5) \qbezier(5,-42.5)(40,-42.5)(40,-10)
    \qbezier(40,-10)(50,-10)(60,-10) \qbezier(60,-10)(60,-10)(60,-5)
  }
}

\begin{picture}(300,120)(-50,-60)
  \put(0,0){
    \put(0,0){\circle*{8}}
    \put(5,0){\circle{12}}
    \put(0,5){\circle{12}}
  }
  \put(120,0) {
    \qbezier(20,0)(0,0)(0,20)  \qbezier(-20,0)(0,0)(0,-20)
    \thicklines
    \qbezier(60,-5)(-20,0)(60,5) \qbezier(-55,-5)(17,0)(-55,5)
    \qbezier(5,47.5)(0,-10)(-5,47.5) \qbezier(5,-42.5)(0,5)(-5,-42.5)
    \vtrefoilSpiralTemplate
  }
  \put(280,0) {
    \thicklines
    \qbezier(60,-5)(60,-5)(30,-5) \qbezier(60,5)(60,5)(30,5)
    \qbezier(-55,-5)(-55,-5)(-30,-5) \qbezier(-55,5)(-55,5)(-30,5)
    \qbezier(5,47.5)(5,47.5)(5,30) \qbezier(-5,47.5)(-5,47.5)(-5,30)
    \qbezier(5,-42.5)(5,-42.5)(5,-30) \qbezier(-5,-42.5)(-5,-42.5)(-5,-30)
    \qbezier(20,5)(0,0)(5,20) \qbezier(20,-5)(0,0)(-5,20)
    \qbezier(-20,5)(0,0)(5,-20) \qbezier(-20,-5)(0,0)(-5,-20)
    \put(0,25)\smallScross \put(0,-25)\smallScross
    \put(25,0)\smallPosInt  \put(-25,0)\smallPosInt
    \vtrefoilSpiralTemplate
  }
\end{picture}

And applied to twisted unknot fat graph it gives

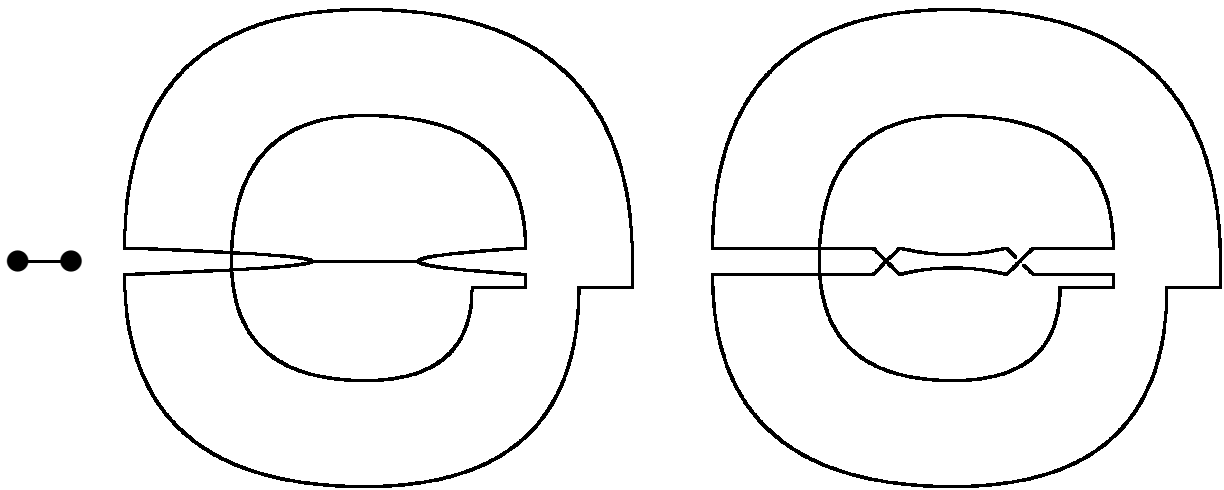
\newcommand\twistedUnknotSpiralTemplate[0] {
  \put(0,0) {
    \thicklines
    \qbezier(60,5)(60,55)(0,55) \qbezier(0,55)(-50,55)(-50,0)
    \qbezier(-50,0)(-50,-45)(0,-45) \qbezier(0,-45)(40,-45)(40,-10)
    \qbezier(40,-10)(50,-10)(60,-10) \qbezier(60,-10)(60,-10)(60,-5)
    \qbezier(100,0)(100,95)(0,95) \qbezier(0,95)(-90,95)(-90,5)
    \qbezier(-90,-5)(-90,-85)(0,-85) \qbezier(0,-85)(80,-85)(80,-10)
    \qbezier(80,-10)(90,-10)(100,-10) \qbezier(100,-10)(100,-10)(100,0)
  }
}

\begin{picture}(300,200)(0,-100)
  \put(0,0){
    \put(0,0){\circle*{8}} \put(20,0){\circle*{8}}
    \qbezier(0,0)(10,0)(20,0)
  }
  \put(130,0) {
    \qbezier(20,0)(0,0)(-20,0)
    \thicklines
    \qbezier(60,-5)(-20,0)(60,5)
    \qbezier(-90,-5)(50,0)(-90,5)
    \twistedUnknotSpiralTemplate
  }

  \put(350,0) {
    \thicklines
    \put(-25,0)\smallScross \put(25,0)\smallPosInt
    \qbezier(-20,5)(0,0)(20,5) \qbezier(-20,-5)(0,0)(20,-5)
    \qbezier(60,5)(60,5)(30,5) \qbezier(60,-5)(60,-5)(30,-5)
    \qbezier(-90,5)(-90,5)(-30,5) \qbezier(-90,-5)(-90,-5)(-30,-5)
    \twistedUnknotSpiralTemplate
  }
\end{picture}

Of course, as number of edges grows, spikes should become sharper and sharper
(as they need to fit inside one $\frac{\pi}{\#\text{edges}}$ sector), which
makes this method of drawing link diagram not very practical. However,
it is good enough to demonstrate \textit{existence} of at least one link diagram,
that maps to any fat graph.

\newpage

\section{Table of dimensions \label{sec:table-of-dimensions}}

This section provides additional set of examples of quantum dimensions.
They can be used as test cases for decomposition rules from section \ref{sec:n-rules},
as well as factorization rules \eqref{fact} \eqref{fact-reduced}

All the expressions in this section can be calculated as follows: for a given knot planar diagram, which
has a fat-graph of interest as the most complicated graph, there exists such a coloring of vertices,
that knot is an unknot (the \textit{unknotting} coloring). Provided quantum dimensions for all lower sub-graphs are already known, this
allows to find the expression for the new one. Then, one can additionally verify with help of the Rolfsen table,
 that expressions for HOMFLY polynomials for the knots, corresponding to different colorings of the planar diagram,
are indeed correct.
Whenever relevant, the corresponding knot planar diagram and unknotting coloring are indicated.

All dimensions below are  {\it un}reduced (i.e. not divided by $[N]$).

\subsection{Extract from \cite{AnoMKhR}}
\label{extract-from-anom}

In \cite{AnoMKhR} the RT formalism was used, i.e. results there are provided only
for ordinary (non-virtual) knots and links.

Dimensions are made from projectors $\pi_1,\ldots,\pi_{m-1}$ and depend on the
words, made from $m-1$ letters, where same letters never stand next to each other
\be
V_{n,\vec a} = [2]^n \Tr\Big(\hat D \, \pi_{a_1}\pi_{a_2}\ldots\Big)
\ee
Here $m$ is the number of strands (i.e. of Seifert cycles and of vertices in our graph)
and $n$ is the number of vertices in link diagram (i.e. the number of edges in the graph).
Operator $\hat D$ is responsible for grading -- it is made out of graded dimensions
of irreducible representations in $[1]^m$, we often include it into the definition of trace.

We write big $V$ to emphasize that in the present text dimensions are {\it un}reduced
(i.e. the common factor $[N]$ is included).

\subsubsection{$m=1$}

There is nothing non-virtual in this case except for the unknot with
\be
V_\emptyset = [N]
\ee

\subsubsection{$m=2$}

There is just one letter and thus just one word (since the letter can not be repeated):
\be
V_n =[2]^n \Tr\pi = [2]^n\frac{[N][N-1]}{[2]}
\label{m2word}
\ee

\subsubsection{$m=3$}

In the case of three strands there are two letters, but the words are still simply classified --
the only type of connected graphs correspond to
\be
V_{n,k} = [2]^n \Tr(\pi_1\pi_2)^k = 2^n\,\frac{[N][N-1]\Big([N-2]+2^{1-2k}[N+1]\Big)}{[2][3]} =\nn \\
= [2]^{n-2k}[N][N-1]\left([N-1]+ [2][N-2]\sum_{i=0}^{k-2}[2]^{2i}\right)
\ee

\subsubsection{$m=4$}

Now there are three letters and classification of words gets messy.
See (6.16)-(6.17) of \cite{AnoMKhR} for some examples (note also misprints there --
the summands in $v_{nk}$ should be $[2]^{2i}$ instead of $[2]^i$).

In particular, the "torus" dimensions
\be
\label{toric-m-4}
V_{n,(123)^k} = [2]^n\Tr (\pi_1\pi_2\pi_3)^k =
[2]^{n-2k-2}[N][N-1]\Big([2][N-1]^2 + 2(2^{k-1}-1)[N-1][N-2]+\xi_k[N-2][N-3]\Big)
\ee
with $\xi_k$ given by
\be
k=1 & 0 \nn\\
k=2 & 1 \nn\\
k=3 & [3]+4 \nn\\
k=4 & [5]+6[3]+12 \nn\\
k=5 & [7]+8[5]+25[3]+33 \nn\\
k=6 & [9]+10[7]+42[5]+91[3]+89 \nn\\
k=7 & [11] + 12[9] +63[7]+185[5]+313[3] + 243 \nn\\
\ldots
\ee
i.e.
\be
\xi_k = \sum_{j=0}^{k-2}  [2k-3-2j] u_{k,j}, \nn \\
u_{k,0} = 1, \nn \\
u_{k,1} = 2k-2, \nn \\
u_{k,2} = k(2k-5), \nn \\
\ldots
\ee

\subsection{Simplest graphs}

\subsubsection{Torus links/knots}

Fat graph for braid representation of torus knot is very easy to draw.
For $[n,m]$-knot it has $n$ braids and $n m$ edges.

Characteristic example of trefoil as $[3,2]$-torus knot

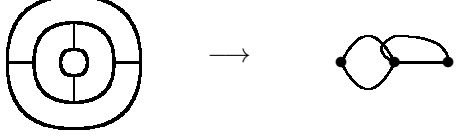
\newcommand\bezierCircle[1]{
  \qbezier(#1,0)(#1,-#1)(0,-#1) \qbezier(0,-#1)(-#1,-#1)(-#1,0)
  \qbezier(-#1,0)(-#1,#1)(0,#1) \qbezier(0,#1)(#1,#1)(#1,0)
}

\begin{picture}(300,50)(-100,-25)
  \thicklines \bezierCircle{25} \thicklines \bezierCircle{15} \thicklines \bezierCircle{5}
  \thinlines \put(15,0){\line(1,0){10}} \put(-15,0){\line(-1,0){10}}
  \put(0,5){\line(0,1){10}} \put(0,-5){\line(0,-1){10}}
  \put(50,0){$\longrightarrow$}
  \put(100,0){
    \put(0,0){\circle*{4}}
    \put(20,0){\circle*{4}}
    \put(40,0){\circle*{4}}
    \qbezier(0,0)(10,20)(20,0) \qbezier(0,0)(10,-20)(20,0)
    \qbezier(20,0)(30,0)(40,0) \qbezier(20,0)(10,5)(20,10)
    \qbezier(20,10)(40,10)(40,0)
  }
\end{picture}

\paragraph{[2,n]-family}

\be
\text{dim}_{[2,n]} = [2]^{n-1}[N][N-1] = [N][N-1] + \Big([2]^n-[2]\Big)\frac{[N][N-1]}{[2]}
\ee

We can write thus, abusing notation of section \ref{extract-from-anom} (ignoring [2] factors there)
\be
  V_{n+1} - [2] V_{n} = 0 \\ \nn
  (e^{\p/\p n} - [2]) V_n = 0,
\ee

where $e^{\p/\p n}$ is, as usual, the ``shift'' operator, which acts on index $n$ here.

\paragraph{[3,n]-family}

\be
\text{dim}_{[3,n]} = [N][N-1]^2 + \Big(\overbrace{[2] + [2]^3 + [2]^5 + ...}^{n-1 \ \text{times}})
[N][N-1][N-2] = \nn \\ =
[N][N-1]^2 + \Big([2]^{2n}-[2]^2\Big)\frac{[N][N-1][N-2]}{[2][3]}
\ = \
\frac{[N+1][N][N-1]}{[3]} + [2]^{2n-1}\frac{[N][N-1][N-2]}{[3]}
\ee

Again, the difference equation is very simple
\begin{equation}
V_{n + 1} - [2]^2 V_n = [N] [N-1](-[2][N] + [N-1]),
\end{equation}

which can be translated into familiar evolution-method second order operator
\be
\label{evolve-2}
V_{n + 2} - ([2]^2 + 1) V_{n+1}  + [2]^2 V_{n} = 0 \\ \nn
(e^{\p/\p n} - 1)(e^{\p/\p n} - [2]^2) V_n = 0
\ee

\paragraph{[4,n]-family}

The general formula can be easily implied from \eqref{toric-m-4}

The corresponding difference equation is
\begin{align}
  V_{k + 1} - [2]^2 V_k = [N][N-1] \Big ( & (2^k - 1) [2] [N]^2 \\ \notag
  & + \left ( (3 \cdot 2^k - 4)[2]^2 - (3 \cdot 2^k - 3) \right ) [N][N-1] \\ \notag
  & + \left ( (2^{k + 1} - 4)[2]^3 - (3 \cdot 2^k - 4)[2] \right ) [N-1]^2 \Big ),
\end{align}

and again 3rd order difference equation annihilating the series is straightforward to write down
\be
  \label{evolve-3}
  V_{n+3} - ([2]^2 + 3) V_{n+2} + (3 [2]^2 + 2) V_{n+1} - 2 [2]^2 V_{n} = 0 \\ \nn
  (e^{\p/\p n} - 1)(e^{\p/\p n} - 2)(e^{\p/\p n} - [2]^3) V_n = 0
\ee

\Theremark {
  Equation \eqref{evolve-2} is easy consequence of rules in section \ref{sec:n-rules}.
  However, \eqref{evolve-3} is not so easy to derive from there
  (at least, we were unable to). But, if one works with higher antisymmetric projectors
  (higher valency vertices), then it's one page computation -- one just needs to express
  projector on [1,1,1,1]-representation through toric diagrams and then utilize the absorption rule.
}

\paragraph{[m,2]-family}

The series $[m,2]$ is represented by graphs $(m,2m-2,m-1,m-1)$

\begin{picture}(200,40)(-100,-15)
\put(-50,-4){\mbox{$A_m \ =$}}
\put(0,0){\line(1,0){105}}
\put(0,0){\circle*{4}}
\put(30,0){\circle*{4}}
\put(60,0){\circle*{4}}
\put(90,0){\circle*{4}}
\put(112,3){\mbox{$\ldots$}}
\put(130,0){\line(1,0){45}}
\put(145,0){\circle*{4}}
\put(175,0){\circle*{4}}
\qbezier(0,0)(-10,10)(15,10)
\qbezier(15,10)(40,10)(30,0)
\qbezier(45,10)(20,10)(30,0)
\qbezier(45,10)(70,10)(60,0)
\qbezier(75,10)(50,10)(60,0)
\qbezier(75,10)(100,10)(90,0)
\qbezier(105,10)(80,10)(90,0)
\qbezier(130,10)(155,10)(145,0)
\qbezier(160,10)(135,10)(145,0)
\qbezier(160,10)(185,10)(175,0)
\end{picture}

\newcommand\binomial[2]{\genfrac{(}{)}{0pt}{}{\displaystyle #1}{\displaystyle #2}}

\noindent
which can be used to test $[N-2]$-rule (\ref{eq:n-2-rule}):
\be
A_{m+1} = [N-2]\cdot A_m + [N-1]^2\cdot A_{m-1}
\ee
i.e.
\be
A_1 = [N], \nn \\
A_2 = [2][N][N-1], \nn \\
A_3 = [N][N-1]\Big([N-1]+[2][N-2]\Big), \nn \\
A_4 = [N][N-1]\Big([2][N-1]^2+[N-1][N-2]+[2][N-2]^2\Big), \nn \\
A_5 = [N][N-1]\Big([N-1]^3+2\cdot [2][N-1]^2[N-2]+[N-1][N-2]^2+[2][N-2]^3 \Big), \nn \\
\ldots
\nn \\
A_{m + 1, m > 0}= [N][N-1] \sum_{j=0}^m [N]^{m-j} [N-1]^j
\genfrac{\{}{.}{0pt}{}
        {\displaystyle j = 2 k - 1: (-)^{m-1} \sum_{i = 0}^k \binomial{k+i}{2 i} \binomial{m - k + i}{k + i - 1} [2]^{2i}}
        {\displaystyle
          j = 2k: (-)^m \sum_{i=0}^k \binomial{k + i + 1}{2 i + 1} \binomial{m - k + i}{k + i} [2]^{2 i + 1}
        }
\ee

\paragraph{[m,3]-family}

The series $[m,3]$, i.e. graphs (m,3m-3,2m-2,m-1):

\be
\l[1,3]: &  [N] \nn \\
\l[2,3]: &   [2]^2[N][N-1] \nn \\
\l[3,3]:  &
 [N][N-1]\Big([N-1] + ([2] + [2]^3) [N-2]\Big),  \nn \\
\l[4,3]:  &
\l[2][N][N-1]\left\{[N][N-1] +  [N-2]\Big(4 [N-1] + 3[2][N-2] + [2]^2[N-3]\Big)\right\}
\nn \\
\l[5,3]:&
\l[N] [N-1] \left\{[N]^2\Big([N-1] - 2 [2] [N-2] - 4 [2]^2 [N-3]
                                  - 3 [2]^3 [N-2]
                                  + 2 [2]^5 [N-4]
                                  + [2]^6 [N-5]\Big)\right.\nn \\
& + [N-1] + 4 [2] [N-2] + [2]^2 [N-3] + 11 [2]^3 [N-2]
  + 4 [2]^4 [N-3] + [2]^5 [N] + [2]^6 [N-3]
     + [2]^7 [N]
   + [2]^9 [N]\Big\}\nn \\
\ldots
\ee
At $q=1$
\be
B_{m+1} = (N-4)B_m + (N-1)B'_{m-1}
\ee
where $B'$ contains and extra double edge, attached with intertwining to a vertex of $B'_{m-1}$,
it can be evaluated with the help of \eqref{eq:n-2-rule}:

\begin{picture}(200,60)(-100,-40)
\put(-70,-2){\mbox{$B'_{m-1}\ =$}}
\put(0,0){\circle*{4}}
\put(30,0){\circle*{4}}
\put(0,0){\line(1,0){30}}
\qbezier(0,0)(-20,20)(30,0)
\qbezier(0,0)(50,20)(30,0)
\put(50,0){\mbox{$\ldots$}}
\put(-15,-2){\circle*{4}}
\qbezier(-15,-2)(2,15)(0,0)
\qbezier(-15,-2)(2,-15)(0,0)
\qbezier[30](10,-10)(45,-30)(80,-10)
\put(37,-30){\mbox{$B_{m-1}$}}
\put(100,-2){\mbox{$=[N-2]B_{m-1} +[N-1]B'_{m-2}$}}
\end{picture}

At $q \neq 1$ computer experiments show (tested up to $m = 25$)
\begin{equation}
A_n^{(3)} = [2] [N-2] A_{n-1}^{(3)} + (2 + [2] [N-1][N-2]) A_{n-2}^{(3)}
+ [N-1][N-2]^2 A_{n-3}^{(3)} - [N-1]^4 A_{n-4}^{(3)}
\end{equation}

However, so far we were unable to derive this recursion from either rules
from section \ref{sec:n-rules} or higher-valency vertices,
but to be honest we didn't try very hard.

\subsubsection{Twist knots}

Seifert resolutions for odd and even twist knots, respectively, are %

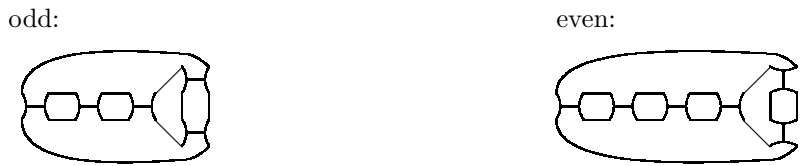
\begin{picture}(300,80)(-100,-30)
\newsavebox{\horcrux}
\savebox{\horcrux}(10,10)[bl]{
  \thinlines
  \qbezier(0,0)(3, 5)(0,10) \qbezier(10,0)(7, 5)(10,10)
  \thicklines
  \put(1.5,5){\line(1,0){7}}}
\newsavebox{\vercrux}
\savebox{\vercrux}(10,10)[bl]{
  \thinlines
  \qbezier(0,0)(5, 3)(10,0) \qbezier(0,10)(5, 7)(10,10)
  \thicklines
  \put(5,1.5){\line(0,1){7}}}

\put(-5,35){odd:}
\multiput(0,0)(20,0){3}{\usebox{\horcrux}}
\multiput(60,-10)(0,20){2}{\usebox{\horcrux}}
\thinlines
\put(10,10){\line(1,0){10}} \put(10,0){\line(1,0){10}}
\put(30,10){\line(1,0){10}} \put(30,0){\line(1,0){10}}
\put(50,10){\line(1,1){10}} \put(50,0){\line(1,-1){10}}
\put(60,10){\line(0,-1){10}} \put(70,10){\line(0,-1){10}}
\qbezier(70,20)(65,25)(60,25) \qbezier(0,10)(0,30)(60,25)
\qbezier(70,-10)(65,-15)(60,-15) \qbezier(0,0)(0,-20)(60,-15)

\put(200,35){even:}
\multiput(200,0)(20,0){4}{\usebox{\horcrux}}
\multiput(280,-10)(0,20){2}{\usebox{\vercrux}}
\thinlines
\multiput(210,10)(20,0){3}{\line(1,0){10}}
\multiput(210,0)(20,0){3}{\line(1,0){10}}
\put(270,10){\line(1,1){10}} \put(270,0){\line(1,-1){10}}
\put(280,10){\line(0,-1){10}} \put(290,10){\line(0,-1){10}}
\qbezier(290,20)(285,25)(280,25) \qbezier(200,10)(200,30)(280,25)
\qbezier(290,-10)(285,-15)(280,-15) \qbezier(200,0)(200,-20)(280,-15)
\end{picture}

\noindent
so the corresponding fat graphs can be drawn as the following decorated necklaces

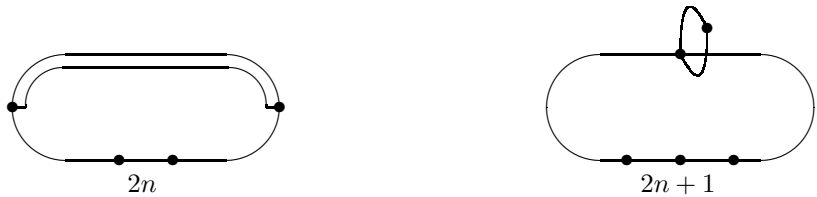
\begin{picture}(300,80)(-100,-35)
\newsavebox{\vertex}
\savebox{\vertex}(0,0)[bl]{
  \put(0,0){\circle*{4}}}
\put(0,0){\oval(100,40)}
\multiput(-10,-20)(20,0){2}{\usebox\vertex}
\put(0,0){\oval(90,30)[t]} \put(-50,0){\line(1,0){5}} \put(50,0){\line(-1,0){5}}
\put(-50,0){\usebox\vertex} \put(50,0){\usebox\vertex}
\put(-7,-32){$2n$}

\put(200,0){\oval(100,40)}
\multiput(180,-20)(20,0){3}{\usebox\vertex}
\put(200,20){\usebox\vertex}
\put(210,30){\usebox\vertex}
\qbezier(200,20)(200,50)(210,30)
\qbezier(200,20)(210,0)(210,30)
\put(185,-32){$2n + 1$}
\end{picture}

\noindent
the odd dessin is very easy -- a glance at formulas (9.3) from \cite{AnoMKhR}
reveals simple connection to the formula for necklace graph

\begin{picture}(300,80)(-50,-40)
\savebox{\vertex}(0,0)[bl]{
  \put(0,0){\circle*{4}}}
\put(0,0){\oval(100,40)}
\multiput(-10,-20)(20,0){2}{\usebox\vertex}
\put(0,0){\oval(90,30)[t]} \put(-50,0){\line(1,0){5}} \put(50,0){\line(-1,0){5}}
\put(-50,0){\usebox\vertex} \put(50,0){\usebox\vertex}
\put(-10,-30){$2n$}
\put(70,0){$= [2] \cdot$}
\put(120,0){\circle{40}}
\put(105,-30){$2n + 2$}
\put(100,0){\usebox\vertex} \put(120,20){\usebox\vertex}
\put(120,-20){\usebox\vertex}\put(140,0){\usebox\vertex}
\end{picture}

The graph for even twist knots is more tricky -- we may relate it to values
of other graphs in two ways

\begin{picture}(300,80)(-100,-40)
\savebox{\vertex}(0,0)[bl]{
  \put(0,0){\circle*{4}}}
\put(000,0){\oval(100,40)}
\multiput(-20,-20)(20,0){3}{\usebox\vertex}
\put(0,20){\usebox\vertex}
\put(10,30){\usebox\vertex}
\qbezier(0,20)(0,50)(10,30)
\qbezier(0,20)(10,0)(10,30)
\put(-10,-30){$2n + 1$}
\put(70,0){$= [N-2] \cdot$}
\put(140,0){\circle{40}}
\put(125,-30){$2n + 2$}
\put(120,0){\usebox\vertex} \put(140,20){\usebox\vertex}
\put(140,-20){\usebox\vertex}\put(160,0){\usebox\vertex}
\put(170,0){$+$}
\multiput(180,-20)(5,10){5}{\usebox\vertex}
\multiput(180,-20)(5,10){4}{\put(0,0){\line(1,2){5}}}
\put(180,-30){$2n + 3$}
\end{picture}

or

\begin{picture}(300,80)(-100,-40)
\savebox{\vertex}(0,0)[bl]{
  \put(0,0){\circle*{4}}}
\put(0,0){\oval(100,40)}
\multiput(-20,-20)(20,0){3}{\usebox\vertex}
\put(0,20){\usebox\vertex}
\put(10,30){\usebox\vertex}
\qbezier(0,20)(0,50)(10,30)
\qbezier(0,20)(10,0)(10,30)
\put(-10,-30){$2n + 1$}
\put(70,0){$= [N-2] \cdot$}
\put(130,0){\circle{20}} \put(120,0){\usebox\vertex} \put(140,0){\usebox\vertex}
\put(140,0){\line(1,0){10}}
\multiput(150,0)(10,0){2}{\put(0,0){\line(1,0){10}} \put(10,0){\usebox\vertex}}
\put(160,-10){$2n$}
\put(180,0){$+$}

\put(250,0){\oval(100,40)}
\multiput(250,-20)(20,0){1}{\usebox\vertex}
\put(250,20){\usebox\vertex}
\put(260,30){\usebox\vertex}
\qbezier(250,20)(250,50)(260,30)
\qbezier(250,20)(260,0)(260,30)
\put(240,-30){$2n - 1$}
\end{picture}

\subsubsection{Pretzel knots/links}

All parallel:  even (number of vertices) cycle with edges of arbitrary multiplicity

\be
\begin{picture}(300,130)(-150,-53)
\put(-235,-20){\line(0,1){60}}
\put(-220,-20){\line(0,1){60}}
\put(-235,-20){\line(1,0){15}}
\put(-235,40){\line(1,0){15}}
\put(-232,5){\mbox{$n_1$}}
\put(-195,-20){\line(0,1){60}}
\put(-180,-20){\line(0,1){60}}
\put(-195,-20){\line(1,0){15}}
\put(-195,40){\line(1,0){15}}
\put(-192,5){\mbox{$n_2$}}
\put(-95,-20){\line(0,1){60}}
\put(-80,-20){\line(0,1){60}}
\put(-95,-20){\line(1,0){15}}
\put(-95,40){\line(1,0){15}}
\put(-91,5){\mbox{$n_{g+1}$}}
\qbezier(-85,-20)(-85,-50)(-120,-50)          \qbezier(-230,-20)(-230,-50)(-160,-50)
\qbezier(-225,-20)(-207.5,-50)(-190,-20)
\qbezier(-90,-20)(-100,-50)(-120,-30)  \qbezier(-185,-20)(-170,-50)(-160,-30)
\qbezier(-85,40)(-85,70)(-120,70)          \qbezier(-230,40)(-230,70)(-160,70)
\qbezier(-225,40)(-207.5,70)(-190,40)
\qbezier(-90,40)(-100,70)(-120,50)  \qbezier(-185,40)(-170,70)(-160,50)
\put(-120,-50){\vector(-1,0){2}}
\put(-207,-35){\vector(-1,0){2}}
\put(-105,-37.5){\vector(-1,0){2}}
\put(-168,-38){\vector(1,0){2}}
\put(-145,5){\mbox{$\ldots$}}
\put(-60,0) {
  \put(0,0){\line(1,0){50}}
  \put(60,-10){\mbox{$\ldots$}}
  \put(80,0){\line(1,0){50}}
  \put(0,0){\line(2,1){65}}
  \put(130,0){\line(-2,1){65}}
  \put(0,0){\circle*{4}} \put(40,0){\circle*{4}}
  \put(90,0){\circle*{4}} \put(130,0){\circle*{4}}
  \put(65,32.5){\circle*{4}}
  \qbezier(0,0)(0,20)(65,32.5) \put(20,17){\mbox{$\ldots$}}  \qbezier(0,0)(40,10)(65,32.5)
  \qbezier(130,0)(130,20)(65,32.5) \put(100,17){\mbox{$\ldots$}}  \qbezier(130,0)(90,10)(65,32.5)
  \put(15,-7){\mbox{$\ldots$}} \put(105,-7){\mbox{$\ldots$}}
  \qbezier(0,0)(20,-30)(40,0)
  \qbezier(40,0)(45,-8)(50,-10)
  \qbezier(90,0)(110,-30)(130,0)
  \qbezier(90,0)(85,-8)(80,-10)
}
\put(80,-2){\mbox{$= [2]^{n_1 + \dots n_{g+1} - (g + 1)}$}}
\savebox{\vertex}(0,0)[bl]{
  \put(0,0){\circle*{4}}}
\put(200,60) {
  \put(0,0){\circle{40}}
  \put(-15,-30){$g + 1$}
  \put(-20,0){\usebox\vertex} \put(0,20){\usebox\vertex}
  \put(0,-20){\usebox\vertex}\put(20,0){\usebox\vertex}
}
\end{picture} \\ \nn
\mathop{=}_{\text{non-virtual}} [2]^{n_1 + \dots n_{g+1} - (g + 1)} \left ( [N-1]^{g+1} - 1 + [N]^2 \right )
\ee

\noindent
For ordinary (non-virtual) knots
the number of vertices $g+1$ ($g$ being the genus of the knot's surface) is even, i.e. the cycle has even length.
Twist knots $(2g-2)_1$ are particular case when all $n_1=\ldots=n_g=1$
and $n_{g+1}=2$.

All antiparallel: two vertices connected by $g+1$ simple lines

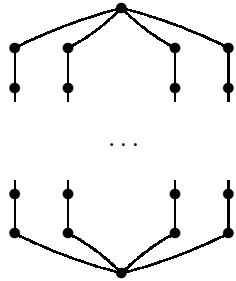
\begin{picture}(300,120)(-150,-10)
\put(0,0){\circle*{4}} \put(0,100){\circle*{4}}
\put(-40,15){\circle*{4}} \put(-40,30){\circle*{4}} \put(-40,70){\circle*{4}} \put(-40,85){\circle*{4}}
\put(-20,15){\circle*{4}} \put(-20,30){\circle*{4}} \put(-20,70){\circle*{4}} \put(-20,85){\circle*{4}}
\put(-5,48){\mbox{\ldots}}
\put(20,15){\circle*{4}} \put(20,30){\circle*{4}} \put(20,70){\circle*{4}} \put(20,85){\circle*{4}}
\put(40,15){\circle*{4}} \put(40,30){\circle*{4}} \put(40,70){\circle*{4}} \put(40,85){\circle*{4}}
\qbezier(-40,15)(-20,5)(0,0) \qbezier(40,15)(20,5)(0,0)
\qbezier(-40,85)(-20,95)(0,100) \qbezier(40,85)(20,95)(0,100)
\qbezier(-20,15)(-10,10)(0,0) \qbezier(20,15)(10,10)(0,0)
\qbezier(-20,85)(-10,90)(0,100) \qbezier(20,85)(10,90)(0,100)
\put(-40,15){\line(0,1){20}}
\put(-20,15){\line(0,1){20}}
\put(20,15){\line(0,1){20}}
\put(40,15){\line(0,1){20}}
\put(-40,85){\line(0,-1){20}}
\put(-20,85){\line(0,-1){20}}
\put(20,85){\line(0,-1){20}}
\put(40,85){\line(0,-1){20}}
\end{picture}

\noindent
For ordinary (non-virtual) knots/links $n_i$'s are either all odd
or all even;
$2$-strand torus knot/link arises in particular case of all $n_i=1$.

\paragraph{``all-even'' case}
The answers for first few $g$'s are
\be
f_0(x_1) & = [N] x_1 \\ \nn
f_1(x_1, x_2) & = x_1 x_2 + y \\ \nn
f_2(x_1, x_2, x_3) & = \frac{1}{[N]} \left ( x_1 x_2 x_3 + y (x_1 + x_2 + x_3) + y^2(1-1/y) \right ) \\ \nn
f_3(x_1, x_2, x_3, x_4) & = \frac{1}{[N]^2} \left ( x_1 x_2 x_3 x_4 + y \sum_{i < j} x_i x_j
+ y^2(1-1/y) \sum_i x_i + y^3(1 - 1/y + 1/y^2) \right ) \\ \nn
f_4(x_1, \dots , x_5) & = \frac{1}{[N]^3} \left ( x_1 x_2 x_3 x_4 x_5 + y \sum_{i < j < k} x_i x_j x_k
+ y^2(1 - \frac{1}{y}) \sum_{i < j} x_i x_j \right. \\ \nn
& + \left. y^3(1 - \frac{1}{y} + \frac{1}{y^2}) \sum_i x_i
+ y^4(1 - \frac{1}{y} + \frac{1}{y^2} - \frac{1}{y^3}) \right ),
\ee
where $x_i$ = $[N-1]^{n_i}$ and $y = [N-1][N+1]$. We hope that general answer
is clear from these examples (it's written in terms of $q$-numbers with $y$ playing role of $q$).

\paragraph{``all-odd'' case}
Again, answers for first few $g$'s are
\be
f_0(x_1) & = [N] x_1 \\ \nn
f_1(x_1, x_2) & = x_1 x_2 + y \\ \nn
f_2(x_1, x_2, x_3) & = \frac{1}{[N]} \left ( x_1 x_2 x_3 + y (x_1 + x_2 + x_3)
+ y^2 \left (\frac{1}{[N-1]}-\frac{1}{[N+1]} \right ) \right ) \\ \nn
f_3(x_1, x_2, x_3, x_4) & = \frac{1}{[N]^2} \left ( x_1 x_2 x_3 x_4 + y \sum_{i < j} x_i x_j
+ y^2 \left (\frac{1}{[N-1]}-\frac{1}{[N+1]} \right ) \sum_i x_i
+ y^3 \left (\frac{1}{[N-1]^2} - \frac{1}{[N-1][N+1]} + \frac{1}{[N+1]^2} \right ) \right ) \\ \nn
\ee
and the general formula should be pretty clear from them.

\paragraph{Pretzel genus $g=2$:} two vertices, connected by two multiple edges and by a simple odd line,
winded between them.

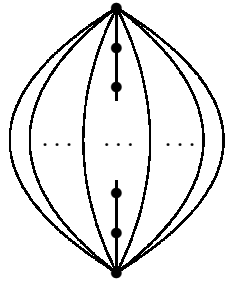
\begin{picture}(300,120)(-150,-10)
\put(0,0){\circle*{4}} \put(0,100){\circle*{4}}
\put(0,15){\circle*{4}} \put(0,30){\circle*{4}} \put(0,70){\circle*{4}} \put(0,85){\circle*{4}}
\put(-5,48){\mbox{\ldots}}
\put(0,0){\line(0,1){35}}  \put(0,100){\line(0,-1){35}}
\qbezier(0,0)(-80,50)(0,100)
\qbezier(0,0)(-65,50)(0,100)
\put(-28,48){\mbox{$\ldots$}}
\qbezier(0,0)(-25,50)(0,100)
\qbezier(0,0)(80,50)(0,100)
\qbezier(0,0)(65,50)(0,100)
\put(18,48){\mbox{$\ldots$}}
\qbezier(0,0)(25,50)(0,100)
\put(0,50){\put(80,-2){\mbox{$= [2]^{n_1 + n_3 - 1} \left ([N-1]^{n_2} + [N+1] \right ) [N-1]$}}}
\end{picture}

\noindent
Multiplicities of left multi-edge and right multi-edge are denoted by $n_1$ and $n_3$ respectively,
and the number of edges in antiparallel central odd line by $n_2$.

\subsection{Miscellaneous}

\subsubsection{Cycles}

Cycles with even and odd number of vertices are equal, respectively
\be
\begin{picture}(50,50)(-25,-2)
  \savebox{\vertex}(0,0)[bl]{
    \put(0,0){\circle*{4}}}
  \put(0,0){\circle{40}}
  \put(-5,-30){$2 k$}
  \put(-20,0){\usebox\vertex} \put(0,20){\usebox\vertex}
  \put(0,-20){\usebox\vertex}\put(20,0){\usebox\vertex}
\end{picture} = [N-1]^{2k} + [N+1][N-1] \\ \nn
\begin{picture}(50,50)(-25,-2)
  \savebox{\vertex}(0,0)[bl]{
    \put(0,0){\circle*{4}}}
  \put(0,0){\circle{40}}
  \put(-15,-30){$2 k + 1$}
  \put(0,20){\usebox\vertex} \put(-17,-10){\usebox\vertex} \put(17,-10){\usebox\vertex}
\end{picture} = [N-1]^{2k+1} - [N-1] - [N][N-1] \\ \nn
\ee

\subsubsection{Simple lines}

Simple lines of 2-valent vertices, with even and odd number of internal vertices
(denoted by rectangle) are equal, respectively

\be
\begin{picture}(50,50)(-25,-2)
  \qbezier(-10,-10)(-10,0)(-10,10) \qbezier(-10,10)(0,10)(10,10)
  \qbezier(10,-10)(10,0)(10,10) \qbezier(-10,-10)(0,-10)(10,-10)
  \put(-5,-2){\mbox{$2 k$}}
  \thicklines
  \put(-7,-20){\qbezier(0,0)(0,5)(0,10)} \put(7,-20){\qbezier(0,0)(0,5)(0,10)}
  \put(-7,10){\qbezier(0,0)(0,5)(0,10)} \put(7,10){\qbezier(0,0)(0,5)(0,10)}
  \put(-7,-12){\vector(0,1){0}} \put(7,-20){\vector(0,-1){0}}
  \put(-7,20){\vector(0,1){0}} \put(7,12){\vector(0,-1){0}}
\end{picture}
= \frac{1}{[N]} \left ([N-1]^{2k} - 1 \right )
\begin{picture}(20,50)(-10,-2)
  \thicklines
  \qbezier(-10,-20)(0,-10)(10,-20) \qbezier(-10,20)(0,10)(10,20)
  \put(4,-15){\vector(1,0){0}} \put(-4,15){\vector(-1,0){0}}
\end{picture}
\ \ + \ \
\begin{picture}(30,50)(-15,-2)
  \thicklines
  \qbezier(-10,-20)(-10,0)(-10,20) \qbezier(10,-20)(10,0)(10,20)
  \put(-10,5){\vector(0,1){0}} \put(10,-5){\vector(0,-1){0}}
\end{picture}
\\ \nn
\begin{picture}(50,50)(-25,-2)
  \qbezier(-15,-10)(-15,0)(-15,10) \qbezier(-15,10)(0,10)(15,10)
  \qbezier(15,-10)(15,0)(15,10) \qbezier(-15,-10)(0,-10)(15,-10)
  \put(-12,-2){\small \mbox{$2 k + 1$}}
  \thicklines
  \put(-7,-20){\qbezier(0,0)(0,5)(0,10)} \put(7,-20){\qbezier(0,0)(0,5)(0,10)}
  \put(-7,10){\qbezier(0,0)(0,5)(0,10)} \put(7,10){\qbezier(0,0)(0,5)(0,10)}
  \put(-7,-12){\vector(0,1){0}} \put(7,-20){\vector(0,-1){0}}
  \put(7,20){\vector(0,1){0}} \put(-7,12){\vector(0,-1){0}}
\end{picture}
= \frac{1}{[N]} \left ([N-1]^{2k+1} - [N-1] \right )
\begin{picture}(20,50)(-10,-2)
  \thicklines
  \qbezier(-10,-20)(0,-10)(10,-20) \qbezier(-10,20)(0,10)(10,20)
  \put(4,-15){\vector(1,0){0}} \put(4,15){\vector(1,0){0}}
\end{picture}
\ \ +\ \
\begin{picture}(20,50)(-10,-2)
  \thicklines
  \qbezier(-10,-20)(0,-10)(10,-20) \qbezier(-10,20)(0,10)(10,20)
  \put(4,-15){\vector(1,0){0}} \put(4,15){\vector(1,0){0}}
  \thinlines
  \qbezier(0,-15)(0,0)(0,15)
\end{picture}
\ee

\bigskip

\bigskip

\section*{Acknowledgements}

We thank Yegor Zenkevich for stimulating discussions
and especially for the idea to look at higher antisymmetric projectors.
This work was performed at the
Institute for Information Transmission Problems with the financial support of the Russian Science
Foundation (Grant No.14-50-00150).

\end{document}